\documentclass[a4paper,10pt,twocolumn, superscriptaddress, nofootinbib, amsmath,amssymb, aps, prd, floatfix]{revtex4-2}

\usepackage{graphicx}
\usepackage{amsfonts}
\usepackage{amsmath}
\usepackage{hyperref}
\usepackage{float}
\usepackage{amssymb,amsbsy}
\usepackage{epstopdf}
\usepackage{color}
\usepackage[normalem]{ulem}
\usepackage{braket}
\usepackage{combelow}
\usepackage{xhfill}% http://ctan.org/pkg/xhfill

\def\bk{{\mathbf{k}}}
\def\feq{f_{0\bk }}

\begin{document}

\title{Relativistic second-order dissipative and anisotropic fluid dynamics in the relaxation-time approximation for an ideal gas of massive particles}

\author{Victor E. Ambru\cb{s}}
\affiliation{Department of Physics, West University of Timi\cb{s}oara, \\
	Bd.~Vasile P\^arvan 4, Timi\cb{s}oara 300223, Romania}

\author{Etele Moln\'ar}
\affiliation{Department of Physics, West University of Timi\cb{s}oara, \\
	Bd.~Vasile P\^arvan 4, Timi\cb{s}oara 300223, Romania}
\affiliation{Incubator of Scientiﬁc Excellence--Centre for Simulations of Superdense Fluids,\\
	University of Wroc{\l}aw, pl. M. Borna 9, PL-50204 Wroc{\l}aw, Poland}
\affiliation{Institut f\"ur Theoretische Physik,
	Johann Wolfgang Goethe--Universit\"at,
	Max-von-Laue-Str.\ 1, D--60438 Frankfurt am Main, Germany}

\author{Dirk H.\ Rischke}
\affiliation{Institut f\"ur Theoretische Physik,
	Johann Wolfgang Goethe--Universit\"at,
	Max-von-Laue-Str.\ 1, D--60438 Frankfurt am Main, Germany}
\affiliation{Helmholtz Research Academy Hesse for FAIR, Campus Riedberg,\\
	Max-von-Laue-Str.~12, D-60438 Frankfurt am Main, Germany}

\begin{abstract}
{
In this paper, we study all transport coefficients of second-order dissipative fluid dynamics derived by 
V. E. Ambrus et al. [Phys. Rev. D 106, 076005 (2022)] from the relativistic Boltzmann equation in the relaxation-time approximation for the collision integral.
These transport coefficients are computed for a classical ideal gas of massive particles, with and without taking into account the conservation of intrinsic quantum numbers.
Through rigorous comparison between kinetic theory, second-order dissipative fluid dynamics, and leading-order anisotropic fluid dynamics for a (0+1)--dimensional boost-invariant flow scenario, we show that both fluid-dynamical theories describe the early far-from-equilibrium stage of the expansion reasonably well.
}
\end{abstract}

%\date{\today } 19.03.2024
\maketitle

%%%
\section{Introduction}

During the last decades, relativistic fluid dynamics has assumed an important role in describing the space-time evolution of matter created in ultrarelativistic heavy-ion collisions, in binary mergers of neutron stars, as well as in the early Universe \cite{Rezzolla.2013}.
Relativistic fluid dynamics is an effective field theory based on the local conservation of energy and momentum, $\partial_\mu T^{\mu\nu} = 0$, where $T^{\mu \nu}$ is the energy-momentum tensor of the fluid, and of multiple conserved charges, e.g., electric charge, baryon number, strangeness, etc.,
$\partial_\mu N_i^\mu =0$, where $N_i^\mu$ is the four-current associated with the $i$th charge.

For the sake of simplicity, in this paper we will consider a single species of particle of rest mass $m_0$.
In this case, there is at most one independent conserved charge, to which -- in a slight abuse of notation -- we refer to as ``particle number'' in the following.
Therefore, the 5 conservation equations contain in general $14$ dynamical fields, 5 of which occur for dissipative as well as for ideal fluids: the particle-number density $n$, the energy density $e$, and the fluid four-velocity $u^\mu$, chosen for instance as the time-like eigenvector of the energy-momentum tensor.
The pressure $P$ is not an independent field, as it is given by an equation of state, $P(e,n)$, for the matter under consideration.
For an ideal fluid, i.e., a fluid in local thermodynamical equilibrium, the 5 conservation equations contain 5 dynamical fields and are thus closed.
For dissipative fluids, however, there are $9$ additional fields that account for irreversible processes: the bulk viscous pressure $\Pi$, the particle diffusion current $V^\mu$, and the shear-stress tensor $\pi^{\mu\nu}$.
In order to close the system of equations of motion, additional equations, sometimes called \textit{constitutive relations}, have to be specified.
The simplest example is Navier-Stokes theory, where $\Pi$, $V^\mu$, and $\pi^{\mu \nu}$ are proportional to first-order gradients of $e$, $n$, and $u^\mu$, and which therefore belongs to the class of so-called first-order fluid-dynamical theories.
The proportionality coefficients are the 3 first-order transport coefficients related to different nonequilibrium transport phenomena: the bulk-viscosity coefficient $\zeta$, the particle-diffusion coefficient $\kappa$, and the shear-viscosity coefficient $\eta$.
Relativistic Navier-Stokes theory is, however, acausal and unstable \cite{Hiscock.1985,Denicol:2008ha,Pu:2009fj}.
One way to cure this problem is to derive fluid dynamics from the relativistic Boltzmann equation, applying Grad's method of moments \cite{Grad:1949}, leading to the 14-moment approximation of Israel and Stewart for relativistic systems \cite{Israel.1979}.

The moment equations up to tensor-rank 2 are truncated based on a power-counting scheme in Knudsen and inverse Reynolds numbers \cite{Denicol:2012cn}.
The Knudsen number $\textrm{Kn}$ is the ratio of the particle mean free path and a characteristic macroscopic scale, while the inverse Reynolds number $\mathrm{Re}^{-1}$ is the ratio of an out-of-equilibrium and a local-equilibrium macroscopic field.
The resulting equations of motion contain terms up to second-order in Knudsen and/or inverse Reynolds numbers.

In these so-called second-order theories of relativistic dissipative fluid dynamics, dynamical equations of motion for the dissipative fields provide closure for the conservation laws and, under certain conditions, also ensure causality and stability \cite{Denicol:2008ha,Pu:2009fj,Bemfica:2020xym}.
These equations of motion are of relaxation type, i.e., the dissipative fields relax onto their respective values given by first-order Navier-Stokes theory on certain time scales.
In addition to the first-order transport coefficients $\zeta$, $\kappa$, and $\eta$, these relaxation equations contain  additional second-order transport coefficients: $5$ in the equation for bulk viscous pressure, $8$ in the equation for the particle diffusion current, and $6$ in the equation for the shear-stress tensor.

In this paper, we will study all first- and second-order transport coefficients appearing in relativistic second-order dissipative fluid dynamics with $14$ dynamical moments obtained from the Boltzmann equation in the Anderson-Witting relaxation-time approximation (RTA) of the collision integral \cite{Anderson:1974a}, using the method of moments coined DNMR~\cite{Denicol:2012cn}.
These transport coefficients explicitly depend on the approximations made for the moments which are nondynamical and lie outside of the truncation.
In order to compare the magnitude of various terms and the coefficients accompanying them we assume that the Knudsen and the inverse Reynolds numbers are of the same magnitude, $\mathrm{Kn} \sim \mathrm{Re}^{-1}$, as for instance is the case in Navier-Stokes theory.
This has lead to the so-called order-of-magnitude approximation, where nondynamical moments are replaced by dynamical moments of order $\mathrm{Re}^{-1}$~\cite{Struchtrup.2004,Molnar.2014,Fotakis:2022usk,Wagner:2022ayd,Ambrus:2022vif}.

When approximating the collision term in RTA all equations of motion up to tensor-rank 2 contain first-order terms $\sim O(\mathrm{Kn})$ or $\sim O(\mathrm{Re}^{-1})$, while all second-order terms in these equations are of order $O(\mathrm{Kn}\mathrm{Re}^{-1})$; i.e., there are no terms of order $O(\mathrm{Kn}^2)$ or $O(\mathrm{Re}^{-2})$.
Moreover, the diagonal nature of the RTA collision term allows the negative-order nondynamical moments to be represented by dynamical moments of order $\mathrm{Re}^{-1}$, without reference to a specific basis of moments.
This leads to the so-called basis-free (BF) approximation of Ref.~\cite{Ambrus:2022vif}, which we also adopt here.

For the sake of completeness here we will review and compare both the BF and the standard DNMR approximations for the transport coefficients for a classical ideal gas of massive particles.
Furthermore, with these new results we will also inspect the evolution of the bulk viscous and shear-stress pressure components, including all cross-coupling coefficients, the so-called bulk-shear coupling in second-order fluid dynamics.
Earlier studies \cite{Denicol:2014vaa} only computed the second-order coefficients of the bulk viscous pressure without taking into account the consequences of particle-number conservation and particle diffusion.
Similarly, the effects of bulk-shear coupling was discussed for nonconformal fluids without explicit particle-number conservation in Ref.~\cite{Denicol:2014mca}.
Therefore, here we also aim to fill these gaps and study all transport coefficients both with and without explicit particle-number conservation in both the BF and the DNMR approximations.

For systems with large initial momentum anisotropy, the framework of anisotropic fluid dynamics was recently developed~\cite{Florkowski:2010cf,Ryblewski:2010bs,Ryblewski:2012rr,Martinez:2010sc,Martinez:2010sd,Martinez:2012tu,Bazow:2013ifa,Bazow:2015cha,Tinti:2013vba,Tinti:2015xwa,Molnar:2016vvu,Molnar:2016gwq,Alqahtani:2017mhy}.
This framework implicitly includes the bulk viscous and shear-stress viscous pressure components; hence we will also study the bulk-shear coupling in leading-order anisotropic fluid dynamics.
The results are also compared to an exact solution of Boltzmann equation in the context of a (0+1)--dimensional boost-invariant expansion both with and without explicit particle-number conservation.

This paper is organized as follows. For reasons of completeness, in Sec.~\ref{sec:fluid_dynamics_general} we review the DNMR method of moments to obtain the equations of relativistic dissipative second-order fluid dynamics in the $14$-moment approximation from the Boltzmann equation.
The first- and second-order transport coefficients are listed in both the BF and the DNMR approximations in Sec.~\ref{sec:with_particle_conservation}.
In Sec.~\ref{sec:without_particle_conservation}, we obtain the transport coefficients in case there are no conserved charges, while for the purpose of supplementing Ref.~\cite{Ambrus:2022vif}, in Sec.~\ref{sec:RTA:MHD} we also list the transport coefficients of magnetohydrodynamics for a massive ideal gas.
Next, in Sec.~\ref{sec::Anisotropic}, we present the equations of leading-order anisotropic fluid dynamics.
The explicit calculations of the transport coefficients for a classical ideal gas of massive particles are presented in Sec.~\ref{sec:massive_gas}.
Their properties and graphical representations are discussed in Sec. \ref{sec:tcoeffs}.
The methods and applications in the case of a (0+1)--dimensional boost-invariant expansion are described in Sec.~\ref{sec:app}.
The results and comparisons to the exact numerical solution of the Boltzmann equation of second-order fluid dynamics as well as to leading-order anisotropic fluid dynamics are discussed in Sec.~\ref{sec:results}.
We conclude this work in Sec.~\ref{sec:conclusions}.
For reasons of brevity additional computations and useful relations are relegated to the Appendices.

%%%
\subsection*{Notation, conventions, and definitions}

In this paper, we will work in flat space-time with metric tensor $g_{\mu \nu}=\text{diag}(1,-1,-1,-1)$ and adopt natural units $\hbar=c=k_{B}=1$.
The rank-four Levi-Civit\'{a} symbol is $\epsilon^{\mu \nu \alpha \beta} = \pm 1$ 
for $(\mu \nu \alpha \beta)$ an even/odd permutation of (0123), and zero otherwise.
Symmetrization of a tensor $A^{\mu \nu}$ is denoted as 
$A^{(\mu \nu)} \equiv \frac{1}{2}(A^{\mu \nu} + A^{\nu \mu})$, while antisymmetrization is denoted as 
$A^{[\mu \nu]} \equiv \frac{1}{2}(A^{\mu \nu} - A^{\nu \mu})$.
Symmetrization can also be done with respect to more than two indices, in which case the normalization factor has to  be adjusted accordingly, for details see, e.g., Ref.~\cite{Groot.1980}.
The time-like normalized fluid-flow four-velocity is denoted by $u^{\mu}=\gamma (1,\mathbf{v})$, where $\gamma=(1-\mathbf{v}^{2})^{-1/2}$ and $u^{\mu}u_{\mu}= 1$.
The local rest frame (LRF) of the fluid is defined by $u_{\text{LRF}}^{\mu }=(1,\mathbf{0})$.
The projection operator orthogonal to
$u^{\mu}$ is defined as $\Delta^{\mu \nu } \equiv g^{\mu \nu}-u^{\mu}u^{\nu}$.
The symmetric and traceless projection tensors of rank $2\ell$ orthogonal to $u^{\mu}$, $\Delta_{\nu_{1}\cdots \nu_{\ell }}^{\mu_{1}\cdots\mu_{\ell }}$, are constructed using rank-two projection operators $\Delta^{\mu \nu }$, for details see, e.g., Ref.~\cite{denicol_rischke_2022}.
The respective projection of a tensor $A^{\mu_1 \cdots \mu_\ell}$ is
denoted as $A^{\left\langle \mu _{1}\cdots \mu_{\ell}\right\rangle }
\equiv \Delta_{\nu_{1}\cdots \nu_{\ell }}^{\mu_{1}\cdots \mu_{\ell }}A^{\nu_{1}\cdots \nu_{\ell }}$.

The space-like normalized anisotropy four-vector orthogonal to the fluid four-velocity is denoted by $l^{\mu}$, with $u_\mu l^\mu =0$ and $l_\mu l^\mu = -1$.
The projection operator orthogonal to both $u^{\mu}$ and $l^{\mu}$ is denoted by $\Xi ^{\mu \nu}\equiv g^{\mu \nu} - u^{\mu }u^{\nu}+l^{\mu }l^{\nu }$.
The rank $2\ell$ symmetric and traceless projection tensors orthogonal to both $u^{\mu}$ and $l^{\mu}$, $\Xi_{\nu _{1}\cdots \nu_{\ell }}^{\mu_{1}\cdots\mu _{\ell }}$, are constructed using the projection operators $\Xi^{\mu \nu }$ in a similar way as for $\Delta_{\nu_{1}\cdots \nu_{\ell }}^{\mu_{1}\cdots\mu_{\ell }}$, for details see Refs.~\cite{Molnar:2016vvu,denicol_rischke_2022}.
The respective projection of a tensor $A^{\mu_1 \cdots \mu_\ell}$ is denoted as $A^{\left\{ \mu _{1}\cdots \mu_{\ell}\right\} } \equiv \Xi_{\nu_{1}\cdots \nu_{\ell }}^{\mu_{1}\cdots \mu_{\ell }}A^{\nu_{1}\cdots \nu_{\ell }}$.

The  particle four-momentum is $k^{\mu}=(k^0,\mathbf{k})$, where $k^{0}=\sqrt{\mathbf{k}^{2}+m_{0}^{2}}$ is the on-shell energy with the three-momentum $\mathbf{k}$ and the rest mass $m_{0} = \sqrt{k^{\mu }k_{\mu }}$.
The four-momentum is decomposed as $k^{\mu } \equiv E_{\mathbf{k}u}u^{\mu }+k^{\left\langle \mu \right\rangle } = E_{\mathbf{k}u}u^{\mu }+E_{\mathbf{k}l}l^{\mu }+k^{\left\{ \mu \right\} }$, where $E_{\mathbf{k}u}= k^{\mu }u_{\mu }$ is the energy of the particle, while $k^{\left\langle \mu \right\rangle} = \Delta^{\mu \nu} k_{\nu}$ is the momentum orthogonal to the flow velocity.
Furthermore, $E_{\mathbf{k}l}=-k^{\mu }l_{\mu }$ is the particle momentum in the direction of the anisotropy, and $k^{\left\{ \mu \right\} }=\Xi ^{\mu \nu }k_{\nu }$ are the components of the momentum orthogonal to both $u^{\mu }$ and $l^{\mu }$.

The four-gradient is decomposed as $\partial_{\mu}\equiv u_{\mu }D + \nabla_{\mu}$, where $D\equiv u^{\mu}\partial _{\mu }$ is the comoving derivative (sometimes also denoted by an overdot, $DA \equiv \dot{A}$), and  $\nabla_{\mu} \equiv \Delta_{\mu}^{\alpha}\partial_{\alpha} = \partial_{\left\langle \mu \right\rangle}$ is the gradient operator.
Therefore, $\partial_{\mu} u_{\nu}\equiv u_{\mu}D u_{\nu} +\nabla_{\mu}u_{\nu}=u_{\mu}\dot{u}_{\nu} + \frac{1}{3}\theta \Delta_{\mu \nu } +\sigma_{\mu\nu}+\omega_{\mu \nu}$, where $\theta\equiv \nabla_{\mu}u^{\mu }$ is the expansion scalar, $\sigma^{\mu \nu }\equiv\nabla^{\left\langle \mu \right. } u^{\left. \nu \right\rangle }= \nabla^{(\mu}u^{\nu )} -\frac{1}{3}\theta \Delta^{\mu \nu}$ is the shear tensor, and $\omega^{\mu \nu}\equiv \nabla^{\left[ \mu \right.} u^{\left.\nu \right]}$ is the fluid vorticity.
Similarly, one can decompose $\partial_{\mu }$ with respect to $u^{\mu }$, $l^{\mu }$, and $\Xi^{\mu \nu}$, as $\partial _{\mu }=u_{\mu }D+l_{\mu }D_{l}+\tilde{\nabla}_{\mu }$, where $D_{l} = -l^{\mu }\partial _{\mu }$ and $\tilde{\nabla}_{\mu } \equiv \Xi _{\mu \nu }\partial ^{\nu }=\partial_{\left\{ \mu \right\} }$ are gradient operators.

The state of local equilibrium is specified by the J\"uttner distribution \cite{Juttner},
\begin{equation}
	f_{0\mathbf{k}}=\left[ \exp \left( \beta E_{\mathbf{k} u}-\alpha\right) +a\right] ^{-1}\;,  \label{f_0k}
\end{equation}
with $\alpha=\mu\beta$, where $\mu$ is the chemical potential and $\beta=1/T$ is the inverse temperature, while $a=\pm 1$ for fermions/bosons and $a\rightarrow 0$ for Boltzmann particles.
Furthermore, we define the abbreviation $\tilde{f}_{0\bk} \equiv  1 - a f_{0\bk}$.

The equilibrium moments of tensor-rank $n$ of power $i$ in energy $E_{\mathbf{k}u}$ are defined as~\cite{Israel.1979}:
\begin{align} \nonumber
	\mathcal{I}_{i}^{\mu_{1}\cdots \mu_{n}} &\equiv \left\langle E_{\mathbf{k}u}^{i}\,
	k^{\mu_{1}}\cdots k^{\mu_{n}}\right\rangle _{0}
	=\sum_{q=0}^{\lfloor n/2\rfloor}\left( -1\right)^{q}\, b_{nq}\, I_{i+n,q}\, \\
	&\times \Delta ^{\left( \mu _{1}\mu _{2}\right.}\cdots \Delta^{\mu _{2q-1}\mu _{2q}}
	u^{\mu _{2q+1}}\cdots u^{\left. \mu_{n}\right) } \;,  \label{I_r_tens}
\end{align}
where the angular brackets denote the momentum-space integrals $\left\langle \cdots  \right\rangle _0 \equiv \int \mathrm{d}K \cdots f_{0\mathbf{k}}$ over the local-equilibrium distribution.
Here, $\mathrm{d}K \equiv g \mathrm{d}^{3}\mathbf{k}/[(2\pi )^{3}k^{0}]$ is the Lorentz-invariant measure, while $g$ is the degeneracy factor of a momentum state.

The equilibrium moments~\eqref{I_r_tens} were expanded with the help of $u^{\mu }$, $\Delta^{\mu \nu }$, and the thermodynamic integrals $I_{i+n,q}$, where $n$ and $q$ are natural numbers, while the sum runs up to $\lfloor n/2 \rfloor$ denoting the largest integer which is less than or equal to $n/2$.
The total number of symmetrized tensors $\Delta^(\cdots  u^)$ is given by $b_{nq} = \frac{n!}{2^{q}q!\left( n-2q\right) !} $, while
\begin{equation}
	I_{nq}\left( \alpha,\beta \right) \equiv \frac{\left( -1\right) ^{q}}{\left( 2q+1\right) !!}
	\left\langle E_{\mathbf{k}u}^{n-2q}\left( \Delta^{\alpha \beta }k_{\alpha }
	k_{\beta }\right)^{q}\right\rangle_{0}\;,
	\label{I_nq}
\end{equation}
where $(2q+1)!! = (2q+1)!/(2^q q!)$ is the double factorial of odd numbers.

The derivatives of $I_{nq}$ with respect to $\alpha$ and $\beta$ lead to auxiliary thermodynamic integrals,
\begin{align}
	J_{nq}\left( \alpha,\beta \right) &\equiv \left( \frac{\partial I_{nq}}{\partial \alpha}\right)_{\beta}
	= -\left(\frac{ \partial I_{n-1,q}}{\partial \beta}\right)_\alpha \nonumber\\
	& = \frac{(-1)^q}{(2q+1)!!}
	\int \mathrm{d}K\, E_{\mathbf{k}u}^{n-2q}
	\left( \Delta ^{\alpha \beta }k_{\alpha }k_{\beta}\right)^{q}f_{0\mathbf{k}} \tilde{f}_{0\bk} \nonumber\\
	&= \beta^{-1} \left[ I_{n-1,q-1} + \left(n-2q\right) I_{n-1,q} \right] \;. \label{J_nq}
\end{align}

Following Refs.~\cite{Molnar:2016vvu,Molnar:2016gwq} the local distribution function of an anisotropic state as a function of $\hat{\alpha}$, $\hat{\beta}_{u}$, and $\hat{\beta}_{l}$, as well as of $E_{\mathbf{k}u}$ and $E_{\mathbf{k}l}$, is denoted by $\hat{f}_{0\mathbf{k}}\left( \hat{\alpha},\hat{\beta}_{u} E_{\mathbf{k}u},\hat{\beta}_{l} E_{\mathbf{k}l}\right)$.
In the limit of vanishing anisotropy parameter $\hat{\beta}_{l}$ the anisotropic distribution converges to the distribution function in local equilibrium,
\begin{equation}
	\lim_{\hat{\beta}_{l}\rightarrow 0}\hat{f}_{0\mathbf{k}}\left( \hat{\alpha},\hat{\beta}_{u}E_{\mathbf{k}u},
	\hat{\beta}_{l}E_{\mathbf{k}l}\right) =f_{0\mathbf{k}}\left( \hat{\alpha},\hat{\beta}_{u}E_{\mathbf{k}u}\right) \;.
	\label{hat_f->f_0}
\end{equation}

In analogy to the equilibrium moments \eqref{I_r_tens}, the moments of tensor
rank $n$ of the anisotropic distribution function $\hat{f}_{0\mathbf{k}}$ are defined 
as~\cite{Molnar:2016vvu}
\begin{align}
	\lefteqn{\hat{\mathcal{I}}_{ij}^{\mu _{1}\cdots \mu _{n}} \equiv \left\langle E_{\mathbf{k}u}^{i}E_{\mathbf{k}l}^{j}
	k^{\mu _{1}}\cdots k^{\mu _{n}}\right\rangle _{\hat{0}}} \notag \\
	& =\sum_{q=0}^{\left[ n/2\right] }\sum_{r=0}^{n-2q}\left( -1\right)^{q}b_{nrq}  \hat{I}_{i+j+n,j+r,q} 	\notag \\
	&\times \Xi^{\left( \mu _{1}\mu _{2}\right. }\cdots
	\Xi ^{\mu_{2q-1}\mu_{2q}}  l^{\mu _{2q+1}}\cdots l^{\mu _{2q+r}}
	u^{\mu_{2q+r+1}}\cdots u^{\left. \mu _{n}\right) } \;, \label{I_ij_tens}
\end{align}
where $\left\langle \cdots \right\rangle _{\hat{0}} \equiv \int \mathrm{d}K\left( \cdots\right) \hat{f}_{0\mathbf{k}}$, and the number of permutations leading to the symmetrized tensors $\Xi ^{\left( \right. }\cdots l\cdots u^{\left. \right) }$ is $b_{nrq}=n!\left( 2q-1\right)!!/[\left( 2q\right)! r ! \left( n-2q-r\right) !]$.
The anisotropic thermodynamic integrals $\hat{I}_{nrq}$  are defined as
\begin{equation}
	\hat{I}_{nrq} \left(\hat{\alpha},\hat{\beta}_{u},\hat{\beta}_{l} \right) =\frac{\left( -1\right) ^{q}}{\left( 2q\right) !!}\left\langle
	E_{\mathbf{k}u}^{n-r-2q}E_{\mathbf{k}l}^{r}\left( \Xi ^{\mu \nu }k_{\mu
	}k_{\nu }\right) ^{q}\right\rangle _{\hat{0}} \;,  \label{I_nrq}
\end{equation}
where $\left( 2q\right) !!=2^{q} q !$ is the double factorial of even numbers.

%%%
\section{Fluid dynamics from the Boltzmann equation}
\label{sec:fluid_dynamics_general}

The space-time evolution of the single-particle distribution function without external forces is given by the relativistic Boltzmann equation~\cite{Groot.1980,Cercignani_book},
\begin{equation}
	k^{\mu }\partial_{\mu } f_{\mathbf{k}}= C[f] ,
	\label{BTE}
\end{equation}
where $C[f]$ is the collision term representing the interaction among particles through collisions.
The collision term is a nonlinear momentum integral in the single-particle distribution function with only a few analytical solutions known in the linearized regime~\cite{Bazow:2016oky,Denicol:2022bsq,Wagner:2023joq}.

For the sake of simplicity, from here on we will use the Anderson-Witting relaxation-time approximation (RTA) for the linearized collision integral \cite{Anderson:1974a,Groot.1980,Cercignani_book},
\begin{equation}
	C[f] \equiv -\frac{E_{\mathbf{k}u}}{\tau_R} \delta f_\bk \;,
	\label{AW}
\end{equation}
where the deviation of the single-particle distribution from local equilibrium is defined as
\begin{equation}
	\delta f_\bk = f_\bk - \feq \;,
\end{equation}
and where the relaxation time $\tau_R \equiv \tau_R(x^{\mu})$ is a momentum-independent parameter proportional to the mean free time between collisions.
The relaxation time allows to introduce a power-counting scheme in terms of the Knudsen number ${\rm Kn} \equiv \tau_R/L$, where $L$ is a typical fluid-dynamical length or time scale, for the derivation of fluid dynamics from the equations of motion for the irreducible moments
\begin{equation}
	\rho_{r}^{\mu _{1}\cdots \mu _{\ell}}\equiv \left\langle
	E_{\mathbf{k} u}^{r}k^{\left\langle \mu_{1}\right. }\cdots
	k^{\left. \mu_{\ell}\right\rangle}\right\rangle_{\delta} \;,
	\label{rho_r_general}
\end{equation}
where $\langle \cdots \rangle_\delta \equiv \int \mathrm{d}K \cdots \delta f_\bk$, and $r$ denotes the power of energy, while the irreducible tensors, $k^{\left\langle \mu _{1}\right. }\cdots k^{\left. \mu _{\ell}\right\rangle } =\Delta_{\nu_{1}\cdots \nu_{\ell }}^{\mu_{1}\cdots \mu_{\ell }}k^{\nu_{1}}\cdots k^{\nu_{\ell }}$, form an orthogonal basis \cite{Denicol:2012cn,Groot.1980}.

Using the comoving derivative of the irreducible moments, $\dot{\rho}_{r}^{\left\langle \mu _{1}\cdots \mu_{\ell }\right\rangle } \equiv \Delta_{\nu_{1}\cdots \nu _{\ell }}^{\mu_{1}\cdots \mu_{\ell }} D \rho_{r}^{\nu_{1}\cdots \nu_{\ell }}$, the equations of fluid dynamics are derived from the various moments of the Boltzmann equation (\ref{BTE}).
Up to tensor rank $2$, these equations of motion are
\begin{align}
	\dot{\rho}_{r}-C_{r-1}& =\alpha_{r}^{\left( 0\right) }\theta + (\textrm{higher-order terms}) \;, \label{D_rho} \\
	\dot{\rho}_{r}^{\left\langle \mu \right\rangle }
	- C_{r-1}^{\left\langle \mu \right\rangle }& = \alpha_{r}^{\left( 1\right) }\nabla^{\mu }\alpha
	+ (\textrm{higher-order terms}) \;, \label{D_rho_mu} \\
	\dot{\rho}_{r}^{\left\langle \mu \nu \right\rangle } - C_{r-1}^{\left\langle\mu \nu \right\rangle}
		&= 2\alpha _{r}^{\left( 2\right) }\sigma^{\mu \nu } + (\textrm{higher-order terms}) \;,
		\label{D_rho_munu}
\end{align}
where the irreducible moments $C_{r-1}^{\langle \mu_1 \cdots \mu_\ell \rangle} $ of the collision term are computed substituting Eq.~\eqref{AW},
\begin{equation}
	C_{r-1}^{\left\langle \mu _{1}\cdots \mu _{\ell }\right\rangle }
	\equiv \int \! \mathrm{d}K\,E_{\mathbf{k}u}^{r-1}\,k^{\langle \mu_1} \cdots k^{\mu_\ell \rangle}
	C\left[ f\right] = - \frac{1}{\tau_R}
	\rho_r^{\mu_1 \cdots \mu_\ell} \;.
	\label{eq:C_r}
\end{equation}
The higher-order contributions to Eqs.~\eqref{D_rho}--\eqref{D_rho_munu} are found in Eqs.~(35)--(37) of Ref.~\cite{Denicol:2012cn}, but are not listed here for the sake of brevity.
Furthermore, the coefficients $\alpha^{(\ell)}_r$ are defined by, see also Eqs.~(42)--(46) of Ref.~\cite{Denicol:2012cn},
\begin{align}
	\alpha_r^{(0)} &= -\beta J_{r+1,1} - \frac{n}{D_{20}}
	\left(h G_{2r} - G_{3r} \right) \;, \label{alpha0}\\
	\alpha_r^{(1)} &= J_{r+1,1} - \frac{J_{r+2,1}}{h} \;, \label{alpha1}\\
	\alpha_r^{(2)} &= \beta J_{r+3,2} \;, \label{alpha2}
\end{align}
where $h\equiv \left( e+P\right) /n$ is the enthalpy per particle and
\begin{align}
	G_{nm} &= J_{n0} J_{m0} - J_{n-1,0} J_{m+1,0} \;, \label{eq:G_nm}\\
	D_{nq} &= J_{n+1,q} J_{n-1,q} - J_{nq}^2 \;.
	\label{eq:D_nq}
\end{align}

%%%
\subsection{Second-order fluid dynamics with particle-number conservation}
\label{sec:with_particle_conservation}

The conservation laws of fluid dynamics read
\begin{equation}
\partial_{\mu} N^{\mu} = 0 \;, \quad
\partial_{\mu} T^{\mu \nu} = 0 \;, \label{cons_eqs}
\end{equation}
where the particle four-current and energy-momentum tensor are
\begin{align}
	N^{\mu} &\equiv \left\langle k^\mu \right\rangle_0 + \left\langle k^\mu \right\rangle_{\delta}
	= \left(n + \rho_1 \right) u^{\mu} + V^{\mu} \; ,
	\label{N_mu}\\
	T^{\mu \nu} &\equiv \left\langle k^\mu k^\nu\right\rangle_0 + \left\langle k^\mu k^\nu \right\rangle_{\delta} \nonumber \\
	&= \left(e + \rho_2\right) u^{\mu}u^{\nu}
	- \left(P + \Pi \right)\Delta^{\mu\nu} + \pi^{\mu \nu} \; .
	\label{T_munu}
\end{align}
The energy-momentum tensor is defined in the Landau frame~\cite{Landau_book} where the time-like eigenvector of the energy-momentum tensor is $u^\mu = T^{\mu\nu} u_\nu/(u_\alpha T^{\alpha \beta} u_\beta)$ and hence $\rho^{\mu}_1 \equiv \left\langle E_{\mathbf{k} u} k^{\langle \mu \rangle} \right\rangle_{\delta} = 0$.
The chemical potential and the temperature are determined from the Landau matching conditions~\cite{Anderson:1974a},
\begin{align}
	\rho_1 &\equiv \left\langle E_{\mathbf{k} u} \right\rangle_{\delta} = 0 \;,  \quad
	\rho_2 \equiv \left\langle E_{\mathbf{k} u}^{2}\right\rangle_{\delta} = 0 \;.
	\label{delta_n_e}
\end{align}
The particle density, energy density, and isotropic pressure in equilibrium are
\begin{align}
	n &\equiv \left\langle E_{\mathbf{k} u} \right\rangle_{0} \;,  \quad
	e \equiv \left\langle E_{\mathbf{k} u}^{2}\right\rangle_{0} \;,  \quad
	P \equiv -\frac{1}{3}\left\langle \Delta_{\mu \nu }k^{\mu }k^{\nu}\right\rangle_{0}\;. \label{P0}
\end{align}
Since $f_{0 \mathbf{k}}$ only depends on two thermodynamic state variables, $\alpha$ and $\beta$, the first two equations can be solved for the latter, yielding $\alpha(e,n)$ and $\beta(e,n)$.
Inserting this into the last equation defines the equation of state of matter under consideration, $P(\alpha(e,n),\beta(e,n)) \equiv P(e,n)$.

The bulk viscous pressure, the particle diffusion four-current, and the shear-stress tensor are defined by
\begin{align}
	\Pi &\equiv  -\frac{1}{3}\left\langle \Delta_{\mu \nu }k^{\mu }k^{\nu }\right\rangle_{\delta}
	= -\frac{m^2_0}{3} \rho_0 \;, \label{Pi} \\
	V^{\mu} &\equiv  \left\langle k^{\langle \mu \rangle} \right\rangle_{\delta}
	= \rho^{\mu}_0 \;, \label{V_mu} \\
	\pi^{\mu \nu} &\equiv  \left\langle k^{\langle \mu} k^{\nu \rangle}\right\rangle_\delta
	= \rho^{\mu \nu}_0 \;. \label{pi_munu}
\end{align}

Second-order fluid dynamics in the approximation with $14$ dynamical moments contains $1+4$ conservation equations (\ref{cons_eqs}), and additionally $1+3+5$ equations of motion for the lowest-order moments, i.e., $r=0$, in Eqs.~(\ref{D_rho})--(\ref{D_rho_munu}).
These irreducible moments, $\rho_0$, $\rho^{\mu}_0$, and $\rho^{\mu \nu}_0$, identified in Eqs.~(\ref{Pi})--(\ref{pi_munu}), are then \textit{dynamical moments}.
The remaining nondynamical moments for $r \ne 0$ can be determined from the dynamical moments using an expansion of the single-particle distribution function around $f_{0\mathbf{k}}$, see Ref.~\cite{Denicol:2012cn} for details,
\begin{equation}
	\delta f_{\mathbf{k}} = f_{0\mathbf{k}} \tilde{f}_{0\bk}
	\sum_{\ell=0}^{\infty }\sum_{n=0}^{N_{\ell }} \rho_{n}^{\mu_{1}\cdots \mu_{\ell }}
	k_{\left\langle \mu _{1}\right. }\cdots k_{\left.\mu _{\ell }\right\rangle} \mathcal{H}_{\mathbf{k}n}^{(\ell)} \;.
	\label{f_k_expansion}
\end{equation}
The coefficient $\mathcal{H}_{\mathbf{k}n}^{(\ell )}$ is a polynomial in energy of order $N_{\ell}$, where in principle $N_\ell \rightarrow \infty$, see Appendix~\ref{Sec:DNMR_coefficients} for details, and it is used to define,
\begin{align}
	\mathcal{F}_{\mp rn}^{\left( \ell \right) } &=\frac{\ell !}{\left( 2\ell +1\right)!!}
	\int \mathrm{d}K\, E_{\mathbf{k}}^{\pm r} \left( \Delta ^{\alpha \beta }k_{\alpha }k_{\beta }\right)^{\ell }
	\mathcal{H}_{\mathbf{k}n}^{\left( \ell \right)}	f_{0\mathbf{k}}	\tilde{f}_{0\bk} \;.
	\label{F_rn}
\end{align}
Therefore any irreducible moment with tensor rank $\ell$ of arbitrary order $r$ can be expressed as a linear combination of rank-$\ell$ moments with positive order $n\geq0$,
\begin{equation}
	\rho_{\pm r}^{\mu _{1}\cdots \mu _{\ell }}
	= \sum_{n=0}^{N_{\ell }}\rho_{n}^{\mu_{1}\cdots \mu_{\ell }}
	\mathcal{F}_{\mp r,n}^{\left( \ell \right) }  \;.
	\label{F_rn_useful}
\end{equation}

Using these steps and approximations the general form of the second-order equations of motion for $\Pi$, $V^\mu$, and $\pi^{\mu\nu}$ reads, for more details see the derivation of Eqs.~(70)--(72) of Ref.~\cite{Denicol:2012cn},
\begin{align}\label{Pidot}
	\tau_\Pi \dot{\Pi} + \Pi &= -\zeta \theta  -\ell_{\Pi V} \nabla_\mu V^\mu -
	\tau_{\Pi V} V_\mu \dot{u}^\mu \\
	&- \delta_{\Pi\Pi} \Pi \theta - \lambda_{\Pi V} V_\mu \nabla^\mu \alpha
	+ \lambda_{\Pi \pi} \pi^{\mu\nu} \sigma_{\mu\nu} \;, \nonumber \\
	\label{Vdot}
	\tau_V \dot{V}^{\langle \mu \rangle} + V^\mu
	&= \kappa \nabla^\mu \alpha
	-\tau_V V_\nu \omega^{\nu\mu} - \delta_{VV} V^\mu \theta \\
	&- \ell_{V\Pi} \nabla^\mu \Pi
	+ \ell_{V\pi} \Delta^{\mu\nu} \nabla_\lambda \pi^\lambda{}_\nu
	+ \tau_{V\Pi} \Pi \dot{u}^\mu \nonumber\\
	&- \tau_{V\pi} \pi^{\mu\nu} \dot{u}_\nu - \lambda_{VV} V_\nu \sigma^{\mu\nu}  \nonumber \\
	&+ \lambda_{V \Pi} \Pi \nabla^\mu \alpha
	- \lambda_{V\pi} \pi^{\mu\nu} \nabla_\nu \alpha \;,  \nonumber \\
	\label{pidot}
	\tau_\pi \dot{\pi}^{\langle \mu \nu \rangle} + \pi^{\mu\nu}
	&= 2 \eta \sigma^{\mu\nu} + 2\tau_\pi \pi^{\langle\mu}_\lambda \omega^{\nu\rangle \lambda} -
	\delta_{\pi\pi} \pi^{\mu\nu} \theta \\
	&- \tau_{\pi\pi} \pi^{\lambda\langle \mu}\sigma^{\nu\rangle}_\lambda
	+ \lambda_{\pi \Pi} \Pi \sigma^{\mu\nu}
	- \tau_{\pi V} V^{\langle \mu} \dot{u}^{\nu \rangle}  \nonumber \\
	&+ \ell_{\pi V} \nabla^{\langle \mu} V^{\nu \rangle}
	+ \lambda_{\pi V} V^{\langle \mu} \nabla^{\nu \rangle} \alpha \; , \nonumber
\end{align}
where $\tau_\Pi$, $\tau_V$, and $\tau_\pi$ are different relaxation times.
Note that in the RTA the relaxation times are strictly equal to the model parameter, $\tau_R \equiv \tau_\Pi = \tau_V = \tau_\pi$.
However, for the sake of clarity we will use different subscripts for the corresponding relaxation times.

From Eqs.~\eqref{D_rho}--\eqref{D_rho_munu} the first-order transport coefficients for the moments with energy index $r$ are
\begin{align}
	\zeta_r &= \tau_\Pi \frac{m_0^2}{3} \alpha^{(0)}_r \;, &
	\kappa_r &= \tau_V \alpha^{(1)}_r \;, &
	\eta_r &= \tau_\pi \alpha^{(2)}_r \;.
	\label{main_RTA_coeffs}
\end{align}
while $\zeta = \zeta_0$, $\kappa = \kappa_0$, and $\eta=\eta_0$ are the first-order transport coefficients of the bulk viscosity, the particle diffusion, and the shear viscosity, respectively.

For the sake of completeness, we recall all transport coefficients in RTA; see Eqs.~(96)--(112) of Ref.~\cite{Ambrus:2022vif}.
The coefficients appearing in Eq.~\eqref{Pidot} are
\begin{align}
	\zeta &= \tau_\Pi\frac{m_0^2}{3} \alpha^{(0)}_{0} , \label{bf_zeta} \\
	\delta_{\Pi\Pi} &= \tau_\Pi \left(\frac{2}{3} - \frac{m_0^2}{3} \frac{G_{20}}{D_{20}}
	+ \frac{m_0^2}{3} \mathcal{R}^{(0)}_{-2,0}\right), \label{bf_delta_Pi_Pi} \\
	\ell_{\Pi V} &= \tau_\Pi\frac{m_0^2}{3}\left(\frac{G_{30}}{D_{20}} - \mathcal{R}^{(1)}_{-1,0} \right), \\
	\tau_{\Pi V} &= - \tau_\Pi\frac{m_0^2}{3} \left(\frac{G_{30}}{D_{20}}
	- \frac{\partial \mathcal{R}^{(1)}_{-1,0} }{\partial \ln \beta} \right), \\
	\lambda_{\Pi V} &= -\tau_\Pi\frac{m_0^2}{3}
	\left(\frac{\partial \mathcal{R}^{(1)}_{-1,0}}{\partial \alpha} +
	\frac{1}{h} \frac{\partial \mathcal{R}^{(1)}_{-1,0}}{\partial \beta}\right), \\
	\lambda_{\Pi \pi} &=  -\tau_\Pi\frac{m_0^2}{3} \left(\frac{G_{20}}{D_{20}}
	-\mathcal{R}^{(2)}_{-2,0}\right). \label{bf_lambda_Pi_pi}
\end{align}
The transport coefficients in the diffusion equation are
\begin{align}
	\kappa &= \tau_V \alpha^{(1)}_0, \quad \label{bf_kappa}
	\delta_{VV} = \tau_V \left(1 + \frac{m_0^2}{3} \mathcal{R}^{(1)}_{-2,0} \right), \\
	\ell_{V \Pi} &= \frac{\tau_V}{h}\! \left(1 -h \mathcal{R}^{(0)}_{-1,0}\right), \
	\ell_{V \pi} = \frac{\tau_V}{h} \!\left( 1- h\mathcal{R}^{(2)}_{-1,0}\right), \\
	\tau_{V \Pi} &= \frac{\tau_V}{h}\! \left(1 -
	h \frac{\partial \mathcal{R}^{(0)}_{-1,0}}{\partial \ln \beta}\right), \\
	\tau_{V \pi} & = \frac{\tau_V}{h} \! \left(1
	- h\frac{\partial \mathcal{R}^{(2)}_{-1,0}}{\partial \ln \beta}\right), \\
	\lambda_{VV} &= \tau_V \left(\frac{3}{5} + \frac{2m_0^2}{5} \mathcal{R}^{(1)}_{-2,0} \right) , \\
	\lambda_{V \Pi} &= \tau_V \left(\frac{\partial \mathcal{R}^{(0)}_{-1,0}}{\partial \alpha}
	+ \frac{1}{h} \frac{\partial \mathcal{R}^{(0)}_{-1,0}}{\partial \beta} \right) , \\
	\lambda_{V \pi} &= \tau_V \left(\frac{\partial \mathcal{R}^{(2)}_{-1,0}}{\partial \alpha}
	+ \frac{1}{h} \frac{\partial \mathcal{R}^{(2)}_{-1,0}}{\partial \beta} \right). \label{bf_lambda_V_pi}
\end{align}

Finally, the transport coefficients appearing in the equation for the shear-stress
tensor~\eqref{pidot} are
\begin{align}
	\eta &= \tau_\pi \alpha^{(2)}_0, \label{bf_eta} \quad
	\delta_{\pi\pi} = \tau_\pi \left( \frac{4}{3} + \frac{m_0^2}{3} \mathcal{R}^{(2)}_{-2,0} \right),\\
	\tau_{\pi\pi} &= \tau_\pi \left(\frac{10}{7} + \frac{4m_0^2}{7} \mathcal{R}^{(2)}_{-2,0} \right), \label{bf_lambda_pi_pi} \\
	\lambda_{\pi \Pi} &= \tau_\pi\left(\frac{6}{5} + \frac{2m_0^2}{5} \mathcal{R}^{(0)}_{-2,0} \right), \label{bf_lambda_pi_Pi} \\
	\tau_{\pi V} &= -\tau_\pi \frac{2m_0^2}{5}
	\frac{\partial \mathcal{R}^{(1)}_{-1,0}}{\partial \ln \beta}, \quad
	\ell_{\pi V} =  -\tau_\pi\frac{2m_0^2 }{5} \mathcal{R}^{(1)}_{-1,0} , \\
	\lambda_{\pi V} &=  -\tau_\pi\frac{2m_0^2 }{5}
	\left(\frac{\partial \mathcal{R}^{(1)}_{-1,0}}{\partial \alpha} +
	\frac{1}{h} \frac{\partial \mathcal{R}^{(1)}_{-1,0}}{\partial \beta}\right).\label{bf_lambda_pi_V}
\end{align}
Here, the ratio $\mathcal{R}^{(\ell)}_{r0}$ in the so-called basis-free approach of Ref.~\cite{Ambrus:2022vif} is defined as
\begin{equation}
	\mathcal{R}^{(\ell)}_{r0} = \frac{\alpha^{(\ell)}_r}{\alpha^{(\ell)}_0} \;.
	\label{R_def}
\end{equation}
As the name suggests, in this approximation the negative-order moments can be obtained without employing basis-dependent representations as in Eq.~\eqref{f_k_expansion}.

Therefore, in the approximation with 14 dynamical moments, for any $r\neq 0$, both positive and negative, we use the following approximation for the nondynamical moments,
\begin{align}  \label{bf_scalar_matching}
	\rho_{r\neq 0} \simeq -\frac{3}{m_0^2} \mathcal{R}^{(0)}_{r0} \Pi \;, \\
	\rho^\mu_{r \neq 0} \simeq \mathcal{R}^{(1)}_{r0} V^\mu \;, \\
	\rho^{\mu\nu}_{r \neq 0} \simeq \mathcal{R}^{(2)}_{r0} \pi^{\mu\nu} \;, \label{bf_matching}
\end{align}
where
\begin{equation}
	\mathcal{R}^{(0)}_{r0} = \frac{\zeta_r}{\zeta}\, , \quad
	\mathcal{R}^{(1)}_{r0} = \frac{\kappa_r}{\kappa}\, ,\quad
	\mathcal{R}^{(2)}_{r0} = \frac{\eta_r}{\eta}\, .
\end{equation}

Note that these second-order transport coefficients can also be obtained in the case of a linearized binary collision integral \cite{Denicol:2012cn}, where $C_{r-1}^{\left\langle \mu_{1}\cdots \mu_{\ell }\right\rangle } = -\sum_{n=0}^{N_{\ell }} \mathcal{A}_{r,n}^{\left( \ell \right)} \rho_{n}^{\mu_{1}\cdots\mu_{\ell}}$, with the following replacements:
\begin{align}
	\mathcal{R}^{(0)}_{-r,0} &\rightarrow \gamma_{r0}^{(0)} = \sum_{n = 0,\neq 1,2}^{N_0} \! \mathcal{F}^{(0)}_{rn}\Omega^{(0)}_{n0} \;, \\
	\mathcal{R}^{(1)}_{-r,0} &\rightarrow \gamma_{r0}^{(1)} = \sum_{n = 0,\neq 1}^{N_1} \!\mathcal{F}^{(1)}_{rn} \Omega^{(1)}_{n0} \;,  \\
	\mathcal{R}^{(2)}_{-r,0} &\rightarrow \gamma_{r0}^{(2)} = \sum_{n = 0}^{N_2} \mathcal{F}^{(2)}_{rn} \Omega^{(2)}_{n0} \;,
	\label{gamma}
\end{align}
where $\Omega^{(\ell)}_{rn}$ diagonalizes the binary collision matrix $\mathcal{A}_{rn}^{\left( \ell \right) }$.

The RTA collision term (\ref{AW}) leads to $\mathcal{A}_{rn}^{\left( \ell \right)} = \frac{\delta_{rn}}{\tau_R}$ and $\Omega_{rn}^{(\ell)} = \delta_{rn}$, while the moments with negative order, $\rho^{\mu_1 \cdots \mu_\ell}_{r < 0}$, according to the DNMR approach~\cite{Denicol:2012cn} read
\begin{align} \label{DNMR_rho_neg}
	\rho_{-r} &\simeq -\frac{3}{m_0^2}\gamma^{(0)}_{r0} \Pi \;, \\
	\rho_{-r}^\mu &\simeq \gamma^{(1)}_{r0} V^\mu \;, \\
	\rho_{-r}^{\mu\nu} &\simeq \gamma^{(2)}_{r0} \pi^{\mu\nu} \;.
	\label{DNMR_rho_munu_neg}
\end{align}
Without going through the detailed derivation of Ref.~\cite{Ambrus:2022vif}, the difference between these two approaches is due to a slightly different treatment of the negative-order moments.
As such, in the standard DNMR approximation, the basis-free ratios $\mathcal{R}^{(\ell)}_{-r,0}$ are replaced by $\gamma^{(\ell)}_{r,0}$.
The corresponding $\gamma^{(\ell)}_{r0} = \mathcal{F}^{(\ell)}_{r0}$ coefficients are calculated in Appendix \ref{Sec:DNMR_coefficients}.

%%%
\subsection{Second-order fluid dynamics without particle-number conservation}
\label{sec:without_particle_conservation}

The case where there are no conserved quantum charges is equivalent to setting $\mu= \alpha = 0$.
Thus, the particle four-flow $N^\mu$ is no longer conserved, $\partial_\mu N^\mu \neq 0$, while $\alpha^{(1)}_r \nabla^\mu \alpha = 0$ in Eq.~\eqref{D_rho_mu} also vanishes identically.
This implies that the vector moments $\rho^\mu_r$ become of second order with respect to the Knudsen and inverse Reynolds numbers.
This in turn leads to third-order contributions to the equations of motion for the scalar and tensor moments, $\rho_r$ and $\rho^{\mu\nu}_r$, which are then ignored in second-order fluid dynamics.

The conservation equation for energy and momentum, $\partial_\mu T^{\mu \nu} = 0$, is closed by the relaxation equations for $\rho_0 = -\frac{3}{m_0^2} \Pi$ and $\rho^{\mu \nu}_0 = \pi^{\mu \nu}$.
The evolution equation for the tensor moments $\rho^{\mu \nu}_r$ remains unchanged compared to Eq.~(37) of Ref.~\cite{Denicol:2012cn}.
However, the equation for the scalar moments, see Eq.~(35) of Ref.~\cite{Denicol:2012cn}, including that for the bulk viscous pressure changes due to fact that the chemical potential is kept constant; i.e., $d\alpha = d\mu= 0$. Hence this leads to the following equation of motion:
\begin{multline}
	\dot{\rho}_{r}-C_{r-1} = \bar{\alpha}_{r}^{\left( 0\right) }\theta
	+\left[ (r-1)\rho _{r-2}^{\mu \nu } -\frac{J_{r+1,0}}{J_{30}}\pi^{\mu \nu }\right]
	\sigma_{\mu \nu }
	\\
	+ \frac{\theta }{3}\left[m_{0}^{2}(r-1)\rho _{r-2}-(r+2)\rho _{r}
	+ 3\frac{J_{r+1,0}}{J_{30}} \Pi \right] \; . \label{D_rho_mu=0}
\end{multline}
In this case, the definition of the transport coefficient $\alpha_r^{(0)}$  changes from Eq.~(\ref{alpha0}), and it will be denoted by $\bar{\alpha}_r^{(0)}$ in the following. Thus,
\begin{equation}
	\bar{\alpha}_r^{(0)} = -\beta J_{r+1,1}
	+ \bar{c}_s^2  \beta J_{r+1,0}\; , \label{alpha0_mu0}
\end{equation}
where $\bar{c}_s^2 \equiv \left(\frac{\partial P}{\partial e}\right)_\mu = J_{31}/J_{30}$ is the speed of sound squared.
The above equation and the value of the coefficient are obtained by replacing
\begin{equation}
\frac{G_{2r}}{D_{20}} \xrightarrow[]{\mu = 0} -\frac{J_{r+1,0}}{J_{30}} \;, \qquad
\frac{G_{3r}}{D_{20}} \xrightarrow[]{\mu = 0} 0 \;,
\label{repl}
\end{equation}
in Eqs.~(\ref{D_rho}) and (\ref{alpha0}), which also leads to the following replacements
\begin{equation}
 \alpha_r^{(0)} \xrightarrow[]{\mu = 0} \bar{\alpha}_r^{(0)} \;, \quad
 \mathcal{R}^{(0)}_{-r,0} \xrightarrow[]{\mu = 0}
 \bar{\mathcal{R}}^{(0)}_{-r,0} \equiv \frac{\bar{\alpha}^{(0)}_{-r}}{\bar{\alpha}^{(0)}_0} \;.
 \label{repl2}
\end{equation}
We stress that the notation ``$\xrightarrow[]{\mu = 0}$'' employed in Eqs.\ (\ref{repl}) and (\ref{repl2}) does not mean the $\mu \rightarrow 0$ limit of the expressions on the left-hand side, but rather the result when particle number is not conserved.
Note that now the nondynamical scalar moments from Eq.~\eqref{bf_scalar_matching} are expressed through $\rho_{r\neq 0} \simeq -\frac{3}{m_0^2} \bar{\mathcal{R}}^{(0)}_{r0} \Pi$, while Eq.~(\ref{bf_matching}) stays unchanged in the basis-free approximation.
Furthermore, the second-order DNMR coefficients follow using the replacements $\bar{\mathcal{R}}^{(0)}_{-r,0} \rightarrow \gamma_{r0}^{(0)}$ and $\mathcal{R}^{(2)}_{-r,0} \rightarrow \gamma_{r0}^{(2)}$.

The second-order relaxation equations corresponding to bulk viscous pressure and shear-stress tensor are different from Eqs.~\eqref{Pidot}--\eqref{pidot}, and now reduce to
\begin{align}
	\label{Pidot_mu=0_bulk}
	\tau_\Pi \dot{\Pi} + \Pi &= -\bar{\zeta}\theta - \bar{\delta}_{\Pi\Pi} \Pi \theta
	+ \bar{\lambda}_{\Pi \pi} \pi^{\mu\nu} \sigma_{\mu\nu} \;, \\
	\label{pidot_mu=0_shear}
	\tau_\pi \dot{\pi}^{\langle \mu \nu \rangle} + \pi^{\mu\nu}
	&= 2 \eta \sigma^{\mu\nu} + 2\tau_\pi \pi^{\langle\mu}_\lambda \omega^{\nu\rangle \lambda} -
	\delta_{\pi\pi} \pi^{\mu\nu} \theta  \nonumber \\
	&- \tau_{\pi\pi} \pi^{\lambda\langle \mu}\sigma^{\nu\rangle}_\lambda
	+ \bar{\lambda}_{\pi \Pi} \Pi \sigma^{\mu\nu}  \; .
\end{align}
These equations and coefficients were first obtained in Ref.~\cite{Denicol:2014vaa} in the standard 
DNMR approximation.
The transport coefficients in the basis-free approximation in the equation of the bulk viscous 
pressure are defined as
\begin{align}
	\bar{\zeta} &= \tau_\Pi\frac{m_0^2}{3} \bar{\alpha}^{(0)}_{0} \; , \label{bf_zeta_bar} \\
	\bar{\delta}_{\Pi\Pi} &= \tau_\Pi \left(\frac{2}{3} + \frac{m_0^2}{3} \frac{J_{10}}{J_{30}}
	+ \frac{m_0^2}{3} \bar{\mathcal{R}}^{(0)}_{-2,0}\right) \; , \\
	\bar{\lambda}_{\Pi \pi} &= \tau_\Pi\frac{m_0^2}{3} \left(\frac{J_{10}}{J_{30}}
	+ \mathcal{R}^{(2)}_{-2,0}\right) \; . \label{bf_lambda_bar_Pi_pi}
\end{align}
These also follow from Eqs.~\eqref{bf_zeta},~\eqref{bf_delta_Pi_Pi}, and~\eqref{bf_lambda_Pi_pi} under the replacements given in Eqs.~\eqref{repl} and~\eqref{repl2}.

The coefficients in the relaxation equation for the shear-stress tensor remain the same as listed in Eqs.~(\ref{bf_eta}) and (\ref{bf_lambda_pi_pi}), except for the second-order shear-bulk coupling coefficient, which now reads
\begin{equation}
	\bar{\lambda}_{\pi \Pi} = \tau_\pi\left(\frac{6}{5} + \frac{2m_0^2}{5}
	\bar{\mathcal{R}}^{(0)}_{-2,0} \right) \; .\label{bf_lambda_bar_pi_Pi}
\end{equation}
The standard DNMR coefficients of Ref.~\cite{Denicol:2014vaa} are obtained from these formulas using the following replacements, $\bar{\mathcal{R}}^{(0)}_{-r,0} \rightarrow \gamma_{r0}^{(0)}$ and $\mathcal{R}^{(2)}_{-r,0} \rightarrow \gamma_{r0}^{(2)}$.
Also note that in Ref.~\cite{Denicol:2014vaa}, the negative scalar moments 
$\rho_{-r} = \gamma^{(0)}_{r0} \Pi$, differ by a factor of $-3/m^2_0$ from our definitions 
in Eq.~\eqref{DNMR_rho_neg}.

%%%
\subsection{Transport coefficients for magnetohydrodynamics}
\label{sec:RTA:MHD}

For the sake of completeness, here we also list the transport coefficients in the equations of nonresistive and resistive magnetohydrodynamics derived from the Boltzmann-Vlasov equation in Refs.~\cite{Denicol.2018,Denicol:2019iyh}.
The additional $\mathcal{J}_{em}^{\mu_1 \cdots \mu_\ell}$ terms that appear on the right-hand sides of Eqs.~\eqref{Pidot}--\eqref{pidot} due to the coupling of the electric charge $\mathfrak{q}$ to the electromagnetic field are obtained from Eqs.~(24)--(26) of Ref.~\cite{Denicol:2019iyh} in the Landau frame,
\begin{align}
	\mathcal{J}_{em} &= -\mathfrak{q} \delta_{\Pi V E} V^\nu E_\nu\;,\label{MHD_J}\\
	\mathcal{J}_{em}^\mu &= \mathfrak{q}\left(\delta_{VE} E^\mu + \delta_{V \Pi E} \Pi E^\mu
	+ \delta_{V\pi E} \pi^{\mu\nu} E_\nu \right)  \nonumber\\
	&- \mathfrak{q} \delta_{VB} B b^{\mu\nu} V_\nu \;,\label{MHD_J_mu}\\
	\mathcal{J}_{em}^{\mu\nu} &= -\mathfrak{q} \left(\delta_{\pi B} B b^{\alpha\beta}
	\Delta^{\mu\nu}_{\alpha\kappa} \pi^\kappa{}_\beta + \delta_{\pi V E} E^{\langle \mu} V^{\nu \rangle}\right)\;,
\end{align}
where $b^{\mu\nu} = - \epsilon^{\mu\nu\alpha\beta} u_\alpha b_\beta$,  $b^\mu = B^\mu /B $, and $B = \sqrt{-B^\mu B_\mu}$ is the magnitude of the magnetic field.
The electric and magnetic fields $E^\mu$ and $B^\mu$ are defined through the Faraday tensor $F^{\mu\nu}$ and the fluid four-velocity $u^\mu$ via
\begin{equation}
	E^\mu = F^{\mu\nu}u_\nu \;, \quad
	B^\mu = \frac{1}{2} \epsilon^{\mu\nu \alpha\beta} F_{\alpha\beta} u_\nu\;.
\end{equation}

The transport coefficients in the RTA proportional to the electric field, see Eqs.~(117)--(121) of Ref.~\cite{Ambrus:2022vif}, are
\begin{align}
	\delta_{VE} &= \tau_V\left( -\frac{n}{h} + \beta J_{11}\right)\; , \label{bf_delta_V_E} \\
	\delta_{\Pi VE}&= -\tau_\Pi\frac{m_{0}^{2}}{3}
	\left(\frac{G_{20}}{D_{20}} -\mathcal{R}^{(1)}_{-2,0}
	+\frac{1}{h} \frac{\partial \mathcal{R}^{(1)}_{-1,0}}{\partial \ln\beta} \right)\;, \\
	\delta_{V\Pi E} &= -\tau_V \left( \frac{2}{m_{0}^{2}}+\mathcal{R}^{(1)}_{-2,0}
	- \frac{1}{h} \frac{\partial \mathcal{R}^{(0)}_{-1,0}}{\partial \ln\beta} \right)\;, \\
	\delta_{\pi VE} &= \tau_\pi\left( \frac{8}{5} + \frac{2m_{0}^{2}}{5} \mathcal{R}^{(1)}_{-2,0}
	- \frac{2m_{0}^{2}}{5h}
	\frac{\partial \mathcal{R}^{(1)}_{-1,0}}{\partial \ln\beta} \right)\;, \label{bf_delta_pi_V_E} \\
	\delta_{V\pi E} &=\tau_V \left(\mathcal{R}^{(2)}_{-2,0} - \frac{1}{h}
	\frac{\partial \mathcal{R}^{(2)}_{-1,0}}{\partial \ln\beta} \right)\;, \label{bf_delta_V_pi_E}
\end{align}
while the coefficients proportional to the magnetic field, see Eqs.~(122) of Ref.~\cite{Ambrus:2022vif}, are:
\begin{align}
	\delta_{VB} &=\tau_V \left( -\frac{1}{h} + \mathcal{R}^{(1)}_{-1,0}\right)\;, &
	\delta_{\pi B} &= 2\tau_\pi \mathcal{R}^{(2)}_{-1,0}\;. \label{bf_delta_pi_B}
\end{align}

%%%
\section{Anisotropic fluid dynamics}
\label{sec::Anisotropic}

Anisotropic fluid dynamics is based on an expansion of $f_{\mathbf{k}}$ around a local anisotropic distribution function $\hat{f}_{0\mathbf{k}}$, as follows:
\begin{equation}
	f_{\mathbf{k}} \equiv \hat{f}_{0\mathbf{k}}
	+\delta \hat{f}_{\mathbf{k}} = f_{0\mathbf{k}}+\delta f_{\mathbf{k}} \;,  \label{kinetic:f=f0+df}
\end{equation}
where instead of $|\delta f_{\mathbf{k}}|\ll f_{0\mathbf{k}}$, we now assume that $|\delta \hat{f}_{\mathbf{k}}|\ll \hat{f}_{0\mathbf{k}}$.
In the case of a strong anisotropy, a suitable choice of $\hat{f}_{0\bk}$ can lead to $|\delta \hat{f}_{\mathbf{k}}|\ll |\delta f_{\mathbf{k}}|$ and consequently, the convergence properties of a series expansion in $\delta \hat{f}_{\mathbf{k}}$ are significantly improved compared to an expansion in terms of $\delta f_\bk$.

The irreducible moments of $\delta \hat{f}_{\mathbf{k}}$ are defined as
\begin{equation}
	\hat{\rho}_{ij}^{\mu_{1}\cdots \mu_{\ell}}\equiv \left\langle
	E_{\mathbf{k} u}^{i} E_{\mathbf{k} l}^{j} k^{\left\{ \mu_{1}\right. }\cdots
	k^{\left. \mu_{\ell}\right\}}\right\rangle_{\hat{\delta} } \; ,
	\label{a_rho_r_general}
\end{equation}
where $\left\langle \cdots \right\rangle _{\hat{\delta}} \equiv \int \mathrm{d}K\left( \cdots \right) \delta \hat{f}_{\mathbf{k}}$.
Similarly as indicated in the previous section, one can also derive the equations of motion for the comoving derivative of the anisotropic moments, $D\hat{\rho}_{ij}^{\left\{ \mu _{1}\cdots \mu_{\ell }\right\} } \equiv \Xi_{\nu_{1}\cdots \nu _{\ell }}^{\mu_{1}\cdots \mu_{\ell }} D \hat{\rho}_{ij}^{\nu_{1}\cdots \nu_{\ell }}$, from the Boltzmann equation \eqref{BTE}; see Ref.~\cite{Molnar:2016vvu} for more details.

Now, focusing on a simpler case, we are explicitly neglecting the $\delta \hat{f}_{\mathbf{k}}$ corrections by setting $\hat{\rho}_{ij}^{\mu_1 \cdots \mu_\ell}\equiv 0$, such that $f_\bk \equiv \hat{f}_{0\bk}$.
Thus the conservation laws of leading-order anisotropic fluid dynamics are solely based on the moments of the anisotropic distribution function $\hat{f}_{0\mathbf{k}}$ and read
\begin{align}
	\partial_{\mu}\hat{N}^{\mu} = \hat{\mathcal{C}}_{00} \equiv 0 \; , \quad
	\partial_{\mu}\hat{T}^{\mu \nu} &= \hat{\mathcal{C}}_{00}^{\nu} \equiv 0\;.  \label{Aniso_conservation}
\end{align}
Here we once again assume that the collision term is given by the RTA,
\begin{equation}
	C[\hat{f}_{0}] \equiv -\frac{E_{\mathbf{k}u}}{\tau_R} (\hat{f}_{0\mathbf{k}} - \feq) \;.
	\label{a_AW}
\end{equation}
The irreducible moments of the collision term are such that $\hat{\mathcal{C}}_{i-1,j}^{\left\{ \mu _{1}\cdots \mu _{\ell }\right\} } = 0$ for $\ell \geq 1$, while
\begin{align}
	\hat{\mathcal{C}}_{i-1,j} &\equiv \int \! \mathrm{d}K \, E_{\mathbf{k}u}^{i-1}E_{\mathbf{k}l}^{j} C[\hat{f}_0] \nonumber\\
	&= -\frac{1}{\tau _R} \left( \hat{I}_{i+j,j,0} - I_{i+j,j,0}\right) \;,  \label{Coll_Int_RTA}
\end{align}
where $\hat{I}_{nrq}$ was introduced in Eq.~\eqref{I_nrq}.

Using the expansion from Eq.~\eqref{I_ij_tens}, $\hat{N}^{\mu }\equiv \hat{\mathcal{I}}_{00}^{\mu }$ and $\hat{T}^{\mu \nu } \equiv \hat{\mathcal{I}}_{00}^{\mu \nu}$ are tensor-decomposed as
\begin{align}
	& \hat{N}^{\mu }\equiv \langle k^{\mu} \rangle_{\hat{0}}=\hat{n}u^{\mu }
	+ \hat{n}_{l}l^{\mu }\;,  \label{N_mu_aniso} \\
	& \hat{T}^{\mu \nu }\equiv\langle k^{\mu} k^{\nu} \rangle_{\hat{0}} =\hat{e}
	u^{\mu }u^{\nu } +\hat{P}_{l}l^{\mu }l^{\nu }-\hat{P}_{\perp }\Xi ^{\mu \nu }\;, \quad
	\label{T_munu_aniso}
\end{align}
where the particle-number density, the energy density, the particle diffusion current in the direction of the anisotropy, and the pressure components in the direction of and transverse to the anisotropy are
\begin{align} \label{e_hat}
	\hat{n} &\equiv \hat{I}_{100}=\hat{\mathcal{I}}_{10}\;, \quad
	\hat{e} \equiv \hat{I}_{200}=\hat{\mathcal{I}}_{20}\;,   \\ \label{P_l_hat}
	\hat{n}_{l} &\equiv \hat{I}_{110}=\hat{\mathcal{I}}_{01}\;,  \quad
	\hat{P}_{l} \equiv \hat{I}_{220}=\hat{\mathcal{I}}_{02}\;,  \\
	\hat{P}_{\perp } &\equiv \hat{I}_{201}
	=-\frac{1}{2}\left( m_{0}^{2}\hat{\mathcal{I}}_{00} -\hat{\mathcal{I}}_{20}
	+\hat{\mathcal{I}}_{02}\right)\; , \label{P_T_hat}
\end{align}
with $\hat{\mathcal{I}}_{ij}^{\mu _{1}\cdots \mu _{n}}$ defined in Eq.~\eqref{I_ij_tens}.
Here we once again chose the LRF according to Landau's definition, rendering $\hat{M} \equiv -\hat{T}^{\mu \nu }u_{\mu }l_{\nu }=\hat{I}_{210}=\hat{\mathcal{I}}_{11}=0 $.
The isotropic pressure is defined as
\begin{equation}
	\hat{P}(\hat{\alpha},\hat{\beta}_u, \hat{\beta}_l) \equiv
	-\frac{1}{3}\hat{T}^{\mu \nu } \Delta_{\mu \nu }
	=\frac{1}{3} \left( \hat{P}_{l}+2\hat{P}_{\perp }\right) \;,   \label{P_iso_relation}
\end{equation}
and hence the bulk viscous pressure is
\begin{align} \nonumber
	\hat{\Pi}(\hat{\alpha},\hat{\beta}_u, \hat{\beta}_l)
	&\equiv \hat{P}(\hat{\alpha},\hat{\beta}_u, \hat{\beta}_l) - P(\alpha, \beta) \\
	&= \frac{1}{3} \left( \hat{P}_{l}+2\hat{P}_{\perp } - 3P\right) \; , \label{aniso_bulk}
\end{align}
where $P$ is the isotropic pressure in equilibrium defined in Eq.~\eqref{P0}.
Similarly to Eqs.~\eqref{delta_n_e} the Landau matching conditions require that the particle-number density and energy density calculated through $\hat{f}_{0\bk}$ are equal to those of a fictitious local-equilibrium state,
\begin{equation}
	\hat{n}(\hat{\alpha},\hat{\beta}_u, \hat{\beta}_l) = n(\alpha, \beta) \; , \quad
	\hat{e}(\hat{\alpha},\hat{\beta}_u, \hat{\beta}_l) = e(\alpha, \beta) \; , \label{aniso_matching}
\end{equation}
where the equilibrium particle-number density and energy density were defined in Eq.~\eqref{P0}.

For practical purposes and explicit comparisons to the existing literature, we apply the spheroidal distribution function introduced by Romatschke and Strickland (RS) in Ref.~\cite{Romatschke:2003ms},
\begin{equation}
	\hat{f}_{RS}\equiv \left[ \exp \left(
	\frac{\sqrt{ E_{\mathbf{k}u}^2 + \xi E_{\mathbf{k}l}^2}}{\Lambda} -\hat{\alpha}\right) +a\right] ^{-1}\;,  \label{f_RS}
\end{equation}
where $\xi$ denotes the so-called anisotropy parameter.
A direct comparison to Eq.~(\ref{hat_f->f_0}) leads to the identifications $\hat{\beta}_{u}\equiv 1/\Lambda$ and $\hat{\beta}_{l}\equiv \sqrt{\xi}/\Lambda$.
Therefore we introduce a new set of thermodynamic integrals, 
$\hat{I}_{nrq}^{RS}\left( \hat{\alpha},\Lambda,\xi \right)$, with the replacement 
$\hat{f}_{0\mathbf{k}} \rightarrow \hat{f}_{RS} \left( \hat{\alpha},\Lambda,\xi \right)$ 
in Eq.~(\ref{I_nrq}); i.e.,
\begin{equation}
	\hat{I}_{nrq}^{RS}=\frac{\left( -1\right) ^{q}}{\left( 2q\right) !!}\int
	\mathrm{d}K\, E_{\mathbf{k}u}^{n-r-2q}E_{\mathbf{k}l}^{r}\left( \Xi ^{\mu \nu }k_{\mu}k_{\nu }\right)^{q}\ \hat{f}_{RS}\;.  \label{I_nrq_RS}
\end{equation}
Therefore, the first and second moments of the RS distribution function are
\begin{align}
	\hat{N}_{RS}^{\mu } &\equiv \hat{n}u^{\mu } = \hat{I}_{100}^{RS} u^{\mu} \;,  \label{N_mu_RS} \\
	\hat{T}_{RS}^{\mu \nu } &\equiv \hat{e}u^{\mu }u^{\nu }+\hat{P}_{l}l^{\mu }l^{\nu}-\hat{P}_{\perp }\Xi ^{\mu \nu } \nonumber \\
	&=\hat{I}_{200}^{RS} u^{\mu }u^{\nu }+\hat{I}_{220}^{RS} l^{\mu }l^{\nu}
	-\hat{I}_{201}^{RS}\Xi ^{\mu \nu }  \;.  \label{T_munu_RS}
\end{align}
The five conservation equations~(\ref{Aniso_conservation}) must be closed by an additional equation of motion.
In this work, we are following Refs.~\cite{Molnar:2016gwq,Niemi:2017stb} and employ the equation of motion for $\hat{P}_{l}$ to supplement the conservation equations.

%%%
\section{Relativistic ideal gas of classical particles}
\label{sec:massive_gas}

The equilibrium momentum distribution of a classical ideal gas of particles with nonzero mass
$m_0$ is given by the Maxwell-J\"uttner distribution~(\ref{f_0k}),
\begin{equation}
	\feq = e^{\alpha -\beta E_{\bk u}} \; .
\end{equation}
Since $a=0$, we have $\partial_\alpha \feq = \feq$, such that $J_{rq} = I_{rq}$, while $I_{rq}$ from Eq.~(\ref{I_nq}) now reads
\begin{equation}
	I_{rq} = \frac{g e^\alpha }{2\pi^2} \frac{ m_0^{r+2}}{(2q+1)!!}
	\int_1^\infty \!  \mathrm{d}x\, x^{r-2q} (x^2 - 1)^{q + \frac{1}{2}} e^{-z x} \; ,
	\label{eq:ideal_Irq}
\end{equation}
where $z \equiv m_0 / T = m_0 \beta$.
The thermodynamic integrals obey the recursion relation
\begin{equation}
	I_{r+2,q} = m^2_0 I_{r,q} + (2q+3) I_{r+2,q+1} \; .
	\label{I_rq_recursion}
\end{equation}
In analogy to these equilibrium thermodynamic integrals, the integral representation of the modified Bessel functions of the second kind $K_q(z)$ for $q>1/2$, see Eq.~(9.6.23) of Ref.~\cite{Abramowitz_Stegun}, reads
\begin{align} \label{eq:K_n_int}
	K_q(z) &\equiv \frac{z^q}{(2q-1)!!} \int_{1}^{\infty} \mathrm{d}x\,
	(x^2 - 1)^{q - \frac{1}{2}}  e^{-z x}   \nonumber \\
	& = \frac{z^q}{(2q-1)!!} \int_{0}^{\infty} \mathrm{d}x\, \sinh^{2q} x  \, e^{-z \cosh x} \; ,
\end{align}
where the double factorial of odd numbers is defined as $(2q-1)!! = 2^{q} \Gamma(q + 1/2)/\sqrt{\pi}$.
These Bessel functions of second kind satisfy the following recurrence relation for $q > 0$, see Eq.~(9.6.26) of Ref.~\cite{Abramowitz_Stegun},
\begin{equation}
	K_{q+2}(z) = K_{q}(z) + \frac{2q+2}{z} K_{q+1}(z) \; . \label{K_n_recursion}
\end{equation}

Using these formulas and recursive relations we express all thermodynamic integrals of interest in terms of Bessel functions.
The particle-number density is expressed as
\begin{equation}
	n \equiv I_{10} = \frac{ge^\alpha}{2\pi^2}  T^3 z^2 K_2(z) \; ,
	\label{eq:n}
\end{equation}
while the isotropic pressure is
\begin{equation} \label{isotropic_pressure}
	P \equiv I_{21} = n T \; .
\end{equation}
This is the equation of state of an ideal gas of classical particles.
The energy density is given by
\begin{equation}
	e \equiv I_{20} = P \left[3 + z \frac{K_1 (z)}{K_2(z)} \right]
	= P \left[z H(z) - 1\right] \; ,
\end{equation}
and therefore the enthalpy per particle reads
\begin{equation}
	h \equiv \frac{e + P}{n} = m_0 \frac{K_3 (z)}{K_2(z)} \equiv m_0 H(z) \; ,
	\label{h_H}
\end{equation}
where the enthalpy per particle divided by the rest mass is $H(z) \equiv K_3 (z)/K_2(z) = h /m_0$, or equivalently, $H(z) = K_1(z)/K_2(z) + 4/z$.

\begin{figure}
	\begin{center}
		\begin{tabular}{c}
			\includegraphics[width=.9\linewidth]{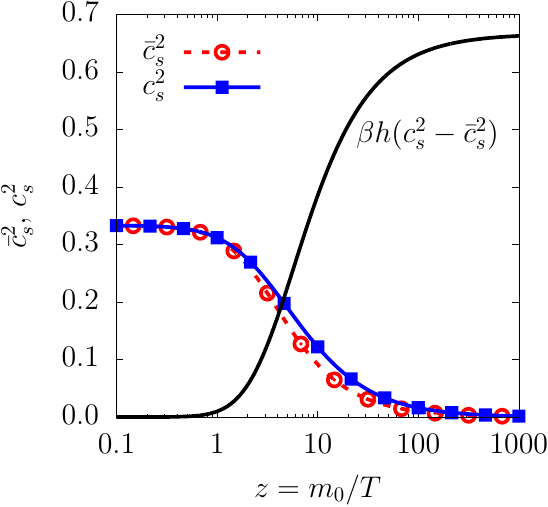}
		\end{tabular}
	\end{center}
	\caption{
		Speed of sound squared, represented as a function of $z = m_0 / T$ for the case of a classical 
		ideal gas without (dashed red line and circles) or with (solid blue line and squares) conserved particle number. The solid line shows $\beta h (c_s^2 - \bar{c}_s^2)$, which approaches $2/3$ 
		as $z \rightarrow \infty$.
		\label{fig:csq}
	}
\end{figure}

Using these results together with the recurrence relations~(\ref{I_rq_recursion}) and~(\ref{K_n_recursion}), it can be shown that
\begin{equation}
	I_{00} \equiv \frac{n}{m_0} \frac{K_1(z)}{K_2(z)} = \frac{e - 3P}{m^2_0} \; ,
\end{equation}
and similarly
\begin{align}
	I_{30} &\equiv m_0 P\left[z + 3H(z)\right] = T \left(c_v P + \frac{e^2}{P}\right) \;, \\
	I_{31} &\equiv  m_0 P H(z) = T \left(e+P \right) \; .
\end{align}
Here the heat capacities at constant volume $c_v$, and at constant pressure $c_p$ are defined as
\begin{align}
	c_v &\equiv \frac{1}{n} \left(\frac{\partial e} { \partial T}\right)_V
	= 3 + z^2 - \frac{e}{P^2} \left(e - 3P \right) \;, \\
	c_p &\equiv \left(\frac{\partial h}{ \partial T}\right)_P = c_v + 1 \; .
\end{align}
The speed of sound squared reads
\begin{equation}
	c_s^2 \equiv \left(\frac{\partial P}{\partial e}\right)_n +
	\frac{1}{h} \left(\frac{\partial P}{\partial n}\right)_e
	=  \frac{c_p P}{c_v(e + P)} \; , \label{c_s2}
\end{equation}
where
\begin{equation}
	\left(\frac{\partial P}{\partial e}\right)_n = \frac{1}{c_v}, \qquad
	\left(\frac{\partial P}{\partial n}\right)_e = T - \frac{e}{n c_v}.
\end{equation}
Without particle-number conservation, the speed of sound squared $\bar{c}_s^2 = \left(\frac{\partial P}{\partial \beta}\right)_\mu \left(\frac{\partial e}{\partial \beta}\right)^{-1}_\mu $, see Eq.~(22) of Ref.~\cite{Denicol:2014vaa}, is given by the following expression
\begin{equation}
	\bar{c}_s^2 \equiv \left(\frac{\partial P}{\partial e}\right)_\mu
	= \frac{I_{31}}{I_{30}} = \frac{P(e + P)}{c_vP^2+e^2} \; , \label{bar_c_s2}
\end{equation}
and hence
\begin{equation}
	c_s^2 \bar{c}_s^2 = \frac{c_p}{c_v(c_v + e^2/P^2)} \; .
\end{equation}
While in the massless limit, both $\bar{c}_s^2$ and $c_s^2$ are equal to $1/3$, at finite values of $z = m_0 / T$, these quantities will in general differ.
As shown in Fig.~\ref{fig:csq}, both $\bar{c}_s^2$ and $c_s^2$ exhibit a similar monotonously decreasing trend, with $\bar{c}_s^2$ approaching $0$ faster than $c_s^2$.
At small values of $z \ll 1$, $c_s^2$ and $\bar{c}_s^2$ have a similar behavior,
\begin{align}
 c_s^2 (z \ll 1) &\simeq \frac{1}{3} - \frac{z^2}{36} + \frac{11z^4}{864} + O(z^6), \nonumber\\
 \bar{c}_s^2 (z \ll 1) &\simeq \frac{1}{3} - \frac{z^2}{36} + \frac{5z^4}{864} + O(z^6).
\end{align}
At large values of $z \gg 1 $, $c_s^2$ and $\bar{c}_s^2$ develop a difference that can be highlighted in the following form:
\begin{equation}
 \beta h (c_s^2 - \bar{c}_s^2)\rfloor_{z \gg 1} \simeq \frac{2}{3} - \frac{11}{3z} + \frac{61}{6z^2} + O(z^3).
 \label{eq:cs_minus_csbar}
\end{equation}
This is illustrated with the solid black line in Fig.~\ref{fig:csq}.

Finally, recalling Eqs.~(\ref{eq:G_nm}) and~(\ref{eq:D_nq}) we express $G_{2r}$, $G_{3r}$, and  $D_{20}$, as follows:
\begin{gather}
	G_{2r} = e I_{r0} - n I_{r+1,0} \;, \quad G_{3r} = I_{30} I_{r0} - e I_{r+1,0} \; , \\
	D_{20} \equiv I_{30} I_{10} - I_{20}^2 = c_v P^2 \;.
\end{gather}
In the particular case where $r = 0$, we find
\begin{equation}
	G_{20} = \frac{P^2}{m^2_0} \left(3 - c_v\right) \; , \quad
	G_{30} = \frac{3T P^2}{m^2_0} \left(\frac{e}{P} - c_v\right) \; ,
\end{equation}
therefore the ratios become
\begin{equation}
	\frac{G_{20}}{D_{20}} = \frac{1}{m^2_0} \left(\frac{3}{c_v} - 1\right) \;, \quad
	\frac{G_{30}}{D_{20}} = \frac{3T}{m^2_0} \left(\frac{e}{Pc_v} - 1\right) \;.
\end{equation}

%%%
%%%

%%%
\subsection{First- and second-order transport coefficients}

In the classical Boltzmann limit, the first-order transport coefficients from Eqs.~(\ref{alpha0})--(\ref{alpha2}) reduce to
\begin{align}
	\alpha_r^{(0)} &=(1 - r) I_{r1} + \frac{n I_{r+1,0} - e I_{r0}}{c_v P} \;, \label{alpha0_mass}\\
	\alpha_r^{(1)} &= I_{r+1,1} - h^{-1}I_{r+2,1} \; ,\label{alpha1_mass} \\
	\alpha_r^{(2)} &= \beta I_{r+3,2} \; . \label{alpha2_mass}
\end{align}
If the particle number is not conserved, i.e., $\mu=0$, then $\alpha_r^{(0)} \xrightarrow[]{\mu = 0} \bar{\alpha}_r^{(0)}$ is given by Eq.~\eqref{alpha0_mu0}, leading to
\begin{equation}
	\bar{\alpha}_r^{(0)} = -\beta I_{r+1,1}
	+ \bar{c}_s^2  \beta I_{r+1,0}\; . \label{alpha0_mu0_mass}
\end{equation}
Here we evaluate the $\alpha^{(0)}_0$ coefficient from Eq.~\eqref{alpha0_mass} proportional to the coefficient of bulk viscosity,
\begin{equation}\label{alpha_0}
	\alpha^{(0)}_0 = \left(1 - 3c_s^2 \right) \frac{e+P}{m_0^2}
	- \frac{2}{3} \frac{e-3P}{m_0^2} - \frac{m_0^2}{3} I_{-2,0} \; ,
\end{equation}
where $c_s^2$ is the speed of sound squared from Eq.~\eqref{c_s2}.
Evaluating Eq.~\eqref{alpha0_mu0_mass} corresponding to the case without particle-number conservation leads to a formally identical expression to that for $\alpha^{(0)}_0$, but with $c_s^2$ replaced by $\bar{c}_s^2$ from Eq.~\eqref{bar_c_s2},
\begin{equation}\label{bar_alpha_0}
	\bar{\alpha}^{(0)}_0 = \left(1 - 3\bar{c}_s^2 \right) \frac{e+P}{m_0^2}
	- \frac{2}{3} \frac{e-3P}{m_0^2} - \frac{m_0^2}{3} I_{-2,0} \;.
\end{equation}
The transport coefficients related to the particle diffusion and shear viscosity are
\begin{equation}\label{alpha_1}
	\alpha^{(1)}_0 = \frac{e-2P}{3h} - \frac{m_0^2}{3} I_{-1,0} \; ,
\end{equation}
and
\begin{equation} \label{alpha_2}
	\alpha^{(2)}_0 = \frac{e + 9P}{15} - \frac{m_0^4}{15} I_{-2,0} \;.
\end{equation}
Using these results we find the following important relationship between $\alpha^{(2)}_0$ and $\alpha^{(0)}_0$ or $\bar{\alpha}^{(0)}_0$:
\begin{equation}
	\alpha^{(2)}_0 = \frac{m_0^2}{5} \alpha^{(0)}_0 +\frac{3c_s^2 \left(e + P\right)}{5}
	=\frac{m_0^2}{5} \bar{\alpha}^{(0)}_0 +\frac{3\bar{c}_s^2 \left(e + P\right)}{5}  \; .
\end{equation}
Recalling the definitions of the first-order transport coefficients from Eqs.~\eqref{main_RTA_coeffs} we obtain, see also Eq.~(41) of Ref.~\cite{Denicol:2014vaa},
\begin{equation}
	\frac{\zeta}{\tau_\Pi} = \frac{5}{3} \frac{\eta}{\tau_\pi} - c^2_s \left(e + P\right) \;, \quad
	\frac{\bar{\zeta}}{\tau_\Pi} = \frac{5}{3} \frac{\eta}{\tau_\pi} - \bar{c}^2_s \left(e + P\right) \;.
\end{equation}
These relate the ratio of the bulk-viscosity coefficient to the bulk relaxation time, $\zeta/\tau_\Pi$ as well as $\bar{\zeta}/\tau_\Pi$, to the ratio of the shear viscosity coefficient to shear relaxation time, $\eta/\tau_\pi$.
Both relations are formally identical but involve different speeds of sound.
In terms of thermodynamic quantities these ratios are expressed as
\begin{align} \label{mass_zeta}
	\frac{\zeta}{\tau_\Pi} &= \left(\frac{1}{3} - c_s^2 \right) (e+P)
	- \frac{2}{9} (e-3P) - \frac{m_0^4}{9} I_{-2,0} \; , \\ \label{mass_bar_zeta}
	\frac{\bar{\zeta}}{\tau_\Pi} &= \left(\frac{1}{3} - \bar{c}_s^2 \right) (e+P)
	- \frac{2}{9} (e-3P) - \frac{m_0^4}{9} I_{-2,0} \;,
\end{align}
and
\begin{equation} \label{mass_eta}
	\frac{\eta}{\tau_\pi} =  \frac{e + 9P}{15} - \frac{m_0^4}{15} I_{-2,0} \;.
\end{equation}
The latter two expressions are identical to Eqs.~(20) and (21) of Ref.~\cite{Denicol:2014vaa}, while
the thermodynamic integrals in these formulas are listed in Appendix~\ref{Sec:Useful_Integrals}.
Also note that all first-order transport coefficients are positive as shown explicitly in Appendix~\ref{Sec:Transport_Coeffs_Positive}.

Furthermore, one can also find simple relations between the second-order transport coefficients of the bulk viscous pressure,
\begin{align} \label{delta_PiPi}
	\frac{\delta_{\Pi\Pi}}{\tau_\Pi} &= 2 - c^2_s\frac{e+P}{P} + \frac{m^2_0}{3} \mathcal{R}^{(0)}_{-2,0} \; , \\ \label{lambda_Pi_pi}
	\frac{\lambda_{\Pi\pi}}{\tau_\Pi} &= \frac{4}{3} - c^2_s\frac{e+P}{P} + \frac{m^2_0}{3} \bar{\mathcal{R}}^{(2)}_{-2,0} \; ,
\end{align}
and similarly,
\begin{align}\label{bar_delta_PiPi}
	\frac{\bar{\delta}_{\Pi\Pi}}{\tau_\Pi} &= 1 - \bar{c}^2_s + \frac{m^2_0}{3} \bar{\mathcal{R}}^{(0)}_{-2,0} \; , \\  \label{bar_lambda_Pi_pi}
	\frac{\bar{\lambda}_{\Pi\pi}}{\tau_\Pi} &= \frac{1}{3} - \bar{c}^2_s + \frac{m^2_0}{3} \bar{\mathcal{R}}^{(2)}_{-2,0} \; .
\end{align}
These latter relations are similar to Eqs.~(33) and (34) of Ref.~\cite{Denicol:2014vaa}.
The remaining second-order transport coefficients divided by the relaxation time are
\begin{align} \label{delta_pi_pi}
	\frac{\delta_{\pi\pi}}{\tau_\pi} &= \frac{4}{3} + \frac{m^2_0}{3}\mathcal{R}^{(2)}_{-2,0} \; , \\
	\label{tau_pi_pi}
	\frac{\tau_{\pi\pi}}{\tau_\pi} &= \frac{10}{7} + \frac{4 m^2_0}{7}\mathcal{R}^{(2)}_{-2,0} \; , \\
	\label{lambda_pi_Pi}
	\frac{\lambda_{\pi\Pi}}{\tau_\pi} &= \frac{6}{5} + \frac{2m^2_0}{5}\mathcal{R}^{(0)}_{-2,0} \; ,
\end{align}
and
\begin{equation} \label{bar_lambda_pi_Pi}
	\frac{\bar{\lambda}_{\pi\Pi}}{\tau_\pi} = \frac{6}{5}
	+ \frac{2m^2_0}{5}\bar{\mathcal{R}}^{(0)}_{-2,0} \; .
\end{equation}
In the DNMR approximation these were also first reported in Eqs.~(35)--(37) of Ref.~\cite{Denicol:2014vaa}.
One also finds the following useful relations between these coefficients
\begin{align}
	\frac{\delta_{\pi\pi}}{\tau_\pi} = \frac{7}{12}  \frac{\tau_{\pi\pi}}{\tau_\pi}
	+ \frac{1}{2} \; ,
\end{align}
and
\begin{align}
	\frac{\delta_{\Pi \Pi}}{\tau_\Pi} &= \frac{5}{6} \frac{\lambda_{\pi \Pi}}{\tau_\pi} -c^2_s \frac{e+P}{P} + 1 \; , \\
	\frac{\bar{\delta}_{\Pi \Pi}}{\tau_\Pi} &= \frac{5}{6} \frac{\bar{\lambda}_{\pi \Pi}}{\tau_\pi} - \bar{c}^2_s \; ,
\end{align}
as well as
\begin{align}
	\frac{\lambda_{\Pi\pi}}{\tau_\Pi}
	&= \frac{7}{12} \frac{\tau_{\pi\pi}}{\tau_\pi} - c^2_s\frac{e+P}{P} + \frac{1}{2}
	= \frac{\delta_{\pi\pi}}{\tau_\pi} - c^2_s \frac{e+P}{P}  \; , \\
	\frac{\bar{\lambda}_{\Pi\pi}}{\tau_\Pi}
	&= \frac{7}{12} \frac{\tau_{\pi\pi}}{\tau_\pi} - \bar{c}^2_s - \frac{1}{2}
	= \frac{\delta_{\pi\pi}}{\tau_\pi} - \bar{c}^2 _s - 1  \; .
\end{align}

%%%-Bulk-figures%%%
\begin{figure*}[!hbt]
	\begin{center}
		\begin{tabular}{ccc}
			\includegraphics[width=.33\linewidth]{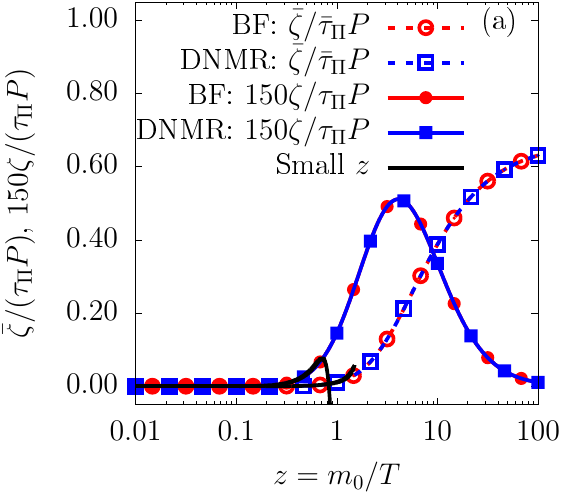} &
			\includegraphics[width=.33\linewidth]{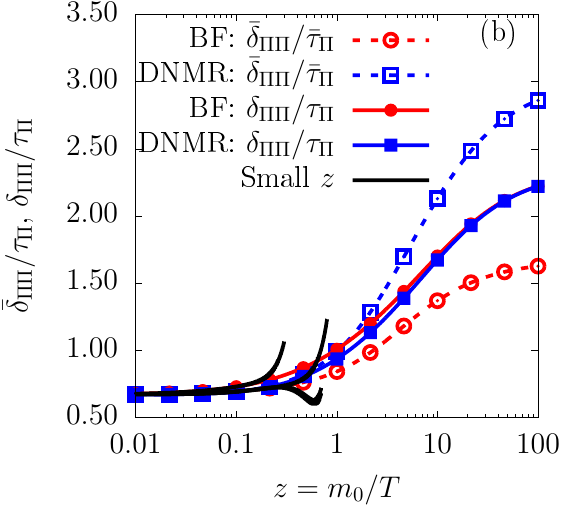} &
			\includegraphics[width=.33\linewidth]{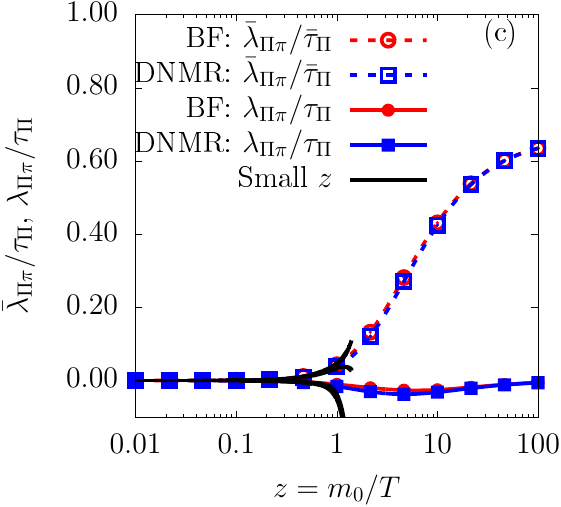} \\
			\includegraphics[width=.33\linewidth]{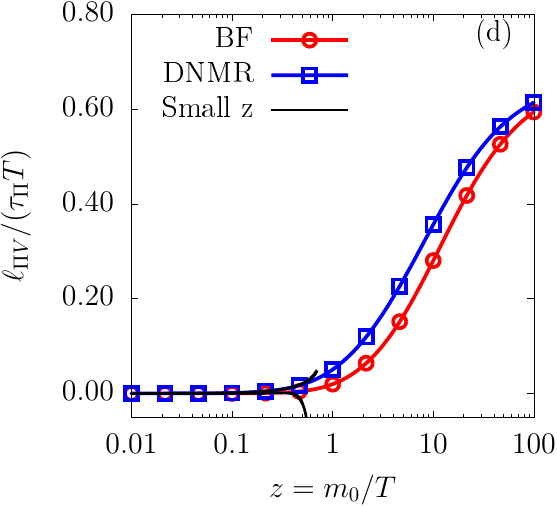} &
			\includegraphics[width=.33\linewidth]{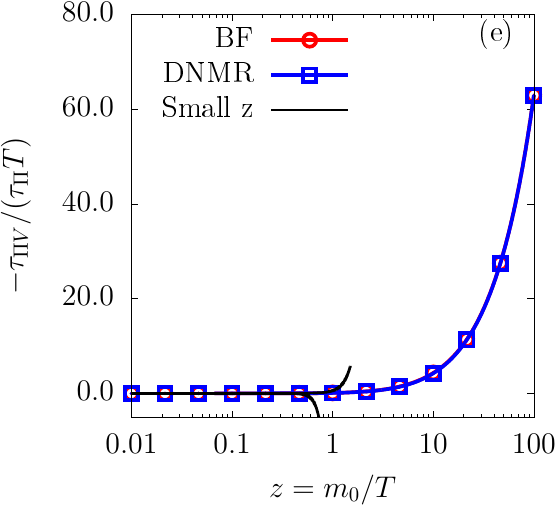} &
			\includegraphics[width=.33\linewidth]{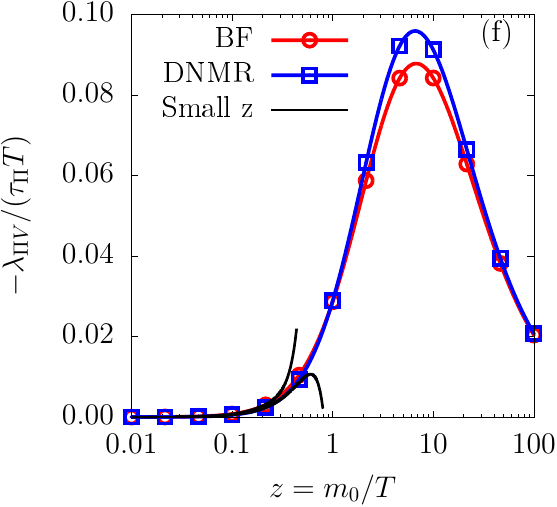}
		\end{tabular}
	\end{center}
	\caption{
		The bulk-viscosity coefficients $\zeta$ and $\bar{\zeta}$, and the second-order transport coefficients, $\delta_{\Pi\Pi}$, $\bar{\delta}_{\Pi\Pi}$, $\lambda_{\Pi\pi}$, $\bar{\lambda}_{\Pi\pi}$, $\ell_{\Pi V}$, $\tau_{\Pi V}$, $\lambda_{\Pi V}$, from Eqs.~(\ref{bf_zeta})--(\ref{bf_lambda_Pi_pi}) and Eqs.~(\ref{bf_zeta_bar})--(\ref{bf_lambda_bar_Pi_pi}), for a classical ideal gas with conserved particle number (solid lines and solid symbols) and without conserved particle number (dashed lines and empty symbols), as a function of $z=m_0/T$.
		The coefficient $\zeta$ is multiplied by $150$ for better visibility.
		All coefficients corresponding to the basis-free (BF) approximation are represented by red lines and circles, those computed within the standard DNMR approximations are represented by blue lines and squares, respectively.
		The solid black lines show the series approximation of the respective coefficients.
				\label{fig:tcoeffs_scalar}
	}
\end{figure*}
%%%-Bulk-figures%%%

%%%-Diffusion-figures-%%%
\begin{figure*}[!htb]
	\begin{center}
		\begin{tabular}{ccc}
			\includegraphics[width=.33\linewidth]{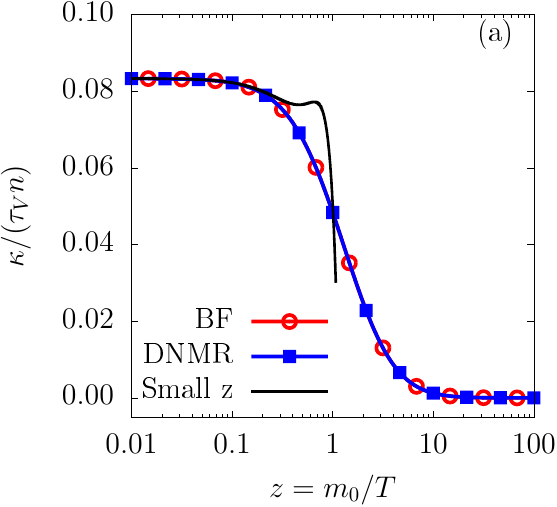} &
			\includegraphics[width=.33\linewidth]{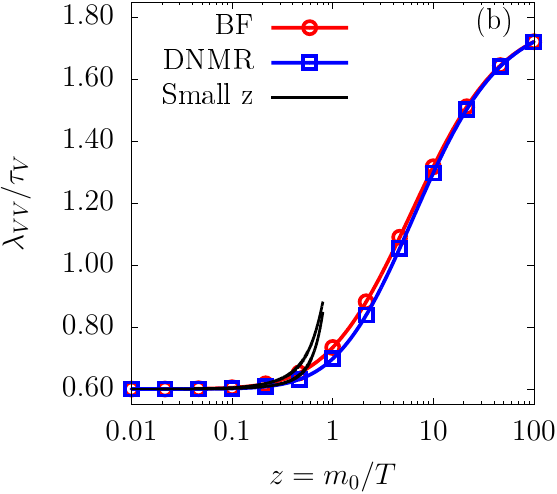} &
			\includegraphics[width=.33\linewidth]{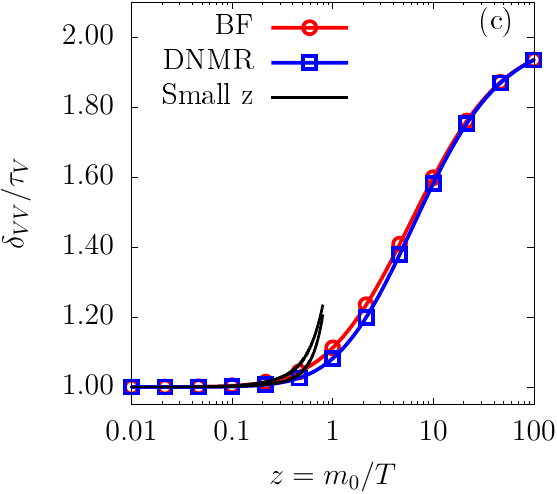} \\
			\includegraphics[width=.33\linewidth]{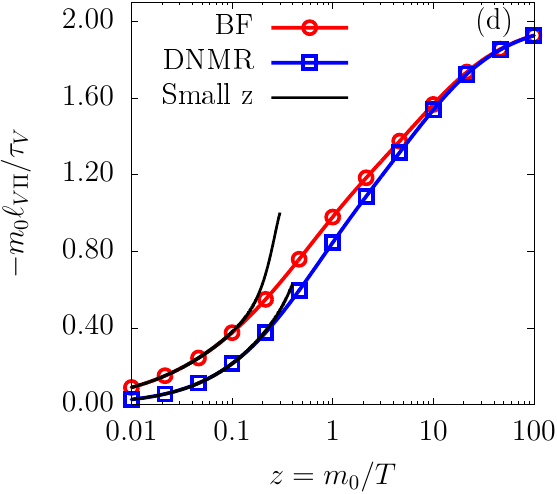} &
			\includegraphics[width=.33\linewidth]{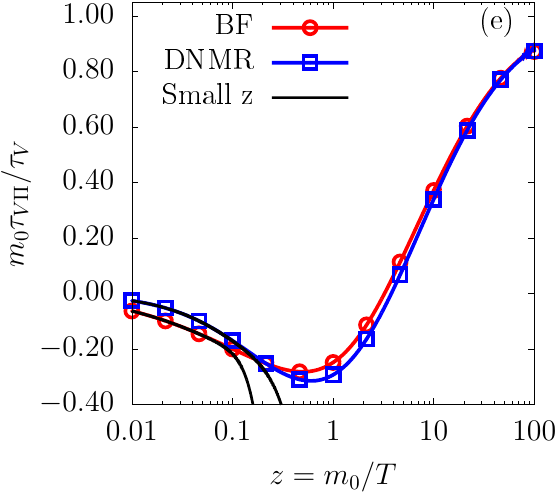} &
			\includegraphics[width=.33\linewidth]{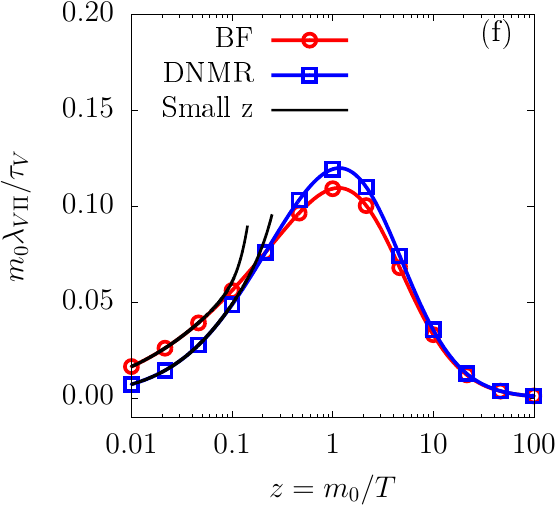} \\
			\includegraphics[width=.33\linewidth]{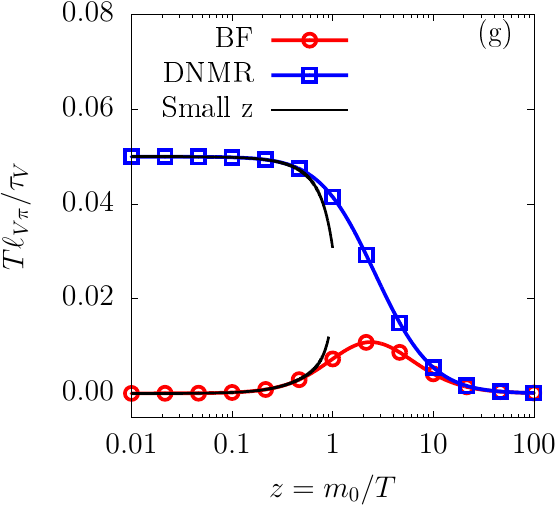} &
			\includegraphics[width=.33\linewidth]{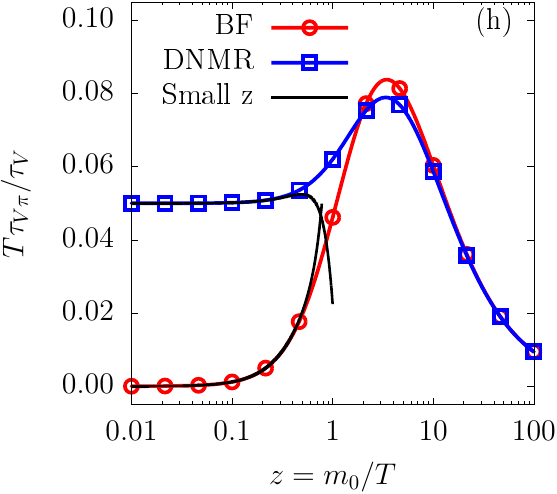} &
			\includegraphics[width=.33\linewidth]{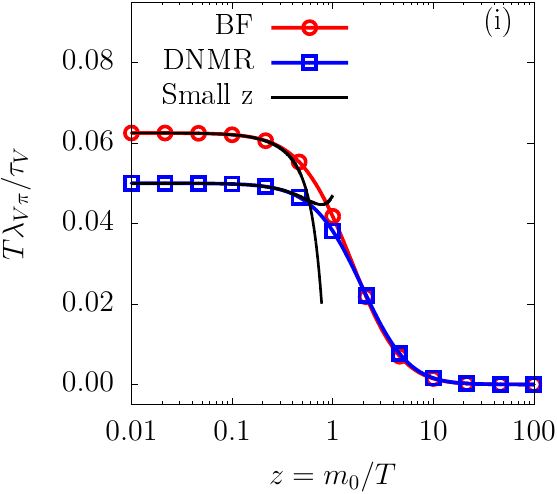}
		\end{tabular}
	\end{center}
	\caption{
		Same as Fig.~\ref{fig:tcoeffs_scalar} for the transport coefficients $\kappa$, $\lambda_{VV}$, $\delta_{VV}$, $\ell_{V\Pi}$, $\tau_{V\Pi}$, $\lambda_{V\Pi}$,
		$\ell_{V\pi}$, $\tau_{V\pi}$, and $\lambda_{V\pi}$ appearing in the particle-diffusion equation \eqref{Vdot} and defined in Eqs.~(\ref{bf_kappa})--(\ref{bf_lambda_V_pi}).
		\label{fig:tcoeffs_diff}
	}
\end{figure*}
%%%-Diffusion-figures-%%%

%%%-Shear-figures-%%%
\begin{figure*}[!htb]
	\begin{center}
		\begin{tabular}{cc}
			\includegraphics[width=.33\linewidth]{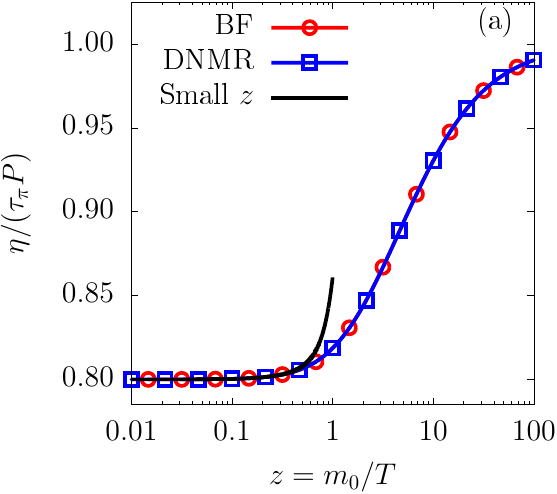} &
			\includegraphics[width=.33\linewidth]{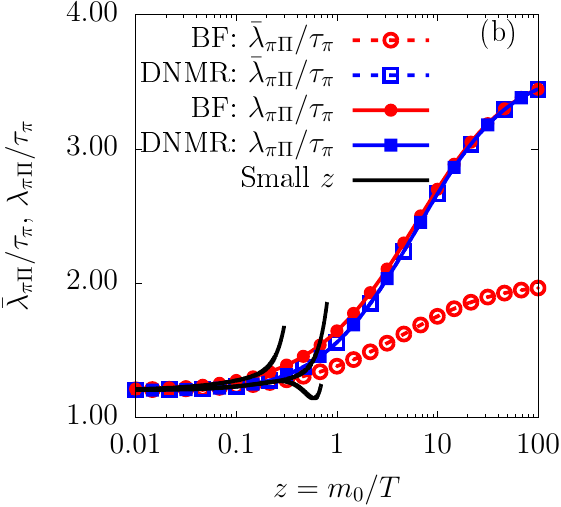} \\
			\includegraphics[width=.33\linewidth]{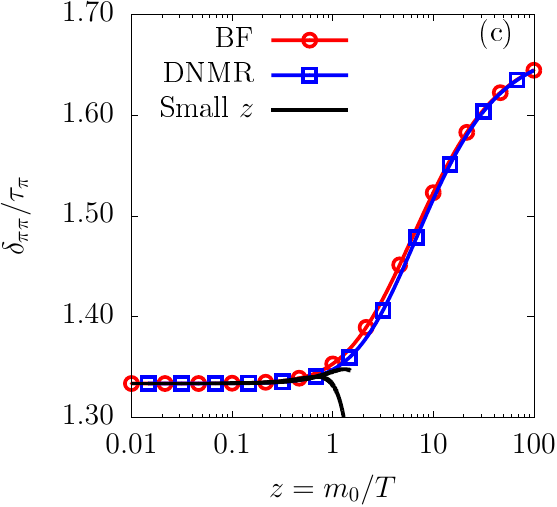} &
			\includegraphics[width=.33\linewidth]{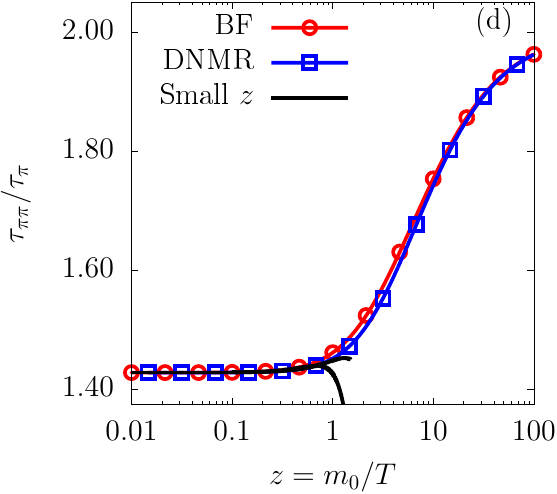}
		\end{tabular}
		\begin{tabular}{ccc}
			\includegraphics[width=.33\linewidth]{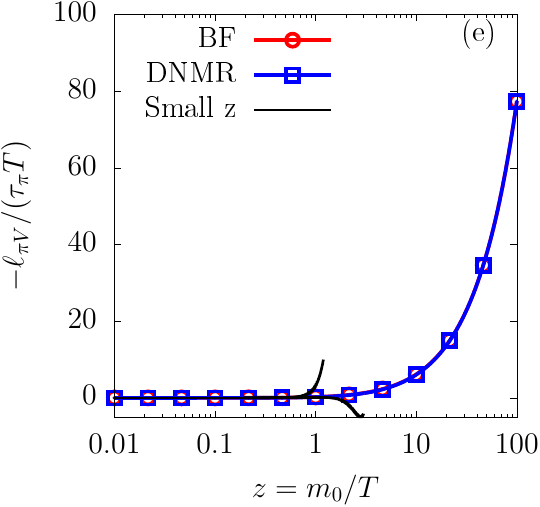} &
			\includegraphics[width=.33\linewidth]{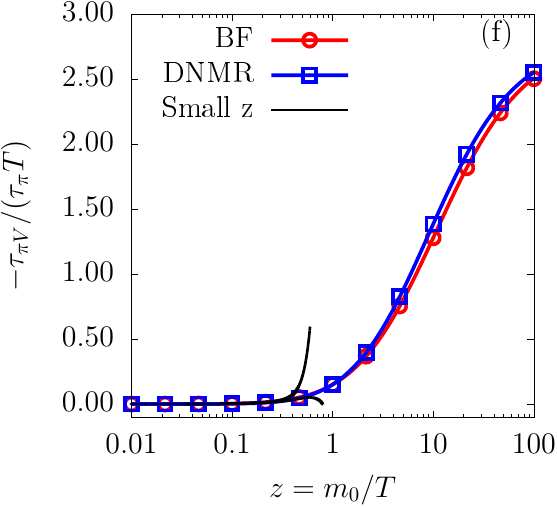} &
			\includegraphics[width=.33\linewidth]{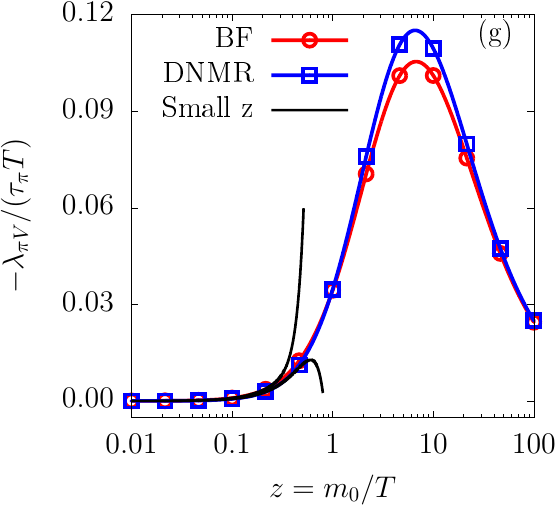}
		\end{tabular}
	\end{center}
	\caption{
		Same as Fig.~\ref{fig:tcoeffs_scalar} for the transport coefficients $\eta$, $\lambda_{\pi\Pi}$, $\bar{\lambda}_{\pi\Pi}$, $\delta_{\pi\pi}$, $\tau_{\pi\pi}$, $\ell_{\pi V}$, $\tau_{\pi V}$, and $\lambda_{\pi V}$ listed in Eqs.~(\ref{bf_eta})--(\ref{bf_lambda_pi_V}) and Eq.~(\ref{bf_lambda_bar_pi_Pi}).
		\label{fig:tcoeffs_shear}
	}
\end{figure*}
%%%-Shear-figures-%%%

%%%-MHD-figures-%%%
\begin{figure*}[t]
\begin{center}
\begin{tabular}{cc}
\includegraphics[width=.33\linewidth]{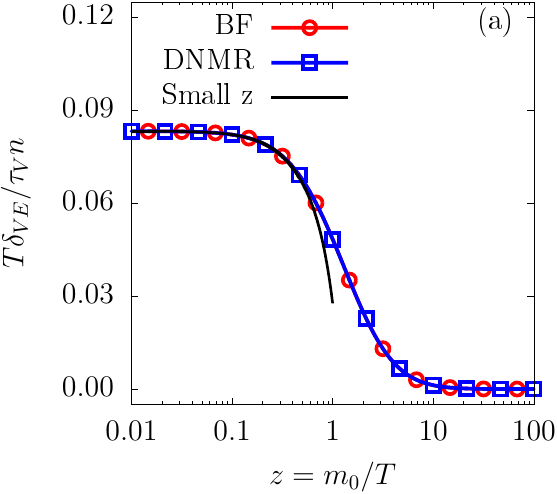} &
\includegraphics[width=.33\linewidth]{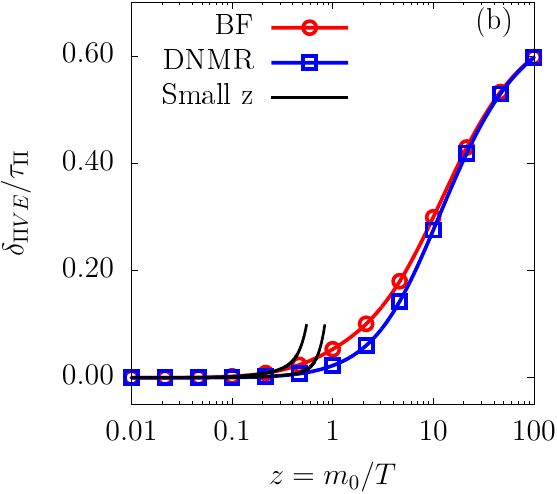} \\
\includegraphics[width=.33\linewidth]{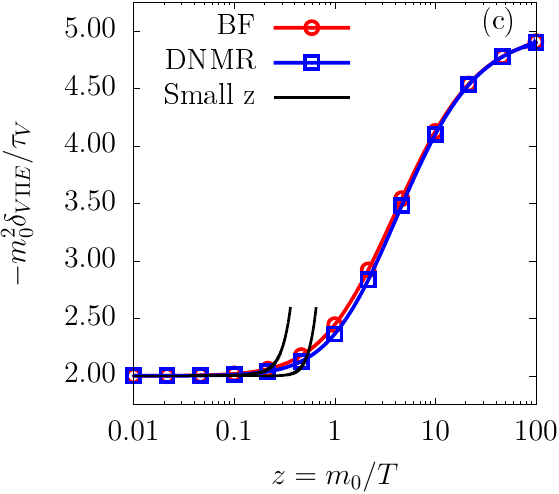} &
\includegraphics[width=.33\linewidth]{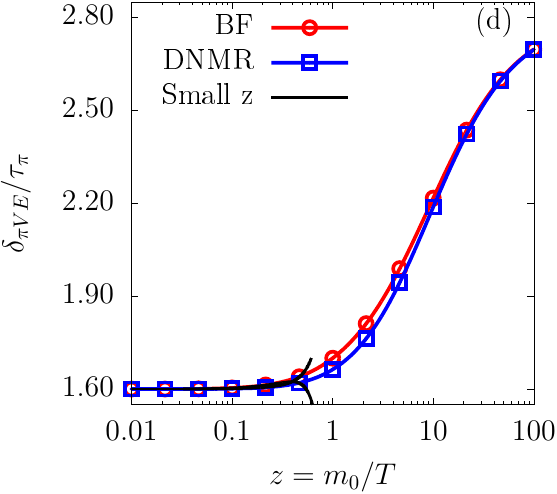}
\end{tabular}
\begin{tabular}{ccc}
 \includegraphics[width=.33\linewidth]{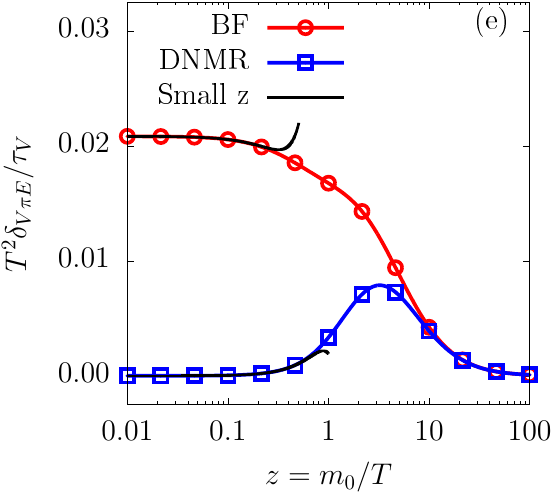} &
 \includegraphics[width=.33\linewidth]{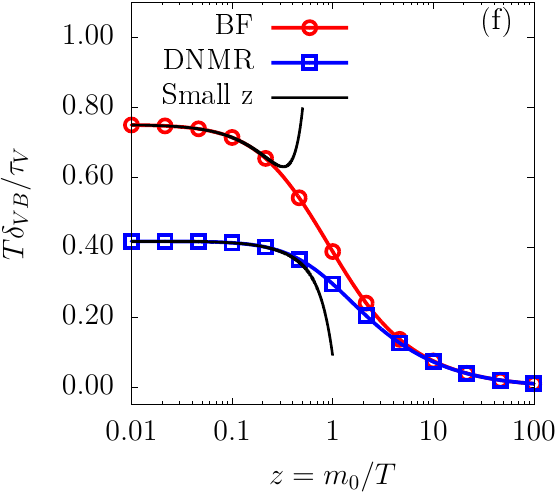} &
 \includegraphics[width=.33\linewidth]{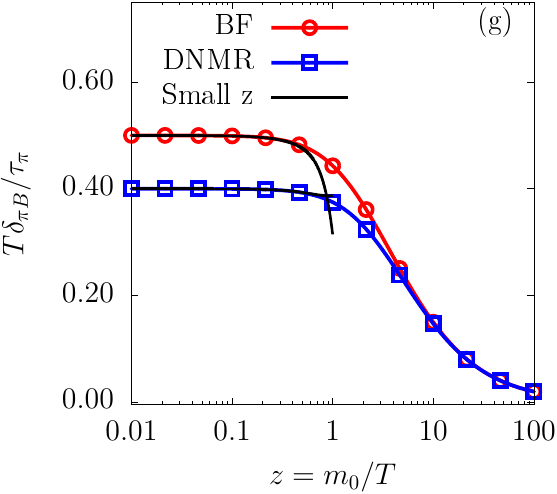}
\end{tabular}
\end{center}
\caption{
Same as Fig.~\ref{fig:tcoeffs_scalar} for the transport coefficients $\delta_{VE}$, $\delta_{\Pi V E}$, $\delta_{V\Pi E}$, $\delta_{\pi V E}$,  $\delta_{V \pi E}$, $\delta_{VB}$, and $\delta_{\pi B}$,
of a charged fluid in an external electromagnetic field.
\label{fig:tcoeffs_mhd}
}
\end{figure*}
%%%-MHD-figures-%%%

%%%
\section{The transport coefficients of the relaxation equations}\label{sec:tcoeffs}

In this section we will investigate all first- and second-order transport coefficients found in the equations of motion for the bulk viscous pressure (\ref{Pidot}), the diffusion current (\ref{Vdot}), and the shear-stress tensor (\ref{pidot}).
Based on the results of Sec.~\ref{sec:with_particle_conservation} we compare the coefficients obtained from the basis-free approximation with those obtained using the standard DNMR approximation.
These choices will be denoted by BF and DNMR in the legends, and plotted with solid red and solid blue lines with symbols, respectively.

The DNMR functions $\gamma^{(\ell)}_{r0}$ do not explicitly depend on whether the particle number is conserved or not.
This is also true for the basis-free coefficients $\mathcal{R}^{(1)}_{-r,0}$ and $\mathcal{R}^{(2)}_{-r,0}$.
However, the coefficients $\mathcal{R}^{(0)}_{-r,0}$ and $\alpha^{(0)}_{-r}$ computed for a gas with conserved particle number change into $\bar{\mathcal{R}}^{(0)}_{-r,0}$ and $\bar{\alpha}^{(0)}_{-r}$ when the particle number is not conserved, as shown in Eqs.~(\ref{repl2}).
Therefore, even the first-order transport coefficients $\zeta_r$ change into $\bar{\zeta}_r$, since they explicitly depend on $\alpha^{(0)}_{r}$ or $\bar{\alpha}^{(0)}_{r}$; see Sec.~\ref{sec:without_particle_conservation}.
The results for the case without particle-number conservation, obtained within the BF and the DNMR 
approaches, are plotted with dashed red and dashed blue lines and empty symbols, respectively.

Furthermore, we also present the small $z=m_0/T \ll 1$ approximations of all transport coefficients in the second-order equations of motion (\ref{Pidot})--(\ref{pidot}).
Using the series expansion of the Bessel functions for $z \ll 1$, we obtain all transport coefficients 
up to $O(z^4)$.
Note that in most cases the series approximations lose validity beyond $z > 0.5$.
The relevant terms appearing in these series approximations are summarized in Tables~\ref{tbl:tcoeffs_Pi_L}--\ref{tbl:tcoeffs_mhd} of Appendix~\ref{app:num} and are plotted with solid black lines in all figures.

The results for all transport coefficients will be shown as dimensionless ratios, by dividing by the corresponding relaxation times (and sometimes functions of $T$ and $z$), such that they become functions of $z$ only.
Therefore, even if the relaxation times are computed not in RTA but in some other approximation, e.g., using the binary collision integral for a constant cross section, the results for the transport coefficients, 
scaled in the same way, are identical.

%%%
\subsection{The coefficients of the equation for the bulk viscous pressure}

Figure~\ref{fig:tcoeffs_scalar}(a) shows the dimensionless ratios $\zeta /[\tau_\Pi P]$ and $\bar{\zeta}/[\tau_\Pi P]$.
Since these are first-order transport coefficients, both the BF and the DNMR approaches lead to the same result.
However, comparing the results corresponding to $\mu \neq 0$ and $\mu = 0$, i.e., with and without particle-number conservation, we see that $\zeta$ is smaller than $\bar{\zeta}$ by more than 2 orders of magnitude.
Note that we multiply $\zeta$ by a factor of $X = 150$ to show it on the same scale as $\bar{\zeta}$.
This order-of-magnitude discrepancy develops around $z \simeq 1$, where neither the small-$z$ nor the large-$z$ expansions are reliable. Indeed, Table~\ref{tbl:tcoeffs_Pi_L} indicates that for small $z$, 
the difference $(\bar{\zeta} - \zeta) / \tau_\Pi P \simeq z^4 / 36$; i.e., it increases $\sim z^4$. 
Using Eqs.~\eqref{mass_zeta} and \eqref{mass_bar_zeta} we obtain
\begin{equation}
 \frac{\bar{\zeta}}{\tau_\Pi P} = \frac{\zeta}{\tau_\Pi P} + h\beta (c_s^2 - \bar{c}_s^2),
\end{equation}
where $\beta h = 1+ e/P$, while the large-$z$ behaviours of $\bar{\zeta}$ and $\zeta$ are fundamentally different:
\begin{align}
 \left.\frac{\zeta}{\tau_\Pi P}\right\rfloor_{z \gg 1} &= \frac{5}{6z^2} + O(z^{-3}), \nonumber\\
 \left.\frac{\bar{\zeta}}{\tau_\Pi P}\right\rfloor_{z \gg 1} &= \frac{2}{3} - \frac{11}{3z} + \frac{11}{z^2} + O(z^{-3}).
 \label{eq:zeta_zetabar_ratios}
\end{align}
Panel (a) of Fig.~\ref{fig:tcoeffs_scalar} confirms that $\bar{\zeta} / (\tau_\Pi P)$ indeed approaches $2/3$ as $z \rightarrow \infty$, while $\zeta / (\tau_\Pi P)$ drops to $0$.

The  coefficients $\delta_{\Pi\Pi}/\tau_\Pi$ and $\bar{\delta}_{\Pi\Pi}/\tau_\Pi$, given in Eqs.~\eqref{delta_PiPi} and \eqref{bar_delta_PiPi}, respectively, are shown in 
Fig.~\ref{fig:tcoeffs_scalar}(b).
The results corresponding to the BF and the DNMR approaches are in very good agreement when $\mu \neq 0$, 
but they differ substantially when $\mu = 0$ and the particle number is not conserved.
Furthermore, Fig.~\ref{fig:tcoeffs_scalar}(c) shows $\lambda_{\Pi\pi}/\tau_\Pi$ and $\bar{\lambda}_{\Pi\pi}/\tau_\Pi$ from Eqs.~\eqref{lambda_Pi_pi} and \eqref{bar_lambda_Pi_pi}.
Significant differences can be seen in the behavior between the cases $\mu \neq 0$ and $\mu = 0$.
The absolute values of $\lambda_{\Pi\pi}$ are almost an order of magnitude smaller than those of $\bar{\lambda}_{\Pi\pi}$, and their sign is different, the former is negative and the latter positive.
The sign mismatch is consistent between the two approaches, as both the basis-free and the DNMR expressions for $\lambda_{\Pi\pi}$ ($\bar{\lambda}_{\Pi\pi}$) yield a negative (positive) value in the $z \rightarrow 0$ limit. Please note that the sign of $\lambda_{\Pi\pi}$ is negative in the $z \rightarrow 0$ limit also in the more realistic case of $2\rightarrow 2$ hard-sphere interaction model; see last column in Table~V of Ref.~\cite{Wagner:2023joq}.

Figures~\ref{fig:tcoeffs_scalar}(d)--\ref{fig:tcoeffs_scalar}(f) show the second-order coefficients from Eq.~\eqref{Pidot} that couple to the diffusion current, $\ell_{\Pi V}$, $\tau_{\Pi V}$, and $\lambda_{\Pi V}$, respectively.
Here, the coefficients $\tau_{\Pi V}$ computed in  the BF and the DNMR approaches are in very good agreement with each other, while $\ell_{\Pi V}$ and  $\lambda_{\Pi V}$ differ in the two approaches.

The series approximations for the first- and second-order transport coefficients in the equation for the bulk viscous pressure, i.e., Eqs.~(\ref{bf_zeta})--(\ref{bf_lambda_Pi_pi}) and Eqs.~(\ref{bf_zeta_bar})--(\ref{bf_lambda_bar_Pi_pi}) are presented in Table~\ref{tbl:tcoeffs_Pi_L} for the BF (middle column) and the DNMR (right column) approaches, and are plotted with solid black lines in Fig.~\ref{fig:tcoeffs_scalar}.

The bulk-viscosity coefficient, $\zeta$ or $\bar{\zeta}$, is of order $O(z^4)$, while $\delta_{\Pi \Pi}$ and $\bar{\delta}_{\Pi \Pi}$ have well-defined nonvanishing massless limits up to $O(z)$ or $O(z^2\ln z)$, depending on the method employed to obtain this second-order coefficient.
Note that those limits are identical for $\delta_{\Pi \Pi}$ and $\bar{\delta}_{\Pi \Pi}$.
All other coefficients vanish in the massless limit.
Similarly, $\ell_{\Pi V}$ and $\tau_{\Pi V}$ show differences already at $O(z^2)$ due to the choice of negative moments in these transport coefficients.
The remaining transport coefficients, $\lambda_{\Pi V}$,  $\lambda_{\Pi \pi}$, and $\bar{\lambda}_{\Pi \pi}$ are of order $O(z^2)$ and exhibit similar characteristics.

%%%
\subsection{The coefficients of the particle diffusion equation}

The dimensionless ratio of the particle diffusion coefficient and the product of relaxation time and particle density, $\kappa/[\tau_V n]$, is shown in Fig.~\ref{fig:tcoeffs_diff}(a).
One observes perfect agreement between the BF and DNMR approaches.

The second-order coefficients, $\lambda_{VV}$ and $\delta_{VV}$, shown in 
Figs.~\ref{fig:tcoeffs_diff}(b) and \ref{fig:tcoeffs_diff}(c), are in  reasonably good agreement between the BF and the DNMR approaches.
Similarly, this conclusion also holds true for the coefficients $\ell_{V \Pi}$, $\tau_{V \Pi}$, and $\lambda_{V \Pi}$, presented in Figs.~\ref{fig:tcoeffs_diff}(d)--\ref{fig:tcoeffs_diff}(f), although they have slightly different values in the massless limit.
The remaining second-order coefficients $\ell_{V\pi}$, $\tau_{V\pi}$, and $\lambda_{V\pi}$, presented
in Figs.~\ref{fig:tcoeffs_diff}(g)--\ref{fig:tcoeffs_diff}(i), start from very different values at $z=0$,
but approach similar values for $z \gg 1$.
The differences seen in the massless limits of $\ell_{V\pi}$, $\tau_{V\pi}$, and  $\lambda_{V\pi}$, also shown in Table~\ref{tbl:tcoeffs_V}, were already clarified in Ref.~\cite{Ambrus:2022vif}, where it was concluded that the proper massless limits of these coefficients are those of the basis-free approximation, and not of the DNMR approach.

The small-$z$ approximations for the transport coefficients appearing in the particle diffusion equation, Eqs.~(\ref{bf_kappa})--(\ref{bf_lambda_V_pi}), are presented in Table~\ref{tbl:tcoeffs_V}.
All transport coefficients have well-defined numerical values in the massless limit $z \rightarrow 0$.
The coefficients $\ell_{V\pi}$ and $\tau_{V\pi}$ differ by $O(z^2)$, hence, in the basis-free approximation
these coefficients vanish when $z \rightarrow 0$ as they should, see Ref.~\cite{Ambrus:2022vif} for more details, while the DNMR results stay finite.

%%%
\subsection{The coefficients of the equation for the shear-stress tensor}

We now discuss the transport coefficients of the equation for the shear-stress tensor.
Figure~\ref{fig:tcoeffs_shear}(a) shows the dimensionless ratio $\eta/[\tau_\pi P]$, which is identical in the BF and the DNMR approaches, also independently on whether particle number is conserved or not.
Figure~\ref{fig:tcoeffs_shear}(b) shows $\lambda_{\pi\Pi}$ and $\bar{\lambda}_{\pi\Pi}$.
We observe that the BF and the DNMR approximations give similar results for $\mu \neq 0$.
On the other hand, the coefficient $\lambda_{\pi\Pi}$ in the basis-free approximation is significantly different without particle-number conservation, being about a factor of $2$ lower at $z = 100$ than its value with conserved particle number.

Figures~\ref{fig:tcoeffs_shear}(c) and \ref{fig:tcoeffs_shear}(d) show $\delta_{\pi\pi} /\tau_\pi$ and $\tau_{\pi\pi} / \tau_\pi$, respectively.
These ratios are independent of particle-number conservation, since they do not involve the coefficients $\alpha^{(0)}_r$ or $\mathcal{R}^{(0)}_{-r,0}$, therefore the results coincide when $\mu = 0$ and $\mu \neq 0$, as expected.
Both coefficients are also in an excellent agreement when comparing the BF and the DNMR approaches.
Figures~\ref{fig:tcoeffs_shear}(e), \ref{fig:tcoeffs_shear}(f),  and \ref{fig:tcoeffs_shear}(g) show 
the second-order coefficients $\ell_{\pi V}$, $\tau_{\pi V}$, and $\lambda_{\pi V}$, which couple to 
the diffusion current.
Similarly as before, these last three coefficients are also in good agreement between the BF and the DNMR approximations.

Finally, the small-$z$ approximations for the transport coefficients from Eqs.~(\ref{bf_eta})--(\ref{bf_lambda_pi_V})
and Eq.~(\ref{bf_lambda_bar_pi_Pi}) are presented in Table~\ref{tbl:tcoeffs_pi_L}.
Most coefficients related to the evolution of the shear-stress tensor differ by a few percent
at $O(z^2)$, while $\lambda_{\pi \Pi}$ and $\bar{\lambda}_{\pi \Pi}$ are different at $O(z)$ when comparing
the BF and the DNMR approximations.

%%%
\subsection{The magnetohydrodynamic coefficients}

Here we provide the magnetohydrodynamic coefficients listed in Sec.~\ref{sec:RTA:MHD}.
Figure~\ref{fig:tcoeffs_mhd} (a) shows the dimensionless ratio $\delta_{VE}/[\tau_V n\beta]$.
The coefficients $\delta_{\Pi V E}$, $\delta_{V\Pi E}$, and $\delta_{\pi V E}$, shown in panels (b)--(d) of the same figure, corroborate the good agreement between the BF and the DNMR approximations.

Figures~\ref{fig:tcoeffs_mhd}(e)--\ref{fig:tcoeffs_mhd}(g) show the second-order coefficients
$\delta_{V \pi E}$, $\delta_{VB}$, and $\delta_{\pi B}$
These coefficients have very different values at $z=0$
but approach the same values at large $z\gg 1$, similarly to the lower row of Fig.~\ref{fig:tcoeffs_diff}.
The small-$z$ approximations for these transport coefficients are presented in Table~\ref{tbl:tcoeffs_mhd}.

%%%
\section{Applications}
\label{sec:app}

In this section, we will consider the (0+1)--dimensional boost-invariant expansion, also known as the Bjorken-flow solution \cite{Bjorken:1982qr}.
We will study the properties of this system using various approaches discussed below.
The relativistic Boltzmann equation in RTA and the details of our numerical solver are discussed in Sec.~\ref{sec:app:boltz} and in Appendix~\ref{app:num:DVM}, respectively.
The equations of second-order fluid dynamics with transport coefficients derived using both the BF and the DNMR approximations are summarized in Sec.~\ref{sec:app:hydro}.
The equations of leading-order anisotropic fluid dynamics (aHydro), based on the Romatschke-Strickland anisotropic distribution function are discussed in Sec.~\ref{sec:app:aniso_hydro}, while details of our numerical implementation are given in Appendix~\ref{app:num:aH}.

The transformation from the usual space-time coordinates $x^\mu = (t,x,y,z)$ to the proper
time and space-time rapidity coordinates $\tilde{x}^\mu = (\tau,x,y,\eta_s)$ reads
$\tau =\sqrt{t^{2}-z^{2}}$ and
$\eta_s =\frac{1}{2}\ln [(t+z)/(t-z)]$, while the inverse transformation corresponds to
$t=\tau \cosh \eta_s $ and $z=\tau \sinh\eta_s $.

For a longitudinally boost-invariant expansion, $v^z_s\equiv z/t=\tanh \eta_s $, such that
\begin{eqnarray}  \label{u_BJ}
	u^{\mu } &\equiv&\left( \frac{t}{\tau} ,0,0,\frac{z}{\tau}\right)
	= \left(\cosh \eta_s ,0,0,\sinh \eta_s \right) \;, \\
	l^{\mu } &\equiv&\left( \frac{z}{\tau} ,0,0,\frac{t}{\tau} \right)
	= \left(\sinh \eta_s ,0,0,\cosh \eta_s \right) \; ,  \label{l_BJ}
\end{eqnarray}
and all thermodynamic quantities are independent of $\eta_s$.

In this case, $D\equiv u^{\mu }\partial _{\mu } =\frac{\partial }{\partial \tau }$, $D_{l}\equiv -l^{\mu }\partial _{\mu } =-\frac{\partial }{\tau \partial \eta_s }$, while $Du^{\mu }=Dl^\mu = 0$, $D_{l}u^{\mu }=-\frac{1}{\tau } l^{\mu }$, $D_{l}l^{\mu }=-\frac{1}{\tau }u^{\mu }$.

Furthermore, the expansion rates are $\theta \equiv \nabla_\mu u^\mu = 1/\tau$, $\tilde{\theta} \equiv \tilde{\nabla}_\mu u^\mu = 0$, and $\tilde{\theta}_l \equiv \tilde{\nabla}_\mu l^\mu = 0$.
The vorticity tensors also vanish, $\omega^{\mu \nu} \equiv  \nabla^{\left[ \mu \right.} u^{\left.\nu \right]} = 0$ and $\tilde{\omega}^{\mu \nu}  \equiv  \tilde{\nabla}^{\left[ \mu \right.} u^{\left.\nu \right]} = 0$.

The four-momentum is expressed in terms of the rapidity variable, $y \equiv \frac{1}{2}\ln [(k^0 + k^z)/(k^0 - k^z)]$, hence the longitudinal velocity of the particle is $v^z_{p} = \tanh y$.
Thus, the four-momentum $k^{\mu} = (k^0,k^x,k^y,k^z)$ is expressed in terms of the particle rapidity as
\begin{equation}
	k^{\mu} \equiv \left(m_\perp \cosh y, k^x, k^y, m_\perp \sinh y \right) \; ,
\end{equation}
where $m_\perp \equiv \sqrt{m^2_0 + k^2_\perp} = \sqrt{m^2_0 + k^2_x + k^2_y} $ denotes the transverse mass.
Therefore, the energy of the particle is $E_{\bk u} \equiv k^\mu u_\mu = m_\perp \cosh (y -\eta_s)$, while its momentum in the direction of the anisotropy is $E_{\bk l} \equiv - k^\mu l_\mu = m_\perp \sinh (y -\eta_s)$.

The coordinate transformation $\tilde{k}^{\mu} = \left(\frac{\partial \tilde{x}^\mu}{\partial x^\nu }\right) k^\nu $ of the four-momentum to proper time and space-time rapidity coordinates $\tilde{k}^{\mu} = (k^\tau,k^x,k^y,k^{\eta_s})$ leads to
\begin{equation}
	\tilde{k}^{\mu} \equiv \left(m_\perp \cosh (y -\eta_s), k^x, k^y, \frac{m_\perp}{\tau} \sinh (y - \eta_s) \right) \; ,
\end{equation}
hence $k^\tau = m_\perp \cosh(y - \eta_s)$ and $k^{\eta_s} = \frac{m_\perp}{\tau} \sinh(y - \eta)$.

%%%
\subsection{The Boltzmann equation for the (0+1)--dimensional boost-invariant expansion}
\label{sec:app:boltz}

The relativistic Boltzmann equation~\eqref{BTE} in RTA~\eqref{AW} for a longitudinally
boost-invariant system reduces to
\begin{equation}
	\frac{\partial f_\bk}{\partial \tau} - \frac{v^z}{\tau} (1 - v_z^2)
	\frac{\partial f_\bk}{\partial v_z} = -\frac{1}{\tau_R} (f_\bk - f_{0\bk}) \; ,
	\label{eq:boltz_tau}
\end{equation}
where we assumed homogeneity in the transverse plane, such that $f_\bk \equiv f_\bk(\tau; m_\perp, \varphi_\perp, v^z)$ is a function of the proper time $\tau$.
The momentum space is parametrized using $m_\perp = (k_x^2 + k_y^2 + m_0^2)^{1/2}$ and $\varphi_\perp = \arctan(k^y / k^x)$, while the longitudinal component of the three-velocity relative to the flow velocity
$v^z_s =\tanh \eta_s$
is defined as
\begin{equation}
	v^z \equiv \tanh(y - \eta_s) \;.
\end{equation}

We are interested in tracking the evolution of the conserved particle four-current $N^{\mu}$,
and the energy-momentum tensor $T^{\mu\nu}$.
These are obtained via the following momentum-space integrals,
\begin{equation}
	N^{\mu} = \frac{g}{(2\pi)^3} \! \int_{m_0}^\infty \mathrm{d}m_\perp m_\perp \! \int_0^{2\pi} \mathrm{d}\varphi_\perp
	\! \int_{-1}^1 \frac{\mathrm{d}v^z}{1 - v_z^2} k^\mu f_\bk \;,
\end{equation}
and
\begin{equation}
	T^{\mu\nu} = \frac{g}{(2\pi)^3} \! \int_{m_0}^\infty \mathrm{d}m_\perp m_\perp \! \int_0^{2\pi} \mathrm{d}\varphi_\perp
	\! \int_{-1}^1 \frac{\mathrm{d}v^z}{1 - v_z^2} k^\mu k^\nu f_\bk \;,
\end{equation}
where we used $\mathrm{d}^3 \mathbf{k}/k^0 \equiv m_\perp \mathrm{d} m_\perp \mathrm{d}\varphi_\perp \mathrm{d}y $.

Taking into account the above form of the particle-four current and the diagonal structure of the energy-momentum tensor,
$T^{\mu\nu}_{\textrm{LRF}} = {\rm diag} (e, P_\perp, P_\perp, \tau^{-2} P_l)$, the particle number density and energy density can be obtained via
\begin{gather}
	n = \int_{-1}^1 \mathrm{d}v^z\, F_1 \;, \quad
	e = \int_{-1}^1 \mathrm{d}v^z\, F_2 \;, \label{eq:bjork_ne_from_F}
\end{gather}
while the pressure in the longitudinal and transverse directions are given by
\begin{gather}
	P_l = \int_{-1}^1 \mathrm{d}v^z \,v_z^2 F_2 \;, \quad P_\perp = \frac{1}{2}(e - P_l - T^\mu_\mu) \; ,
	\label{eq:Pl_from_F}
\end{gather}
where $	T^\mu_\mu = m_0^2 \int_{-1}^1 \mathrm{d}v^z \,F_0$.

The functions $F_n$ appearing in Eqs.~\eqref{eq:bjork_ne_from_F}--\eqref{eq:Pl_from_F} are obtained by integrating over $f_\bk$,
\begin{equation}
	F_n \equiv \frac{g}{(2\pi)^3} \int_0^{2\pi} \mathrm{d}\varphi_\perp \int_{m_0}^\infty
	\frac{\mathrm{d}m_\perp\, m_\perp^{n+1} }{(1 - v_z^2)^{(n+2)/2}} f_\bk \; ,
	\label{eq:Fn}
\end{equation}
and satisfy the following equation,
\begin{multline}
	\frac{\partial F_n}{\partial \tau} + \frac{1}{\tau}[1+ (n-1) v_z^2] F_n
	- \frac{1}{\tau} \frac{\partial[v^z(1 - v_z^2) F_n]}{\partial v^z} \\
	= -\frac{1}{\tau_R} (F_n - F_n^{\text{eq}}) \; .
	\label{eq:boltz_Fn}
\end{multline}
The functions $F_n^{\text{eq}}$ are obtained substituting the equilibrium distribution function \eqref{f_0k} into Eq.\eqref{eq:Fn}:
\begin{equation}
	F_n^{\text{eq}} = \frac{g e^\alpha}{4\pi^2} T^{n+2} \,
	\Gamma\left(n + 2, \frac{m_0 / T}{\sqrt{1 - v_z^2}}\right) \; ,
\end{equation}
where $\Gamma(n,x) = \int_{x}^{\infty} t^{n-1} e^{-t} \mathrm{d}t$ denotes the incomplete Gamma function.
Note that the numerical algorithm for solving Eq.~\eqref{eq:boltz_Fn} is presented in detail in Appendix~\ref{app:num:DVM}.

%%%
\subsection{Second-order fluid dynamics for the (0+1)--dimensional boost-invariant expansion}
\label{sec:app:hydro}

In the (0+1)--dimensional boost-invariant expansion of matter, the fluid-dynamical equations simplify substantially.
The conservation equations (\ref{cons_eqs}) reduce to
\begin{align}
	& D n + \frac{n}{\tau} = 0 \; , \label{Dn_cons}\\
	& D e + \frac{1}{\tau} \left( e + P_l \right) = 0\ . \label{De_cons}
\end{align}
Here, the particle-number density and energy density were defined in Eq.~\eqref{P0}, while $P_l$ is the pressure component in the longitudinal direction.
The latter and the transverse pressure component $P_\perp$ are related to the thermodynamic pressure $P$,
bulk pressure $\Pi$, and the shear-stress tensor component $\pi \equiv \pi^{\eta_s}_{\eta_s}$ via
\begin{equation}
	P_l \equiv P + \Pi - \pi \; , \quad
	P_\perp \equiv P + \Pi + \frac{\pi}{2} \; . \label{Pl_PT}
\end{equation}

The second-order relaxation equations~\eqref{Pidot} and \eqref{pidot} reduce to,
see Eqs.~(9), (10) in Ref.~\cite{Denicol:2014mca},
\begin{align}
\label{Pidot_BJ_bulk}
\tau_R D \Pi + \Pi &= -\frac{\zeta}{\tau} - \delta_{\Pi\Pi} \frac{ \Pi}{\tau}
+ \lambda_{\Pi \pi} \frac{\pi}{\tau} \; ,  \\
\label{pidot_BJ_shear}
\tau_R D \pi + \pi
&= \frac{4 \eta}{3\tau} - \delta_{\pi\pi} \frac{\pi}{\tau}
- \tau_{\pi\pi} \frac{\pi}{3\tau}
+ \lambda_{\pi \Pi} \frac{2\Pi}{3\tau} \; ,
\end{align}
where we replaced $\tau_\Pi = \tau_\pi = \tau_R$.
The transport coefficients for the above equations are given in Eqs.~(\ref{mass_zeta}), (\ref{mass_eta}), (\ref{delta_PiPi})--(\ref{lambda_Pi_pi}), and (\ref{delta_pi_pi})--(\ref{lambda_pi_Pi}).
We note again that, for a chosen rest mass $m_0$, all transport coefficients only depend on $\mu$ and $T$.

In the case without explicit particle-number conservation, we do not consider Eq.~(\ref{Dn_cons}), but we solve Eq.~(\ref{De_cons}) together with the relaxation equations that follow from Eqs.~\eqref{Pidot_mu=0_bulk} and~\eqref{pidot_mu=0_shear}. For a (0+1)--dimensional boost-invariant expansion these correspond to the relaxation equations \eqref{Pidot_BJ_bulk} and \eqref{pidot_BJ_shear}, but with different coefficients as explained in Sec.~\ref{sec:without_particle_conservation}.
Therefore, in this case the corresponding transport coefficients are given in Eqs.~(\ref{mass_bar_zeta}), (\ref{mass_eta}), (\ref{bar_delta_PiPi}), (\ref{bar_lambda_Pi_pi}), (\ref{delta_pi_pi}), (\ref{tau_pi_pi}), and (\ref{bar_lambda_pi_Pi}).
In Sec.~\ref{sec:res_hydro} we will study the solutions of second-order fluid dynamics in the case of explicit particle-number conservation as well as without it, using both the BF and the DNMR approaches for the transport coefficients.

%%%
\subsection{Anisotropic fluid dynamics for the (0+1)--dimensional boost-invariant expansion}
\label{sec:app:aniso_hydro}

In this section, we turn our attention to study the equations of leading-order anisotropic fluid dynamics in the case of the boost-invariant expansion scenario.
The Romatschke-Strickland distribution function given in Eq.~\eqref{f_RS} reduces to
\begin{equation}
 \hat{f}_{RS} = \exp\left(\hat{\alpha} -\frac{k^\tau}{\Lambda} \sqrt{1 + \xi v_z^2}\right)\;. \label{eq:RS}
\end{equation}
The corresponding equation of motion for the irreducible moments, $\hat{I}_{i+j,j,0}$, was derived in Eqs.\ (53), (54) of Ref.~\cite{Molnar:2016gwq} and is reproduced below,
\begin{multline}
	D \hat{I}_{i+j,j,0} + \frac{1}{\tau }\left[
	\left( j+1\right) \hat{I}_{i+j,j,0}+\left( i-1\right) \hat{I}_{i+j,j+2,0}
	\right] \\ = -\frac{1}{\tau_R} \left( \hat{I}_{i+j,j,0}-I_{i+j,j,0}\right)\; . \label{Main_eq_motion}
\end{multline}
The equations for the particle-number density, the energy density, and the longitudinal pressure component follow by setting $(i,j) = (1,0)$, $(2,0)$, and $(0,2)$ in Eq.~\eqref{Main_eq_motion}, respectively,
\begin{align}
	D \hat{n} + \frac{\hat{n}}{\tau } &= -\frac{1}{\tau_R}(\hat{n} - n)\;,\label{aH_Dn} \\
	D \hat{e} + \frac{1}{\tau }\left( \hat{e} +\hat{P}_{l}\right) &= -\frac{1}{\tau_R}(\hat{e} - e)\;,\label{aH_De}
\end{align}
and
\begin{align}
	 D \hat{P}_{l} + \frac{1}{\tau }\left( 3\hat{P}_{l}-\hat{I}_{240}^{RS}\right)  &= -\frac{1}{\tau_R}\left( \hat{P}_{l}-P\right) \;,  \label{aH_DPl}
\end{align}
where $\hat{I}^{RS}_{240}$ is
\begin{equation}
	\hat{I}_{240}^{RS} = \int
	\mathrm{d}K\, E_{\mathbf{k}u}^{-2} E_{\mathbf{k}l}^4 \, \hat{f}_{RS}\;.  \label{I_240_RS}
\end{equation}
Note that in the case of particle-number conservation, the Landau matching conditions from Eq.~\eqref{aniso_matching} require $\hat{n} = n$, as well as $\hat{e} = e$; hence the right-hand side of Eqs.~\eqref{aH_Dn} and \eqref{aH_De} vanish.
If the particle number is not conserved, then in general $\hat{n}\neq n$.
However, we still impose Landau matching for the energy, $\hat{e} = e$, such that the right-hand side of Eq.~\eqref{aH_De} vanishes.

Note that instead of using Eq.~\eqref{aH_DPl} there are other choices to close the conservation equations, e.g., using higher moments of the Boltzmann equation.
It was shown in Refs.~\cite{Florkowski:2014bba,Molnar:2016gwq,Niemi:2017stb} that some of these choices lead to very similar results as when using Eq.~(\ref{aH_DPl}).

Furthermore, similarly to Eqs.~\eqref{Pl_PT}, the bulk viscous pressure from Eq.~\eqref{aniso_bulk} and the shear-stress tensor component $\hat{\pi}$ are given by
\begin{equation}
 \hat{\Pi} = \frac{1}{3} \left( \hat{P}_{l}+2\hat{P}_{\perp }\right) - P \; , \quad
 \hat{\pi} = \frac{2}{3}(\hat{P}_\perp - \hat{P}_l)\;,
 \label{eq:RS_Pi_and_pi}
\end{equation}
where the transverse pressure component is
\begin{equation}
	\hat{P}_{\perp} \equiv \hat{I}_{201}^{RS}
	= \frac{1}{2}\left(e - \hat{P}_l - m_0^2 \hat{I}_{000}^{RS}\right) \; ,
\end{equation}
where $\hat{I}_{000}^{RS} = \int \mathrm{d}K\, \hat{f}_{RS}$.

Now introducing the integrals of the anisotropic distribution function, $\hat{F}^{RS}_n$, similar to Eq.~\eqref{eq:Fn},
\begin{align}
 \hat{F}^{RS}_n &\equiv \frac{g}{(2\pi)^3} \int_0^{2\pi} \mathrm{d}\varphi_\perp \int_{m_0}^\infty
 \frac{ \mathrm{d}m_\perp\, m_\perp^{n+1} }{(1 - v_z^2)^{(n+2)/2}} \hat{f}_{RS}\; \nonumber\\
 &= \frac{g e^{\hat{\alpha}}}{4\pi^2}
 \left(\frac{\Lambda}{\sqrt{1 + \xi v_z^2}}\right)^{n+2}
\Gamma\left(n + 2, \lambda \right),
 \label{eq:Fn_RS}
\end{align}
where $\lambda = m_0 \sqrt{1 + \xi v_z^2} / \left(\Lambda \sqrt{1 - v_z^2} \right)$, the quantities $\hat{I}^{RS}_{000}$, $\hat{n}$, $\hat{e}$, $\hat{P}_l$, and $\hat{I}^{RS}_{240}$ can be obtained via
\begin{gather}
 \hat{I}^{RS}_{000} = \int_{-1}^1  \mathrm{d}v^z\, \hat{F}^{RS}_0, \quad
 \hat{n} = \int_{-1}^1  \mathrm{d}v^z\, \hat{F}^{RS}_1, \nonumber\\
 \begin{pmatrix}
  \hat{e} \\ \hat{P}_l \\ \hat{I}^{RS}_{240}
 \end{pmatrix} = \int_{-1}^1  \mathrm{d}v^z
 \begin{pmatrix}
  1 \\\ v_z^2 \\ v_z^4
 \end{pmatrix} \hat{F}^{RS}_2.\label{eq:RS_quant}
\end{gather}
The integrals over $v^z$ must be computed numerically, except for the case of $\hat{n}$, for which similarly to Eq.~\eqref{eq:n} an exact analytical result can be derived \cite{Florkowski:2014bba},
\begin{equation}
 \hat{n}= \frac{g e^{\hat{\alpha}}}{2\pi^2} \frac{m_0^2 \Lambda}{\sqrt{1 + \xi}} K_2\left(\frac{m_0}{\Lambda}\right)\;.
 \label{aH_n}
\end{equation}
Note that compared to the above result, Eq.~(25) of Ref.~\cite{Florkowski:2014bba} differs by a factor of 2 accounting for spin degeneracy.
The strategy for solving the equations of anisotropic fluid dynamics is presented 
in Appendix~\ref{app:num:aH}.

%%%-RTA-hydro-figures-%%%
\begin{figure*}
	\begin{tabular}{ccc}
		\hspace{25pt} $m_0 = 0.01\ {\rm GeV}/c^2$ &
		%, $k_B T_0 = 0.5$ GeV &
		%, $\mu_0 = 0$ &
		\hspace{25pt}$m_0 = 1\ {\rm GeV}/c^2$ &
		%, $k_B T_0 = 0.5$ GeV &
		%, $\mu_0 = 0$ &
		\hspace{25pt}$m_0 = 10\ {\rm GeV}/c^2$ \\
		%, $k_B T_0 = 0.5$ GeV \\
		%, $\mu_0 = 0$ \\
		\includegraphics[width=.33\linewidth]{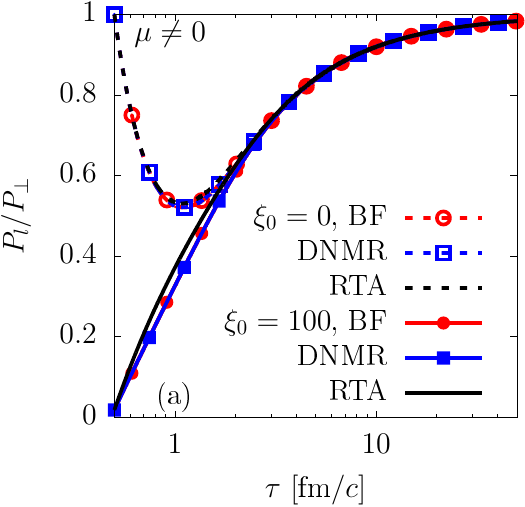} &
		\includegraphics[width=.33\linewidth]{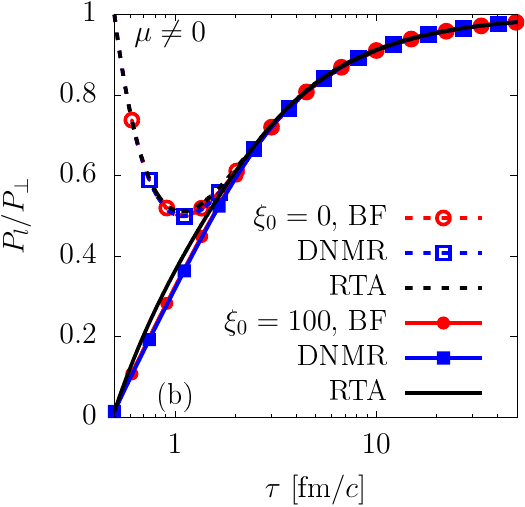} &
		\includegraphics[width=.33\linewidth]{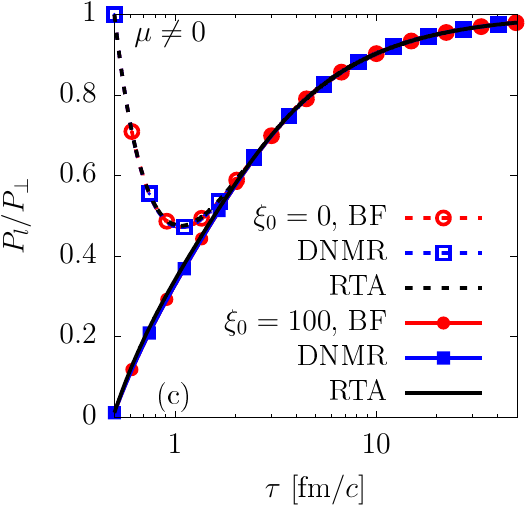} \\
		\includegraphics[width=.33\linewidth]{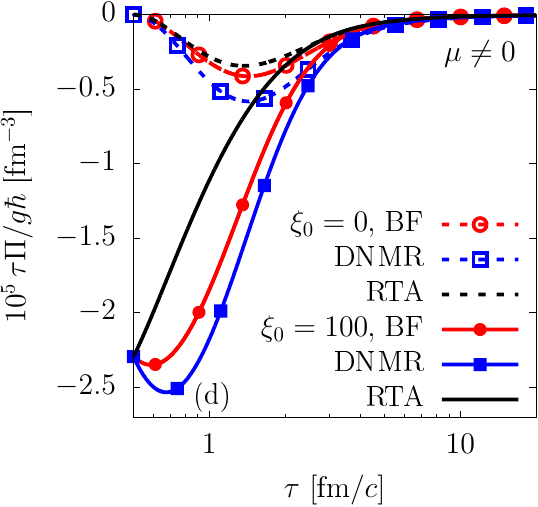} &
		\includegraphics[width=.33\linewidth]{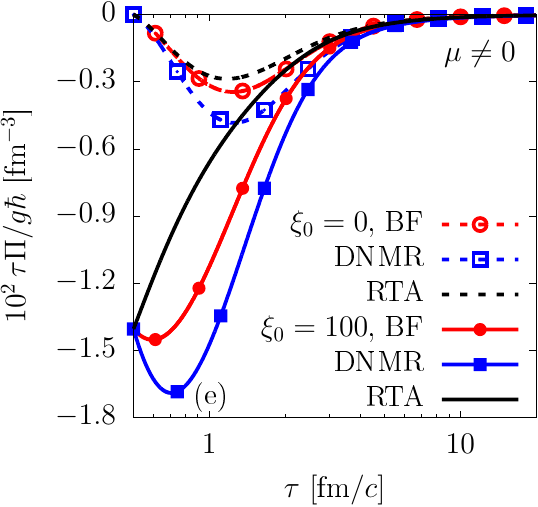} &
		\includegraphics[width=.33\linewidth]{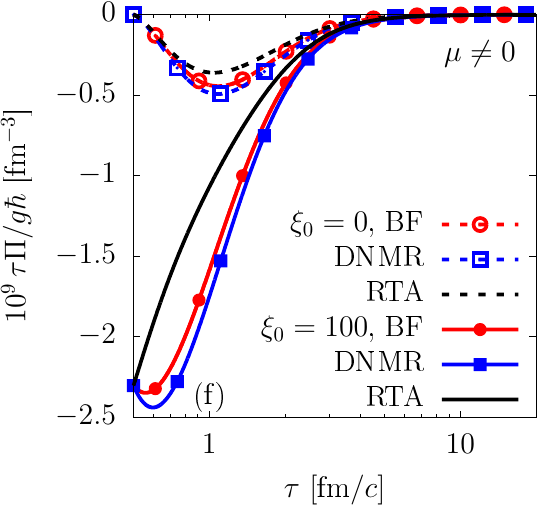} \\
		\includegraphics[width=.33\linewidth]{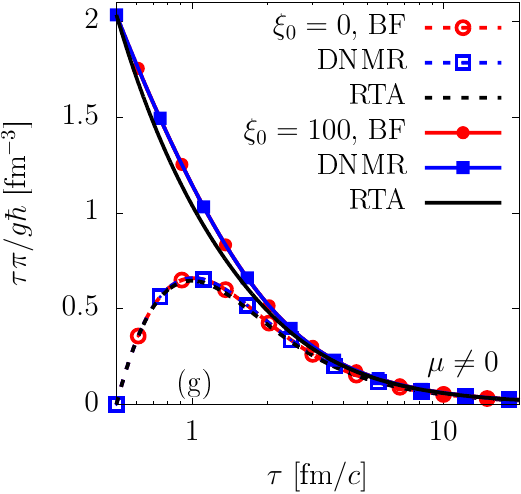} &
		\includegraphics[width=.33\linewidth]{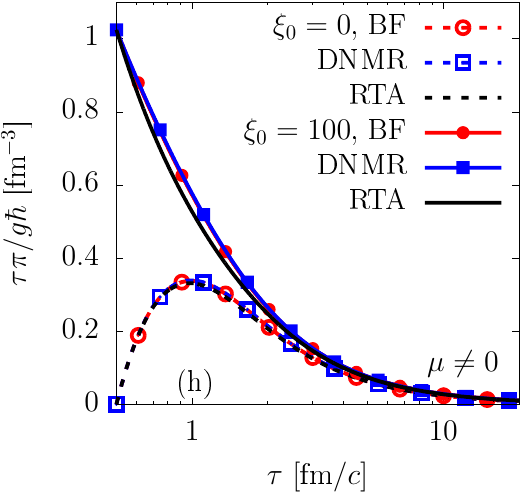} &
		\includegraphics[width=.33\linewidth]{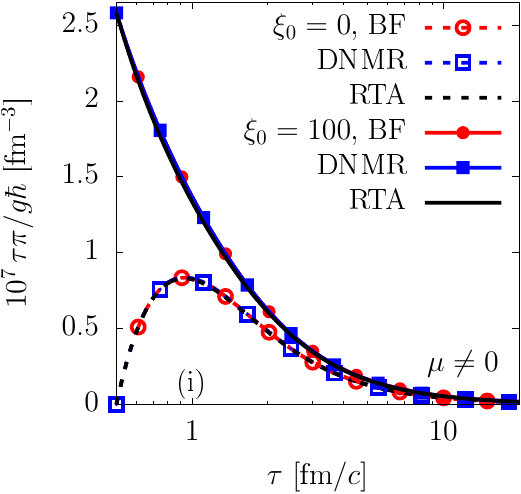}
	\end{tabular}
	\caption{The time evolution of $P_l / P_\perp$ (top row), bulk viscous pressure $\tau \Pi$ (middle row), and shear viscous pressure $\tau \pi$ (bottom row) of an ideal gas with conserved particle number.
		The particle masses correspond to $m_0 = 0.01\ {\rm GeV}/c^2$ (left column), $m_0 = 1\ {\rm GeV}/c^2$ (middle column), and $m_0 = 10\ {\rm GeV}/c^2$ (right column).
		The results with an initial anisotropy parameter $\xi_0 = 0$ and $\xi_0 = 100$ are represented with empty and solid symbols, respectively.
		The exact numerical solutions of the Boltzmann equation in RTA are shown with black lines.
		The solutions of second-order fluid dynamics with transport coefficients derived using the basis-free approach are shown as red lines with circles, while those of the DNMR approach are shown as blue lines with squares, respectively.
		In all cases, we have fixed the relaxation times $\tau_R = \tau_\Pi = \tau_\pi = 0.5$~fm/$c$, while the initial temperature and chemical potential are set to $T_0 = 0.5$~GeV and
		$\mu_0 = 0$~GeV, respectively, at $\tau_0 = 0.5$~fm/$c$.
		\label{fig:bjorken_with_conservation}
	}
\end{figure*}
%%%-RTA-hydro-figures-%%%

%%%-RTA-hydro-mu=0-figures-%%%
\begin{figure*}
	\begin{tabular}{ccc}
		\hspace{25pt} $m_0 = 0.01\ {\rm GeV}/c^2$ &
		\hspace{25pt}$m_0 = 1\ {\rm GeV}/c^2$ &
		\hspace{25pt}$m_0 = 10\ {\rm GeV}/c^2$ \\
		\includegraphics[width=.33\linewidth]{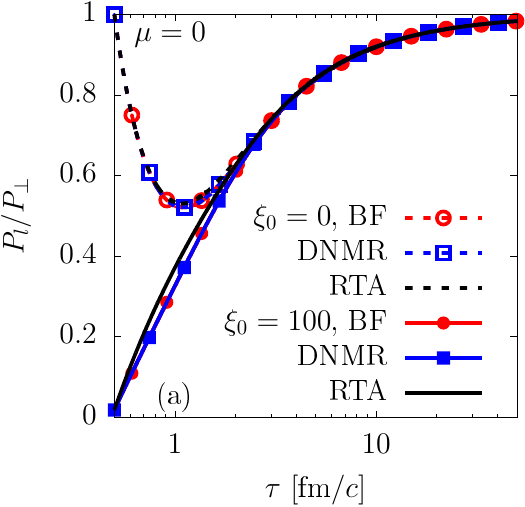} &
		\includegraphics[width=.33\linewidth]{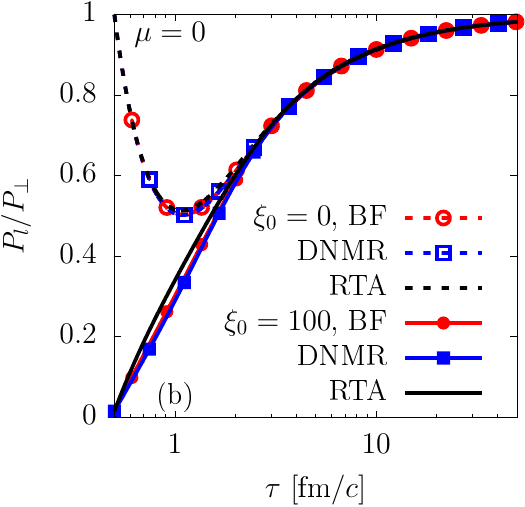} &
		\includegraphics[width=.33\linewidth]{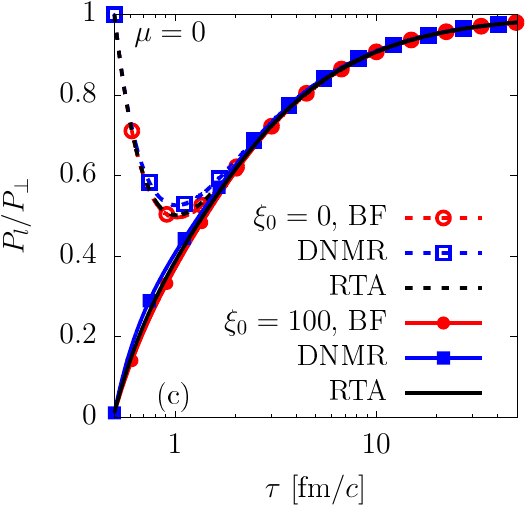} \\
		\includegraphics[width=.33\linewidth]{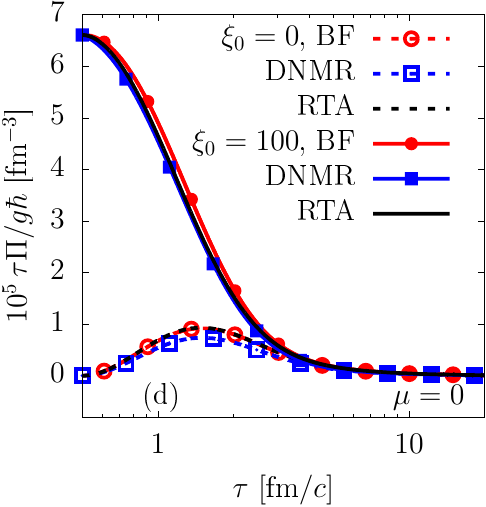} &
		\includegraphics[width=.33\linewidth]{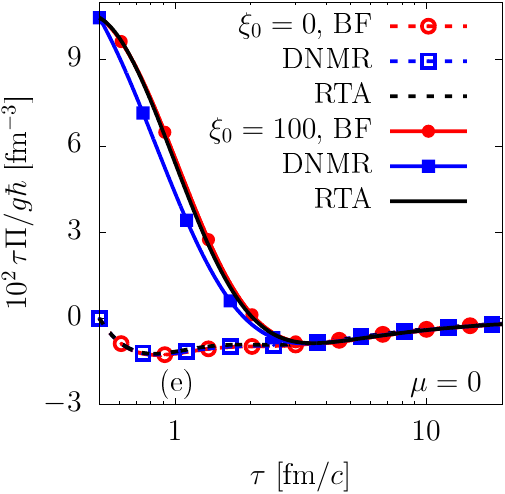} &
		\includegraphics[width=.33\linewidth]{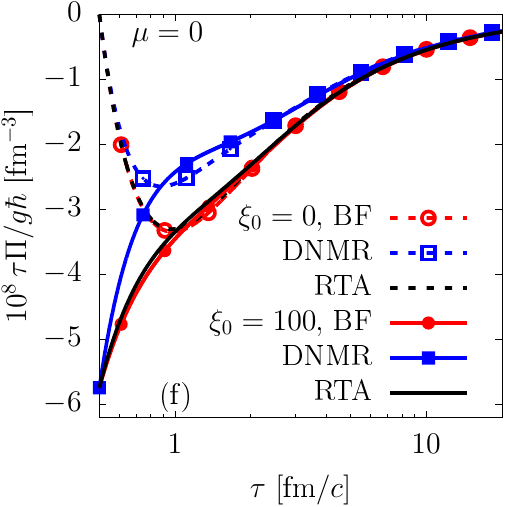} \\
		\includegraphics[width=.33\linewidth]{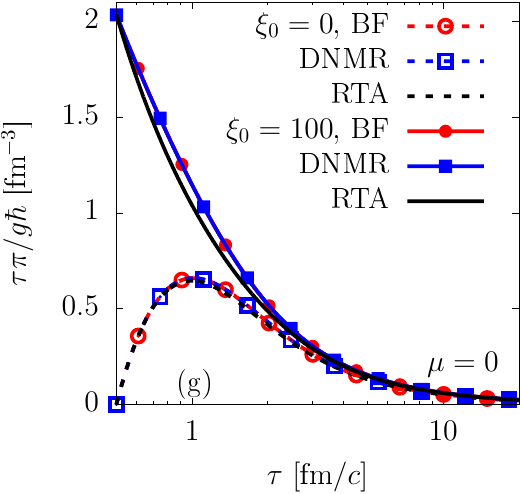} &
		\includegraphics[width=.33\linewidth]{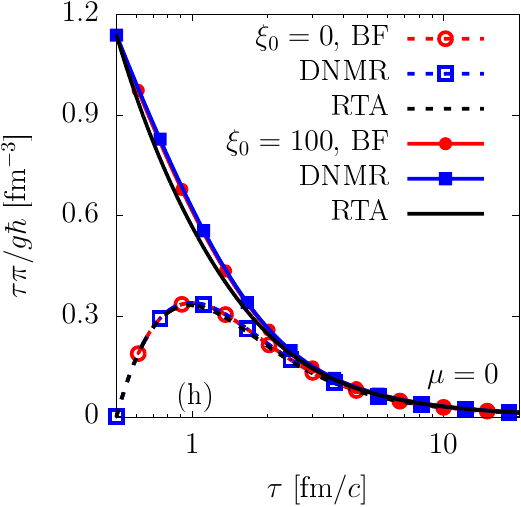} &
		\includegraphics[width=.33\linewidth]{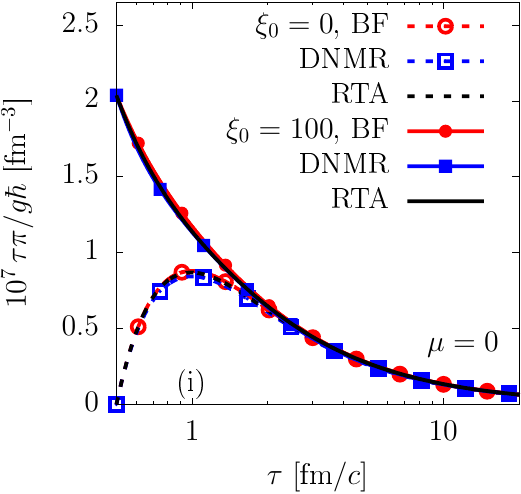}
	\end{tabular}
	\caption{The same as Fig.~\ref{fig:bjorken_with_conservation} but for an ideal gas without
		particle-number conservation, i.e., $\mu(\tau) = 0$~GeV.
		\label{fig:bjorken_without_conservation}
	}
\end{figure*}
%%%-RTA-hydro-mu=0-figures-%%%

%%%-RTA-ahydro-figures-%%%
\begin{figure*}
	\begin{tabular}{ccc}
		\hspace{25pt} $m_0 = 0.01\ {\rm GeV}/c^2$ &
		%, $k_B T_0 = 0.5$ GeV &
		%, $\mu_0 = 0$ &
		\hspace{25pt}$m_0 = 1\ {\rm GeV}/c^2$ &
		%, $k_B T_0 = 0.5$ GeV &
		%, $\mu_0 = 0$ &
		\hspace{25pt}$m_0 = 10\ {\rm GeV}/c^2$ \\
		%, $k_B T_0 = 0.5$ GeV \\
		%, $\mu_0 = 0$ \\
		\includegraphics[width=.33\linewidth]{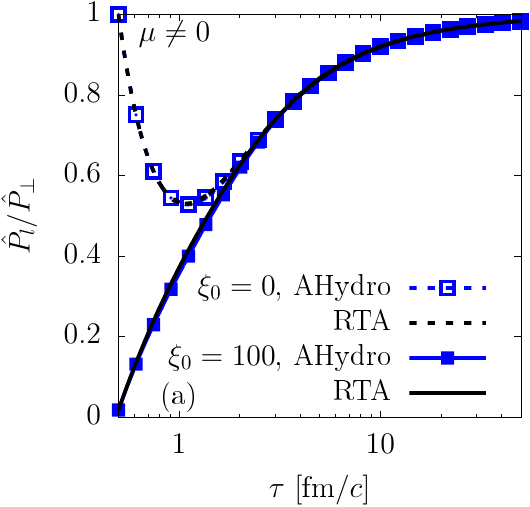} &
		\includegraphics[width=.33\linewidth]{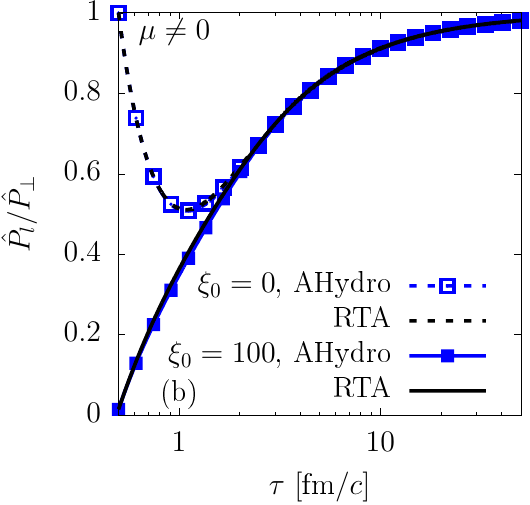} &
		\includegraphics[width=.33\linewidth]{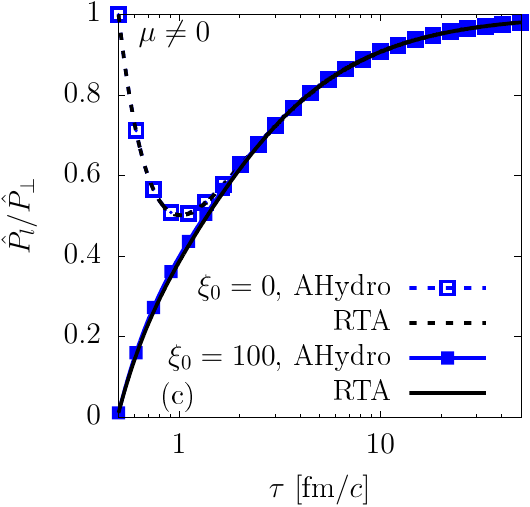} \\
		\includegraphics[width=.33\linewidth]{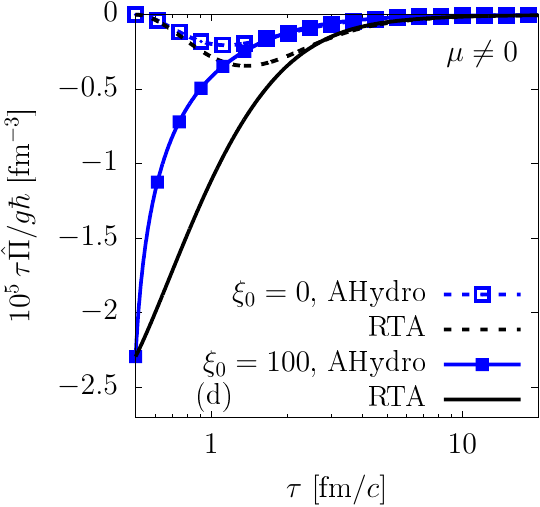} &
		\includegraphics[width=.33\linewidth]{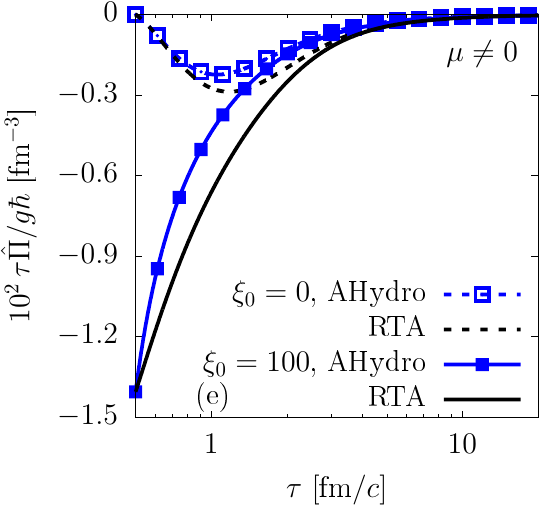} &
		\includegraphics[width=.33\linewidth]{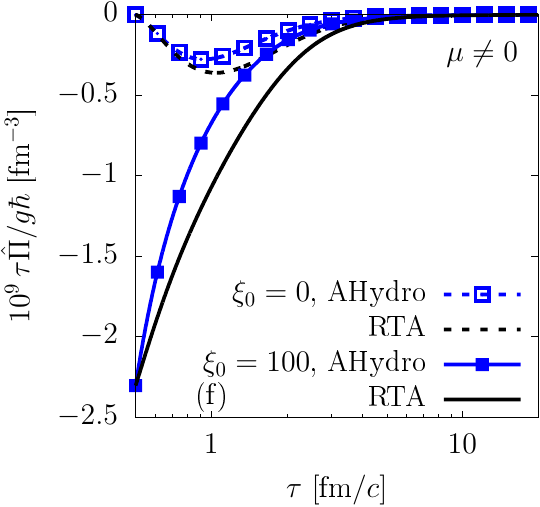} \\
		\includegraphics[width=.33\linewidth]{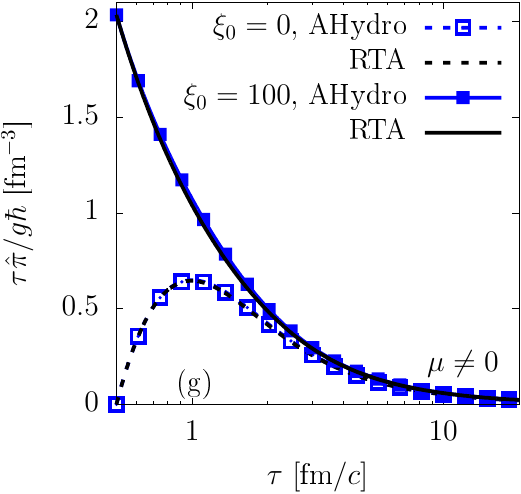} &
		\includegraphics[width=.33\linewidth]{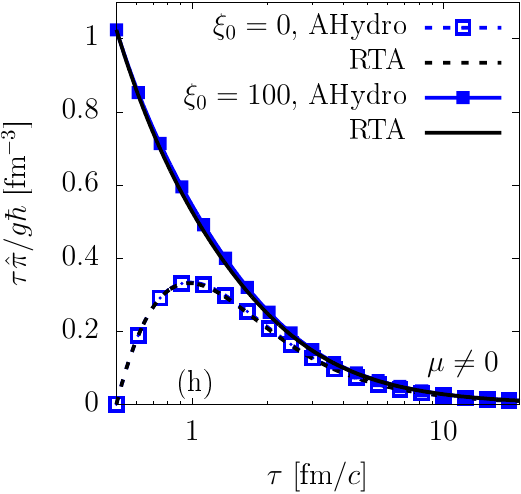} &
		\includegraphics[width=.33\linewidth]{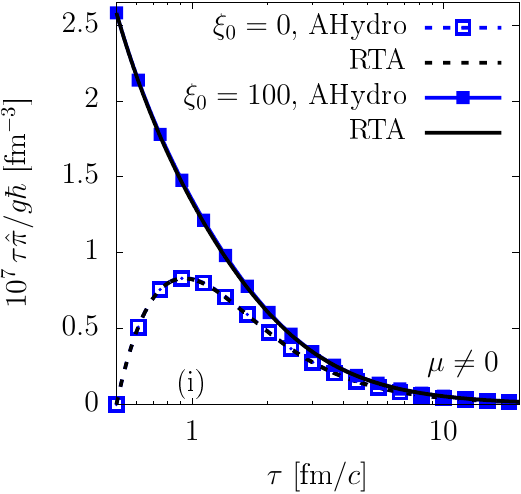}
	\end{tabular}
	\caption{Same as Fig.~\ref{fig:bjorken_with_conservation}.
		Here we compare the results of the Boltzmann equation in RTA (black lines) with the results of leading-order anisotropic fluid dynamics (blue lines with symbols).
		\label{fig:aniso_with_conservation}
	}
\end{figure*}
%%%-RTA-ahydro-figures-%%%

%%%-RTA-ahydro-mu=0-figures-%%%
\begin{figure*}
	\begin{tabular}{ccc}
		\hspace{25pt} $m_0 = 0.01\ {\rm GeV}/c^2$ &
		\hspace{25pt}$m_0 = 1\ {\rm GeV}/c^2$ &
		\hspace{25pt}$m_0 = 10\ {\rm GeV}/c^2$ \\
		\includegraphics[width=.33\linewidth]{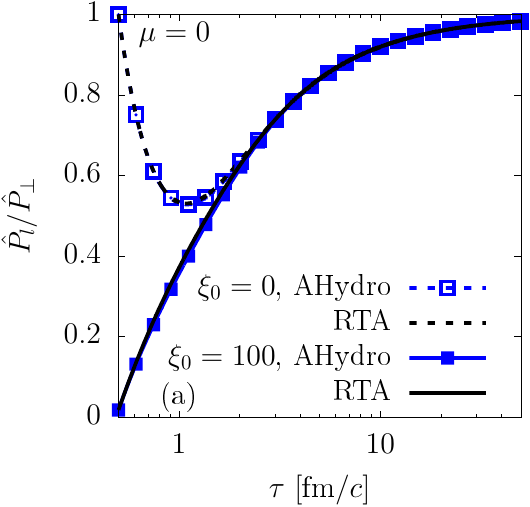} &
		\includegraphics[width=.33\linewidth]{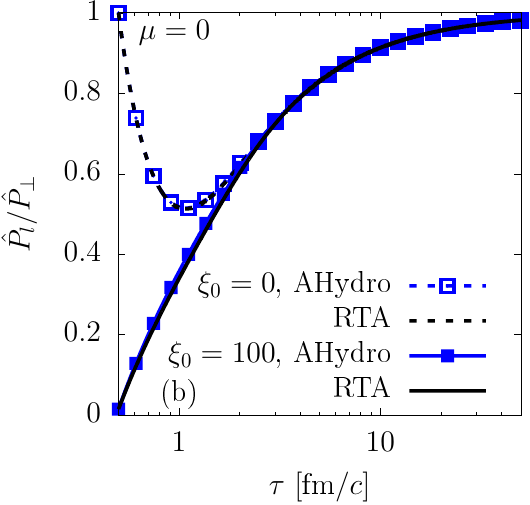} &
		\includegraphics[width=.33\linewidth]{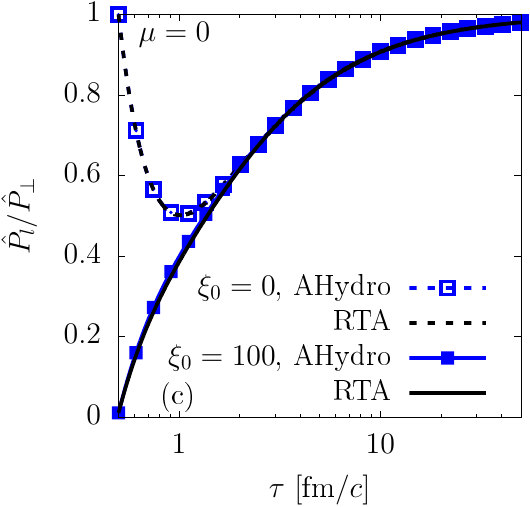} \\
		\includegraphics[width=.33\linewidth]{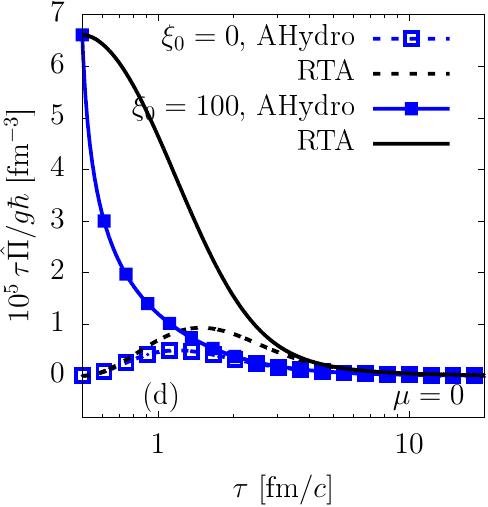} &
		\includegraphics[width=.33\linewidth]{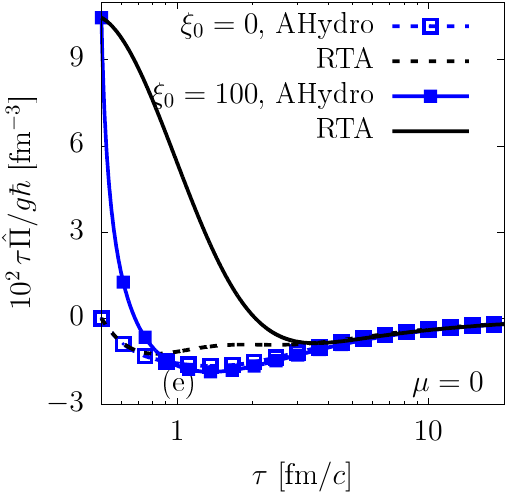} &
		\includegraphics[width=.33\linewidth]{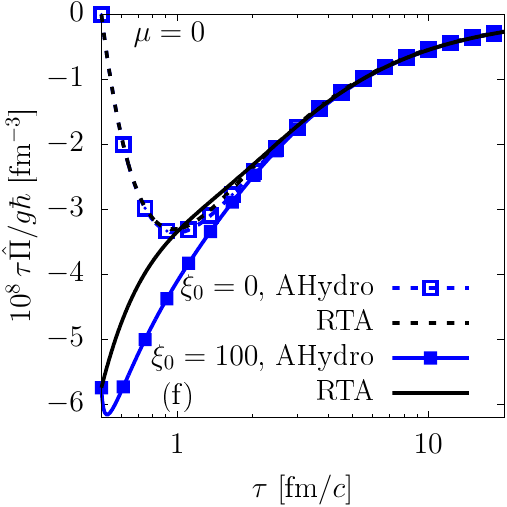} \\
		\includegraphics[width=.33\linewidth]{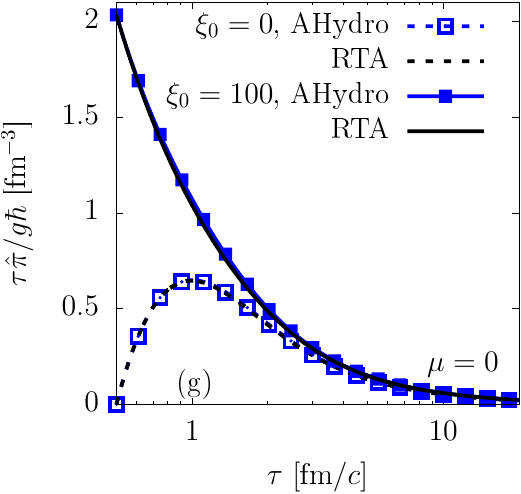} &
		\includegraphics[width=.33\linewidth]{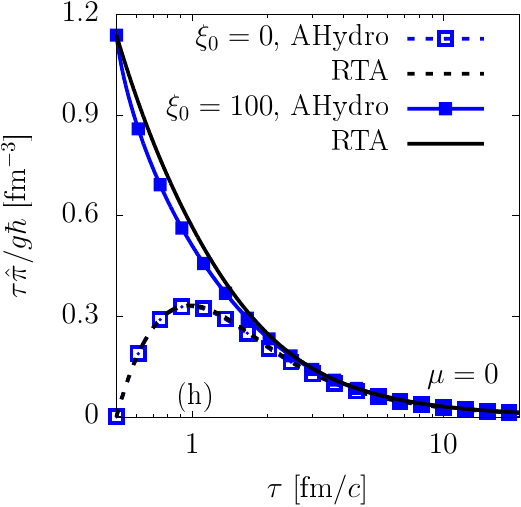} &
		\includegraphics[width=.33\linewidth]{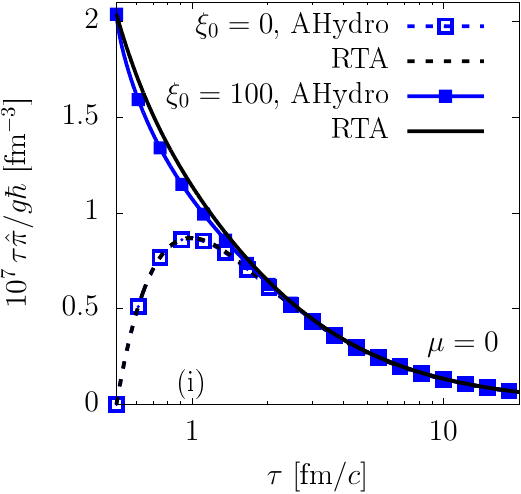}
	\end{tabular}
	\caption{The same as Fig.~\ref{fig:aniso_with_conservation} but without
		particle-number conservation, i.e., $\hat{\alpha}(\tau) = \mu(\tau) = 0$.
		\label{fig:aniso_without_conservation}
	}
\end{figure*}
%%%-RTA-ahydro-mu=0-figures-%%%

%%% Table %%%
\begin{table}
	\begin{tabular}{r|ccc}
		$m_0$ [GeV] & $\overline{\Lambda}_0$ [GeV] & $\Lambda_0$ [GeV] & $\hat{\alpha}_0$ \\\hline
		$0.01$ & $0.945$ & $0.634$ & $1.596$ \\
		$1$ & $0.878$ & $0.653$ & $1.267$ \\
		$10$ & $0.560$ & $0.718$ & $-4.338$
	\end{tabular}
	\caption{
		Values employed for the Romatschke-Strickland distribution for a classical ideal gas without ($\overline{\Lambda}_0$) and with ($\Lambda_0$, $\hat{\alpha}_0$) particle-number conservation.
		The energy density $e_0$ and particle-number density $n_0$ (for the latter case) correspond to an initial temperature $T_0 = 0.5~{\rm GeV}$ and vanishing chemical potential $\mu_0 = 0~{\rm GeV}$ in the case when $\xi_0 = 100$.
		Note that when $\xi_0 = 0$, we have $\Lambda_0 = T_0$.
		\label{tbl:RS}
	}
\end{table}
%%% Table %%%

\subsection{Initial and matching conditions} \label{sec:app:initial_cond}

We now discuss the initial conditions for all approaches presented in this section.
At the initial time, $\tau_0$, we assume that the momentum distribution function $f_\bk = \hat{f}_{RS}$ is given by the Romatschke-Strickland distribution, Eq.~\eqref{eq:RS}, with parameters $(\hat{\alpha}_0, \Lambda_0, \xi_0)$ describing the initial number density (in the case of the fluid with conserved particle number), the energy density, and the pressure anisotropy $P_l / P_\perp$.

Now applying the Landau matching conditions we look for the connection between
the parameters $(\hat{\alpha}, \Lambda, \xi)$ and $(\alpha, T)$ through the initial particle number and energy densities $(n_0, e_0)$, computed via Eq.~\eqref{eq:RS_quant}.
If the particle number is conserved, we use the $e_0/n_0$ ratio in order to eliminate $\hat{\alpha}_0$, leading to
\begin{equation}
 \int_{-1}^1 \frac{ \mathrm{d}v^z \, \Gamma(4,\lambda_0)}{(1 + \xi_0 v_z^2)^2} = \frac{2e_0 (m_0 / \Lambda_0)^2}{n_0 \Lambda_0\sqrt{1 + \xi_0}} K_2\left(\frac{m_0}{\Lambda_0}\right)\;,
	\label{eq:RS_Ideal}
\end{equation}
with $\lambda_0 = m_0 \sqrt{1 + \xi_0 v_z^2} / \left(\Lambda_0 \sqrt{1 - v_z^2} \right)$.
In the case where the particle number is not conserved, $\hat{\alpha}= 0$, while $\Lambda_0$ follows from
\begin{equation}
 e_0 = \frac{g \Lambda_0^4}{4\pi^2} \int_{-1}^1 \frac{ \mathrm{d}v^z\, \Gamma(4, \lambda_0)}{(1 + \xi_0 v_z^2)^2}\;.
 \label{eq:RS_Parton}
\end{equation}

Both Eq.~\eqref{eq:RS_Ideal} and Eq.~\eqref{eq:RS_Parton} are solved using a bisection
algorithm, as follows.
Starting from $\Lambda_0 = T_0$, we look for a valid pair $\Lambda_{\rm min} (\Lambda_{\rm max})$ such that the right-hand sides are smaller (larger) than the left-hand sides when $\Lambda_0 = \Lambda_{\rm min}$ ($\Lambda_{\rm max}$).
We start the search for this window by setting $\Lambda_{\rm min}= \Lambda_{\rm max} = \Lambda_0$ and subsequently halving $\Lambda_{\rm min}$ or doubling $\Lambda_{\rm max}$.
Finally, the integrals with respect to $v_z$ are computed using the adaptive Simpson integration of third order.

%%%
\section{Results and discussions}
\label{sec:results}

In this section, we present and discuss the temporal evolution of a classical ideal gas of massive particles and relevant quantities using second-order fluid dynamics in the 14-moment approximation as well as the leading-order anisotropic fluid-dynamics framework, both with and without particle-number conservation.
These fluid-dynamical results are directly compared to the exact solutions of the Boltzmann transport equation also obtained in RTA.

In what follows, in all cases we have initialized the system at $\tau_0 = 0.5$~fm/$c$, with temperature $T_0 = 0.5$~GeV and chemical potential $\mu_0 = \alpha_0 T_0 = 0$~GeV.
We have fixed the relaxation times in both the fluid-dynamical and transport calculations to
$\tau_R = \tau_\Pi = \tau_\pi = 0.5$~fm/$c$.
For the particle rest mass we will use the following values: $m_0 = 0.01$~GeV to approximate the massless limit, as well as $1$~GeV and $10$~GeV.
At the initial time, $4\pi \eta / s$ evaluates to $3.18$, $2.63$, and $0.68$ for $m_0 = 0.01$, $1$, and $10$~GeV, respectively.
Furthermore, we will consider two different values for the initial anisotropy, $\xi_0 = 0$ and $100$, corresponding to an isotropic and oblate spheroidal distribution in momentum space, respectively.
The initial values for the bulk viscous pressure and shear stress are obtained from the initial Romatschke-Strickland distribution through Eq.~\eqref{eq:RS_Pi_and_pi}.

%%%
\subsection{Second-order fluid dynamics vs.\ Boltzmann equation in RTA}
\label{sec:res_hydro}

In this section we compare the results of second-order fluid dynamics discussed in Sec.~\ref{sec:app:hydro} with the exact numerical solution of the Boltzmann equation.
The results are presented in Figs.~\ref{fig:bjorken_with_conservation} and \ref{fig:bjorken_without_conservation} for fluids with and without particle-number conservation, respectively.
Here, Figs.~\ref{fig:tcoeffs_scalar} and \ref{fig:tcoeffs_shear} can be used to estimate the corresponding transport coefficients as functions of $z$ .

The solutions of the Boltzmann equation are shown with black lines in all figures.
The solutions of second-order fluid dynamics with transport coefficients derived using the BF and the DNMR approaches are plotted using red lines with circles and blue lines with squares, respectively.

The evolution of $P_l/P_\perp$ is shown on the top rows of Figs.~\ref{fig:bjorken_with_conservation} and \ref{fig:bjorken_without_conservation}.
The middle row shows the time evolution of the bulk viscous pressure multiplied by the proper time, $\tau \Pi$, while the bottom row shows the shear-stress pressure multiplied by the proper time, $\tau \pi$.
The particle rest masses from the left column to the right column are fixed to $m_0 = 0.01$, $1$, and $10\ {\rm GeV}/c^2$, respectively.

The case with explicit particle-number conservation is shown in Fig.~\ref{fig:bjorken_with_conservation}.
The evolution of $P_l/P_\perp$ is equally well reproduced by second-order fluid dynamics with either choice of transport coefficients, especially at later times when the gradients have already decreased.
Significant discrepancies can be observed between second-order fluid dynamics and the exact kinetic results for the early-time behavior of the bulk pressure.
This is when the gradients as well as the temperature are largest, while the ratio $z = m_0 / T$ is smallest.
We also note that using the basis-free transport coefficients brings the fluid-dynamical results closer to the exact kinetic solutions than when the standard DNMR transport coefficients are used.

On the other hand, the evolution of the shear-stress tensor component stays in a reasonably good agreement with the exact kinetic results in all cases presented, for both the BF and the DNMR coefficients.
This favorable agreement between these two approaches is not entirely unexpected, since both the BF and the DNMR transport coefficients that govern the shear-stress tensor evolution are very similar during the whole evolution, see Figs.~\ref{fig:tcoeffs_shear}.
Notice also that the bulk viscous pressure is at least 2 orders of magnitude smaller than the shear-stress tensor, hence its contribution to the $P_l/P_\perp$ ratio is less significant for the overall evolution of the system.

We now turn our attention to the case when the particle number is not conserved.
The evolution of the $P_l/P_\perp$ ratio is again in very good agreement between the exact kinetic results and our two variants of second-order transport coefficients, as apparent from the first row of Fig.~\ref{fig:bjorken_without_conservation}.
However, at very large rest mass, the fluid-dynamical results with the BF coefficients show a better agreement with the kinetic results.
Moreover, the early-time behavior of the bulk viscous pressure is in much better agreement now than for the case with particle-number conservation.
The results improve when using the BF coefficients while the deviation of the DNMR results from the exact kinetic results becomes more pronounced at larger particle masses, as can be seen in panels (e) and (f) of Fig.~\ref{fig:bjorken_without_conservation}.
Finally, the evolution of shear-stress tensor is in very good agreement with the exact solution in all cases, as shown in panels (g) and (i) of Fig.~\ref{fig:bjorken_without_conservation}.

%%%
\subsection{Anisotropic fluid dynamics vs. the Boltzmann equation in the RTA}
\label{sec:res_aniso_hydro}

In this section we compare the results of leading-order anisotropic fluid dynamics (aHydro) with the exact numerical solution of the Boltzmann equation in RTA.
These results are presented in Figs.~\ref{fig:aniso_with_conservation} and \ref{fig:aniso_without_conservation} for fluids with and without particle-number conservation, respectively.
Here we also considered two values for the initial anisotropy parameter, namely $\xi_0 = 0$ and $\xi_0 = 100$, represented with empty and filled symbols.
The numerical solutions of the Boltzmann equation are shown with black lines.

Similarly to the second-order fluid-dynamical results, the aHydro results (blue lines with squares) agree very well with exact kinetic results (black lines) at the level of $\hat{P}_l / \hat{P}_\perp$ (top row) and $\tau \hat{\pi}$ (bottom row), in all cases presented.
The agreement with the kinetic results are far better when the particle number is conserved than otherwise.
Comparing Figs.~\ref{fig:aniso_with_conservation} to Figs.~\ref{fig:bjorken_with_conservation} we also observe that the aHydro results are in a better agreement with the exact kinetic solutions than they are with second-order fluid dynamics in case the particle number is conserved.

Furthermore, in case the particle number is not conserved, large deviations can be observed between the exact kinetic and the aHydro results in the bulk viscous pressure $\tau \hat{\Pi}$ shown in the middle rows of Figs.~\ref{fig:aniso_with_conservation} and \ref{fig:aniso_without_conservation}.
Similarly, comparing Figs.~\ref{fig:aniso_without_conservation} to Figs.~\ref{fig:bjorken_without_conservation}, the aHydro results are in a far worse agreement with kinetic theory than those of second-order fluid dynamics.

This was expected since the anisotropic fluid-dynamical framework considered in this paper employs only two or three free parameters, namely $\xi$, $\Lambda$, and/or $\hat{\alpha}$, the latter only in the case with particle-number conservation.
This also means that aHydro describes the evolution of dissipative quantities with one equation of motion fewer than second-order fluid dynamics.
However, as shown in Ref.~\cite{Nopoush:2014pfa}, an additional parameter modifying the distribution function in Eq.~\eqref{f_RS} denoted $\Phi$ may be used to improve the aHydro results for the bulk viscous pressure.

%%%
\section{Conclusions}
\label{sec:conclusions}

In this paper we studied all first- and second-order transport coefficients of second-order fluid dynamics with $14$ dynamical moments.
Through explicit computations for the case of a classical, massive ideal gas, with and without taking into account particle-number conservation, we compared the basis-free and the standard DNMR approximations for the second-order transport coefficients arising from the Boltzmann equation in the Anderson-Witting relaxation-time approximation.
We found that most transport coefficients are insensitive to the chosen approximation for the negative-order moments, but some of them differ in the ultrarelativistic to mildly relativistic regions, as described by the ratio $z=m_0/T$ between the particle rest mass and the temperature.

Using these transport coefficients, we reasserted the validity of second-order fluid dynamics in the well-known (0+1)--dimensional boost-invariant Bjorken expansion scenario for a massive ideal gas.
By contrasting the second-order transport coefficients in the BF and the DNMR approximations, we showed that the agreement between kinetic theory and second-order fluid dynamics is improved in the far-from-equilibrium regime when the BF coefficients are used.
This notable improvement can be traced back to the distinct behavior of the various transport coefficients as function of the $z = m_0/T$ ratio.
Specifically, we observed significant differences in the bulk pressure self-coupling $\bar{\delta}_{\Pi\Pi}$ and shear-bulk coupling $\bar{\lambda}_{\pi\Pi}$ coefficients for a gas without conserved particle number, which in turn lead to slightly different outcomes.

Finally, we also considered the equations of leading-order anisotropic fluid dynamics (aHydro). The formulation considered in this paper employs as degrees of freedom the energy scale $\Lambda$, the anisotropy parameter $\xi$, as well as the term $\hat{\alpha}$ for the case when particle number is conserved.
We found very good agreement between the aHydro and kinetic results at the level of the ratio $P_l / P_\perp$ between the longitudinal and transverse pressure and the shear-stress component $\tau \hat{\pi}$ for a large range of particle masses.
The results for the evolution of the bulk viscous pressure, $\tau \hat{\Pi}$, showed discrepancies compared to the RTA results, as expected since our aHydro implementation does not provide a separate degree of freedom for the bulk viscous pressure.

%%%
\begin{acknowledgments}
	We thank P.~Huovinen for reading the manuscript and him and D.~Wagner for constructive comments and fruitful discussions.
	The authors acknowledge support by the Deutsche Forschungsgemeinschaft (DFG, German Research Foundation) through the CRC-TR 211 ``Strong-interaction matter 	under extreme conditions'' -- Project No. 315477589 -- TRR 211.
	V.E.A.~and E.M.~gratefully acknowledge the support through a grant of the Ministry of Research, Innovation and Digitization, CNCS - UEFISCDI, Project No. PN-III-P1-1.1-TE-2021-1707, within PNCDI III.
	E.M.~was also supported by the program Excellence Initiative--Research University of the University of Wroc{\l}aw of the Ministry of Education and Science. D.H.R.~is supported by the State of Hesse within the Research Cluster ELEMENTS (Project ID 500/10.006).
\end{acknowledgments}

%%%
\appendix

\section{The DNMR coefficients $\gamma^{(\ell)}_{r0}$}
\label{Sec:DNMR_coefficients}

In this appendix we compute the coefficients $\gamma^{(\ell)}_{r0} = \mathcal{F}^{(\ell)}_{r0}$ defined in Eq.~\eqref{F_rn} and applied in the 14-moment approximation.

The coefficient $\mathcal{H}_{\mathbf{k}n}^{(\ell )}$ in Eqs.~(\ref{F_rn}), (\ref{F_rn_useful})
is a polynomial in energy of order $N_\ell \rightarrow \infty$,
\begin{equation}
	\mathcal{H}_{\mathbf{k}n}^{(\ell )} =\frac{(-1)^{\ell}}{\ell! J_{2\ell,\ell}}
	\sum_{m=n}^{N_\ell} a_{mn}^{(\ell)} P_{\mathbf{k} m}^{(\ell)}\;, \label{eq:Hfunction_k}
\end{equation}
where the polynomials of order $m$ in energy are
\begin{equation}
	P_{\mathbf{k} m}^{(\ell)} = \sum_{r=0}^{m}
	a_{mr}^{(\ell)}E_{\mathbf{k}}^{r} \; .
	\label{eq:P_k}
\end{equation}
The coefficients $a_{mn}^{(\ell)}$ are obtained through the Gram-Schmidt procedure imposing the following orthogonality condition:
\begin{equation}
	\int \mathrm{d}K \omega^{(\ell)} P_{\bk m}^{(\ell)}
	P_{\bk n}^{(\ell)} = \delta_{mn}\;,
	\label{eq:P_ortho}
\end{equation}
where the weight $\omega^{(\ell)}$ is
\begin{equation}
	\omega^{(\ell)} = \frac{(-1)^{\ell}}{(2\ell + 1)!!} \frac{ 1}{J_{2\ell,\ell} }
	(\Delta^{\alpha\beta} k_\alpha k_\beta)^\ell f_{0\mathbf{k}} \tilde{f}_{0\bk}\;.
	\label{eq:omega_def}
\end{equation}
Using these results together with Eq.~\eqref{F_rn}, the coefficients $\gamma^{(\ell)}_{r0}$ are defined as
\begin{align}
	\gamma^{(0)}_{r0} &\equiv \mathcal{F}_{r0}^{\left( 0\right) } =  \frac{J_{-r,0}D_{30}+J_{-r+1,0}G_{23}+J_{-r+2,0}D_{20}}
	{J_{20}D_{20}+J_{30}G_{12}+J_{40}D_{10}}\;,  \label{OMG_rho} \\
	\gamma^{(1)}_{r0} &\equiv \mathcal{F}_{r0}^{\left( 1\right) }
	=\frac{J_{-r+2,1}J_{41}-J_{-r+3,1}J_{31}}{D_{31}} \;,  \label{OMG_rho_mu} \\
	\gamma^{(2)}_{r0} &\equiv \mathcal{F}_{r0}^{\left( 2\right) } = \frac{J_{-r+4,2}}{J_{42}}\;.
	\label{OMG_rho_mu_nu}
\end{align}

\section{First-order transport coefficients}
\label{Sec:Transport_Coeffs_Positive}

In this appendix we discuss the positivity of the first-order transport coefficients from Eq.~\eqref{main_RTA_coeffs}.
The shear viscosity coefficient $\eta = \tau_R\alpha^{(2)}_0$, where $\alpha^{(2)}_0 = \beta J_{32}$ 
from Eq.~\eqref{alpha2}. 
Replacing the thermodynamic integral from Eq.~\eqref{J_nq}, using $n = 3$ and $q = 2$, we obtain
\begin{equation}
	\eta = \frac{\beta \tau_R}{15} \int \frac{dK}{E_\bk} f_{0\bk} \tilde{f}_{0\bk} (E_\bk^2 - m_0^2)^2 \;.
\end{equation}
This integrand is positive definite, and hence $\eta > 0$.
	
The diffusion coefficient is given by $\kappa = \tau_R \alpha^{(1)}_0$, where $\alpha^{(1)}_0 = J_{11} - J_{21}/h$ 
from Eq.~\eqref{alpha1}, where $h = (e + P) / n$ is the enthalpy per particle. This can be rewritten as
\begin{align}
	\alpha^{(1)}_0 =
	J_{11} - \frac{2}{h} J_{21} + \frac{1}{h^2} J_{31} \;,
\end{align}
where we used that $J_{21} = nT$ and $J_{31} = T(e + P)$, while we added and subtracted 
$J_{21}/h = J_{31} /h^2$. Now, substituting $J_{nq}$ from Eq.~\eqref{J_nq}, we arrive at
	\begin{equation}
		\kappa = \frac{\tau_V}{3} \int \frac{dK}{E_\bk} f_{0\bk} \tilde{f}_{0\bk} (E_\bk^2 - m_0^2) 
		\left(1 - \frac{ E_\bk}{h}\right)^2 \; ,
	\end{equation}
where the integrand is positive definite and hence $\kappa > 0$.
	
Recalling the thermodynamic relations obtained from integration by parts, see Eq.~\eqref{J_nq},
\begin{align}
	dI_{nq}\left( \alpha,\beta \right) & \equiv \left( \frac{\partial
		I_{nq}}{\partial \alpha}\right)_{\beta} d\alpha +\left( \frac{\partial
		I_{nq}}{\partial \beta}\right)_{\alpha} d\beta  \notag \\
	& =J_{nq}d\alpha - J_{n+1,q}d\beta \;,  \label{d_Inq}
\end{align}
and using this relation for $I_{10}=n$ and $I_{20}= e$, we obtain
\begin{align}
	d\alpha &= \frac{1}{D_{20}} \left(J_{30} dn - J_{20}de \right) \; , \\ 
	d\beta  &= \frac{1}{D_{20}} \left(J_{20} dn - J_{10}de \right) \; .
\end{align}
Using these relations the partial derivatives of the pressure, $P=I_{21}$, at constant particle number density and constant energy density can be written as
\begin{align}
	\left(\frac{\partial P}{\partial e}\right)_n &= \frac{J_{31}}{D_{20}} 
	\left(J_{10} - \frac{1}{h} J_{20}\right)\;, \nonumber\\
	\left(\frac{\partial P}{\partial n}\right)_e &= \frac{J_{31}}{D_{20}} 
	\left(-J_{20} + \frac{1}{h} J_{30}\right)\;.
\end{align}
Therefore the speed of sound squared, cf.\ Eq.~\eqref{c_s2}, reads 
\begin{align}
	c_s^2 &\equiv \frac{J_{31}}{D_{20}}\left(J_{10} - \frac{2}{h} J_{20} + \frac{1}{h^2} J_{30}\right) 
	\notag \\
	&= \frac{1}{3} + \frac{m^2_0}{3 h D_{20}} \left(h G_{20} - G_{30} \right) \; , \label{app:cs2}
\end{align}
where we used that $m^2_0 J_{n0} = J_{n+2,0} - 3 J_{n+2,1}$.
Similarly, in the case without particle-number conservation, the speed of sound squared reads
\begin{equation}
	\bar{c}_s^2 \equiv \frac{J_{31}}{J_{30}} = \frac{1}{3} \left(1 - m^2_0 \frac{J_{10}}{J_{30}} \right) \;.
\end{equation}

The bulk viscosity coefficient, $\zeta = \tau_\Pi m_0^2 \alpha^{(0)}_0 / 3$, with $\alpha^{(0)}_r$ given in Eq.~\eqref{alpha0}, can be written using Eq.~\eqref{app:cs2} as
\begin{equation}
	\zeta =  \frac{\tau_\Pi}{3} \left[-m_0^2 \beta J_{11} + (1 - 3c_s^2)(e + P)\right].
\end{equation}
Replacing $m_0^2 J_{11} = J_{31} - 5 J_{32}$, we arrive at
\begin{equation}
	\zeta = \tau_\Pi \left[\frac{5}{3}\beta J_{32} - c_s^2(e + P)\right] \; ,
	\label{eq:zeta_aux}
\end{equation}
where the first term inside the square brackets is
\begin{equation}
	\frac{5}{3} J_{32} = \frac{1}{9} \int \frac{dK}{E_\bk} f_{0\bk} \tilde{f}_{0\bk} (E_\bk^2 - m_0^2)^2 \; .
\end{equation}
We now consider the relations,
\begin{align}
	\int \frac{dK}{E_\bk} f_{0\bk} \tilde{f}_{0\bk} (E_\bk^2 - m_0^2) T_\bk &= \frac{3(e + P)}{\beta} \;,\\
	\int \frac{dK}{E_\bk} f_{0\bk} \tilde{f}_{0\bk} T_\bk^2 &= \frac{e + P}{\beta c_s^2} \;,
\end{align}
where we introduced the notation,
\begin{equation}
	T_\bk = \frac{E_\bk^2 (J_{10} - \frac{1}{h} J_{20}) - E_\bk (J_{20} - \frac{1}{h} J_{30})}{J_{10} - \frac{2}{h} J_{20} + \frac{1}{h^2} J_{30}} \; .
\end{equation}
Inserting and subtracting the term $c_s^2(e + P)$ in Eq.~\eqref{eq:zeta_aux}, one arrives at
	\begin{align}
		\zeta &= \tau_\Pi  \left[\frac{5}{3} \beta J_{32} - 2c_s^2(e + P) + c_s^2(e + P)\right] \nonumber\\
		&= \frac{\beta \tau_\Pi}{9} \int \frac{dK}{E_\bk} f_{0\bk} \tilde{f}_{0\bk} \left(E_\bk^2 - m_0^2 - 3 c_s^2 T_\bk\right)^2 \;.
	\end{align}
This integrand is always positive and so $\zeta > 0$. Note that this expression agrees with Eq.~(A26) of Ref.~\cite{Albright:2015fpa}.

We also remark that $\bar{\zeta} = \tau_\Pi m_0^2 \bar{\alpha}^{(0)}_0 / 3 > 0$, where 
$\bar{\alpha}^{(0)}_r$ was defined in Eq.~\eqref{alpha0_mu0}.
To show this, we use the recurrence relation 
$3J_{11} = J_{10} - m_0^2 J_{-1,0}$ to rewrite $\bar{\zeta}$ as
	\begin{equation}
		\bar{\zeta} = \tau_\Pi \frac{\beta m_0^2}{9} [(3\bar{c}_s^2 - 1) J_{10} + m_0^2 J_{-1,0}] \; .
		\label{eq:zetabar_aux}
	\end{equation}
Now adding and subtracting $m_0^2 J_{10}(1 - 3\bar{c}_s^2) = J_{30}(1 - 3\bar{c}_s^2)^2$ inside the square brackets leads to
\begin{equation}
	\bar{\zeta} = \tau_\Pi \frac{\beta}{9} [(1 - 3\bar{c}_s^2)^2 J_{30} - 2m_0^2 (1 - 3\bar{c}_s^2) J_{10} + m_0^4 J_{-1,0}] \;.
\end{equation}
Finally, restoring the integral expression for the $J_{nq}$'s, we arrive at
\begin{equation}
	\bar{\zeta} = \frac{\tau_\Pi \beta}{9}
	\int \frac{dK}{E_\bk} f_{0\bk} \tilde{f}_{0\bk} \left[(1-3\bar{c}_s^2)^2 E_\bk^2 - m_0^2\right]^2 \;,
	\label{eq:zetabar_chakra}
\end{equation}
which generalizes Eq.~(58) of Ref.~\cite{Chakraborty:2010fr} to the case of quantum statistics.

%%%
\section{Useful thermodynamic integrals}
\label{Sec:Useful_Integrals}

The thermodynamic integrals with negative power, $I_{-r,0}$, can be obtained in terms of the functions ${\rm Ki}_r$ as:
\begin{equation}
	I_{-r,0} =  \frac{ \beta n  m_0^{-r} }{K_2(z)}  \left[{\rm Ki}_{r-2}(z)
	- {\rm Ki}_{r}(z)\right] \; ,\label{eq:Ir0_Ki}
\end{equation}
where the Bickley function ${\rm Ki}_q(z)$ is defined as the repeated integral of the Bessel function $K_0(z)$, having the following integral representation, see Eq.~(11.2.10) of
Ref.~\cite{Abramowitz_Stegun} and Eq.~(10.43.11) of Ref.~\cite{olver2010nist}:
\begin{equation}
	{\rm Ki}_q(z) = \int_1^\infty \mathrm{d}x\, \frac{e^{-z x}}{x^q \sqrt{x^2 - 1}}
	= \int_{0}^{\infty} \mathrm{d}x\, \frac{e^{-z \cosh x}} {\cosh^q x} \;,
	\label{eq:Ki_int}
\end{equation}
such that
\begin{equation}
	{\rm Ki}_0(z) \equiv K_0(z) = K_2(z) \left[1 - \frac{2}{z} \frac{K_1(z)}{K_2(z)} \right]\;,
\end{equation}
while for the negative orders we have, see Eq.~(11.2.9) of Ref.~\cite{Abramowitz_Stegun},
\begin{align}
	{\rm Ki}_{-q}(z) &= (-1)^q \frac{\mathrm{d}^q}{\mathrm{d}z^q} K_0(z) \; . \label{Ki_-q}
\end{align}
More generally, one can derive the recurrence relation, see
Eq.~(11.2.14) of Ref.\ \cite{Abramowitz_Stegun},
\begin{equation}
	q {\rm Ki}_{q+1}(z) = z {\rm Ki}_{q-2}(z) + (q-1) {\rm Ki}_{q-1}(z) - z {\rm Ki}_{q}(z) \;,
	\label{Ki_n_recursion}
\end{equation}
which holds for any non-negative integer $q \ge 0$ and allows the higher-order functions ${\rm Ki}_n(z)$ to be expressed in terms of ${\rm Ki}_1(z)$.
An integration by parts of $K_0(z)$ leads to
\begin{align}
	K_0(z) &\equiv \int_1^\infty \frac{\mathrm{d}x\, e^{-zx}}{\sqrt{x^2 - 1}} \nonumber \\
	&= z K_1(z) - z{\rm Ki}_1(z) + K_0(z) - {\rm Ki}_2(z)\;,
\end{align}
which then provides in agreement with Eq.\ (\ref{Ki_-q}),
\begin{align}
	{\rm Ki}_{-1}(z) &= {\rm Ki}_1(z) +  \frac{{\rm Ki}_2(z)}{z} \equiv K_1(z) \;.
	\label{K_1}
\end{align}

The function ${\rm Ki}_1(z)$ can be expressed~\cite{Florkowski:2014sfa,Denicol:2014vaa}
in terms of the modified Bessel functions and of the
modified Struve functions $\mathbf{L}_\nu(z)$, see Eq.~(12.2.2) of Ref.~\cite{Abramowitz_Stegun}
\begin{align}
	{\rm Ki}_1(z) &= \frac{\pi}{2} - \frac{\pi z}{2}[K_0(z) \mathbf{L}_{-1}(z)
	+ K_1(z) \mathbf{L}_0(z)] \; .
\end{align}
Introducing a notation similar to that used in Eq.~(\ref{h_H}), ${\rm Hi}(z) = {\rm Ki}_1(z) / K_2(z)$, Eq. (\ref{K_1}) leads to
\begin{equation}
	{\rm Ki}_2(z) = K_2(z) \left[ z H(z) - z {\rm Hi}(z) - 4 \right]\;.
\end{equation}
Similarly, using these formulas together with the recursion relation (\ref{Ki_n_recursion}) we calculate
\begin{align}
	{\rm Ki}_3(z) &= \frac{K_2(z)}{2z} \left[
	8 + 5z^2 - z(2 + z^2) H(z) \right. \nonumber\\
	&+  \left.	z(1 + z^2) {\rm Hi}(z) \right] \;, \\
	{\rm Ki}_4(z) & = -\frac{K_2(z)}{6} \left[
	24 + 5z^2 - z(6+z^2) H(z)\right. \nonumber\\
	&+ \left. z (3 + z^2){\rm Hi}(z) \right] \;.
	\label{eq:Ki_expressions}
\end{align}

Now substituting into Eq.~\eqref{eq:Ir0_Ki} leads to
\begin{align} \label{I_minus10}
	I_{-1,0} &= -\frac{n}{m_0^2} \left[4 - z H(z) + z {\rm Hi}(z) \right] \;, \\ \label{I_minus20}
	I_{-2,0} &= \frac{n T}{m_0^4} \left[8 + 5z^2 - z (2 + z^2) H(z)
	+ z^3 {\rm Hi}(z) \right] \;,  \\ \nonumber
	I_{-3,0} &= -\frac{n}{2m_0^4} \left[8 + 5z^2 - z (2 + z^2) H(z) \right. \\
	&- \left. z (1 - z^2) {\rm Hi}(z) \right] \; , \\
	I_{-4,0} &= \frac{n}{6m_0^4 T} \left[ 5z^2- z^3 H(z) - z (3 -  z^2) {\rm Hi}(z) \right]\;.
\end{align}
Using these results together with Eq.\ (\ref{K_n_recursion}) for $q = 1$ and $-2 \le r \le 1$ we obtain
\begin{align}
	I_{1,1} &= \frac{n}{3} \left[5 - z H(z) + z {\rm Hi}(z) \right]\; , \\
	I_{0,1} &= -\frac{n T}{3 m_0^2} \left[12 + 5z^2 - z(3 + z^2) H(z) + z^3 {\rm Hi}(z) \right]\;, \\
	I_{-1,1} &= \frac{n}{6 m_0^2} \left[5z^2 - z^3 H(z) - z(3 - z^2) {\rm Hi}(z) \right]
	\nonumber  \\
	&= m_0^2 T\, I_{-4,0}\;,\\
	I_{-2,1} &= \frac{n T}{18 m_0^4} \left[48 + 30 z^2 - 5z^4 -
	z (12 + 6z^2 - z^4) H(z) \right. \nonumber \\
	&+ \left. z^3 (9 - z^2) {\rm Hi}(z) \right]\;.
\end{align}
Finally, the functions $I_{rq}$ with $q = 2$ and $0 \le r \le 2$ are
\begin{align}
	I_{22} &= \frac{n T}{15} \left[15 +5 z^2 - z (3 + z^2) H(z) + z^3 {\rm Hi}(z) \right]\;,\\
	I_{12} &= \frac{n}{30} \left[10 - 5z^2 - z (2 - z^2)H(z) + z (5 - z^2)  {\rm Hi}(z) \right]\;, \\
	I_{02} &= -\frac{n T}{90 m_0^2} \left[120 + 60 z^2 - 5z^4  \right. \nonumber\\
	&- \left. z (30 + 12 z^2 - z^4) H(z)
	+  z^3 (15 - z^2) {\rm Hi}(z) \right]\;.
\end{align}
Here we also list the remaining coefficients of interest,
\begin{align}
	\alpha^{(0)}_{-1} &= 2I_{-1,1} - \frac{e I_{-1,0}}{c_v P} + \frac{I_{00}}{c_v T} \; , \\
	\alpha^{(0)}_{-2} &= 3I_{-2,1} - \frac{e I_{-2,0}}{c_v P} + \frac{I_{-1,0}}{c_v T} \; ,
\end{align}
and
\begin{align}
	\bar{\alpha}^{(0)}_{-1} &= 2I_{-1,1} - I_{-1,0} + \bar{c}_s^2 \frac{I_{00}}{T} \; , \\
	\bar{\alpha}^{(0)}_{-2} &= 3I_{-2,1} - I_{-2,0} + \bar{c}_s^2 \frac{I_{-1,0}}{T} \; ,
\end{align}
These are followed by the relations
\begin{align}
	\alpha^{(1)}_{-1} &= I_{0,1} - \frac{I_{1,1}}{h} \; , \quad
	\alpha^{(1)}_{-2} = I_{-1,1} -\frac{I_{0,1}}{h} \; , \\
	\alpha^{(2)}_{-1} &= I_{1,1} - 2I_{1,2} \; , \quad
	\alpha^{(2)}_{-2} = I_{0,1} - 3I_{0,2} \; .
\end{align}

Using the explicit form of the thermodynamic integrals we evaluate the first-order transport coefficients.
Using Eq.~\eqref{alpha_0} together with \eqref{I_minus20} we obtain
\begin{align}
	\frac{\zeta}{\tau_{\Pi}}
	&= z^2 \frac{P}{3} \left[\frac{H(z)}{z} (1 - 3c^2_s) - \frac{1}{3}
	+  \frac{z}{3} \frac{K_1(z) - {\rm Ki}_1(z)}{ K_2(z)} \right] \;,
\end{align}
and similarly
\begin{align}
	\frac{\bar{\zeta}}{\tau_{\Pi}}
	&=  z^2 \frac{P}{3}\left[\bar{c}^2_s -\frac{1}{3}
	+  \frac{z}{3} \frac{K_1(z) - {\rm Ki}_1(z)}{ K_2(z)} \right] \;.
\end{align}
Furthermore, using Eq.~\eqref{alpha_2}, together with Eq.\ \eqref{I_minus20} we obtain
\begin{align} \nonumber
	\frac{\eta}{\tau_{\pi}} &= \frac{e + 9P}{15} - \frac{m_0^4}{15} I_{-2,0} \\
	&= z^3 \frac{P}{15} \left[ \frac{3}{z^2} \frac{K_3(z)}{K_2(z)}
	 - \frac{1}{z} + \frac{K_1(z) - {\rm Ki}_1(z)}{ K_2(z)}\right] \;.
\end{align}
These latter two results were found by Anderson and Witting, see Eqs.~(75), (76) of Ref.~\cite{Anderson:1974a}, as well as by Florkowski et al., see Eqs.\ (37), (45) of Ref.~\cite{Florkowski:2014sfa}.

The ratio of the diffusion coefficient and the relaxation time follows from Eqs.~\eqref{alpha_1} and \eqref{I_minus10},
\begin{align} \nonumber
	\frac{\kappa}{\tau_{V}} &\equiv \frac{e - 2P}{3 h} - \frac{m_0^2}{3} I_{-1,0} \\
	&= z \frac{P}{3T} \left[ \frac{1}{z} - \frac{3}{z^2} \frac{K_2(z)}{K_3(z)}
	- \frac{K_1(z) - {\rm Ki}_1(z)}{ K_2(z)} \right] \;.
\end{align}

%%%
\section{Numerical methods}\label{app:num}

In this section we briefly present the details of the discrete-velocity method employed to solve the Boltzmann equation \eqref{eq:boltz_Fn} of Sec.~\ref{app:num:DVM}.
Similarly, the strategy employed to solve the equations of anisotropic fluid dynamics as well as the Runge-Kutta time-stepping algorithm are presented in Secs.~\ref{app:num:aH} and \ref{app:num:RK}, respectively. A note on the code that we employed can be found in Sec.~\ref{app:num:code}.

%%%
\subsection{Discrete-velocity algorithm} \label{app:num:DVM}

The algorithm employed in this paper to solve the relativistic Boltzmann equation is based on the relativistic lattice-Boltzmann method introduced in Refs.~\cite{Romatschke:2011hm,Ambrus:2016aub} for massless particles and in Refs.~\cite{Gabbana:2017uvc,Gabbana:2019ydb} for the case of massive particles; see also Ref.~\cite{Ambrus:2022adp} for details.

In particular, the only remaining degree of freedom $v^z$ is discretized according to the Gauss-Legendre quadrature method of order $Q$, by which $v^z_{j}$ ($1 \le j \le Q$) are the roots of the Legendre polynomial $P_Q(v^z)$ of order $Q_z$.
This prescription ensures the exact integration of any polynomial in $v^z$ of order less than
or equal to $2Q - 1$, i.e.,
\begin{equation}
 \int_{-1}^1 \mathrm{d}v^z\, (v^z)^n \simeq \sum_{j = 1}^Q w_j (v^z_{j})^n \; ,
\end{equation}
where equality is achieved when $0 \le n < 2Q$.
The quadrature weights $w_j$ are computed via
\begin{equation}
 w_j = \frac{2[1 - (v^z_{j})^2]}{[(Q + 1) P_{Q+1}(v^z_{j})]^2} \; .
\end{equation}

In our case, we have to deal with nonpolynomial functions of $v^z$, however the integration method becomes systematically more accurate as $Q_z$ is increased.
On the other hand, there is a limit on the achievable accuracy due to loss of precision in floating-point arithmetic.
For practical purposes, we employ $Q_z = 200$ for the simulations presented in this paper.
The necessary roots of the Legendre polynomials together with the corresponding Gauss-Legendre quadrature weights can be found in the supplementary material to Ref.~\cite{Ambrus:2016aub}.

Following the discretization of $v^z$, the continuous distributions $F_n(v^z)$ are replaced by a discrete set $F_{n;j}$, defined as \begin{equation}
 F_{n;j} = w_j F_n(v^z_{j}) \;.
\end{equation}
With the above discrete distributions, the macroscopic quantities $e$, $P_l$, and $T^\mu_\mu$ are obtained via
\begin{equation}
 e = \sum_{j = 1}^Q F_{2;j} \;, \quad
 P_l = \sum_{j = 1}^Q (v^z_{j})^2 F_{2;j} \;, \quad
 T^\mu_\mu = m_0^2 \sum_{j = 1}^Q F_{0;j} \;.
\end{equation}
Finally, the gradient with respect to $v^z$ appearing in Eq.~\eqref{eq:boltz_Fn} is evaluated by projecting it onto the space of Legendre polynomials, as described in Ref.~\cite{Ambrus:2016aub}:
\begin{equation}
 \left(\frac{\partial[v^z(1 - v_z^2)F_n]}{\partial v^z}\right)_j =
 \sum_{j' = 1}^Q \mathcal{K}_{j,j'} F_{n;j'} \;,
\end{equation}
where the matrix elements of the kernel $\mathcal{K}_{j,j'}$ can be found in Eq.~(3.54) of Ref.~\cite{Ambrus:2022adp} and are not repeated here for the sake of brevity.
To summarize, Eq.~\eqref{eq:boltz_Fn} becomes
\begin{multline}
 \frac{\partial F_{n;j}}{\partial \tau} + \frac{1}{\tau} \left[1+ (n-1) (v^{z}_{j})^2 \right] F_{n;j}
 - \frac{1}{\tau} \sum_{j' = 1}^Q \mathcal{K}_{j,j'} F_{n;j'} \\
 = -\frac{1}{\tau_R} (F_{n;j} - F_{n;j}^{eq}) \;.
 \label{eq:boltz_Fnj}
\end{multline}

%%%
\subsection{Anisotropic fluid-dynamics algorithm} \label{app:num:aH}

In this section, we discuss the solutions to the equations of anisotropic fluid dynamics, 
Eqs.~\eqref{aH_Dn}--\eqref{aH_DPl}.

While these equations can be directly solved for $\hat{n}$, $\hat{e}$, and $\hat{P}_l$, for the sake of convenience we recast them as evolution equations for the parameters $(\hat{\alpha}, \xi, \Lambda)$.
This is done generically by considering that a function $f \in \{\hat{n}, \hat{e}, \hat{P}_l\}$ describing the fluid properties depends only on the above parameters, such that
\begin{equation}
 Df = \partial_{\hat{\alpha}} f \, D\hat{\alpha} + \partial_\xi f \, D\xi + \partial_\Lambda f \, D\Lambda \;,
 \label{eq:aH_df}
\end{equation}
where $\partial_{A} f = \partial f / \partial A$.
This leads to the matrix equation
\begin{equation}
 \begin{pmatrix}
  \partial_{\hat{\alpha}} \hat{n} & \partial_\xi \hat{n} & \partial_\Lambda \hat{n} \\
  \partial_{\hat{\alpha}} \hat{e} & \partial_\xi \hat{e} & \partial_\Lambda \hat{e}\\
  \partial_{\hat{\alpha}} \hat{P}_l & \partial_\xi \hat{P}_l & \partial_\Lambda \hat{P}_l
 \end{pmatrix}
 \begin{pmatrix}
  D\hat{\alpha} \\ D\xi \\ D\Lambda
 \end{pmatrix} =
 \begin{pmatrix}
  S_n \\ S_e \\ S_l
 \end{pmatrix} \;,
 \label{aH_system_gen}
\end{equation}
with source functions
\begin{gather}
 S_n = -\frac{\hat{n}}{\tau} - \frac{\hat{n} - n_{\rm eq}}{\tau_R} \;, \quad
 S_e = -\frac{\hat{e} +\hat{P}_l}{\tau}  \;, \nonumber\\
 S_l = -\frac{3\hat{P}_l - \hat{I}^{\rm RS}_{240}}{\tau} - \frac{\hat{P}_l - P}{\tau_R}  \;.
\end{gather}

The derivatives with respect to $\hat{\alpha}$ are simply:
\begin{equation}
 \partial_{\hat{\alpha}} \hat{n} = \hat{n}  \;, \quad
 \partial_{\hat{\alpha}} \hat{e} = \hat{e}  \;, \quad
 \partial_{\hat{\alpha}} \hat{P}_l = \hat{P}_l  \;.
\end{equation}
Taking into account the expression \eqref{aH_n},
the derivatives of $\hat{n}$ with respect to $\xi$ and $\Lambda$ can be obtained as
\begin{equation}
 \frac{\mathrm{d}\hat{n}}{\mathrm{d}\xi} = -\frac{\hat{n}}{2(1+\xi)}  \;, \quad
\frac{\mathrm{d}\hat{n}}{\mathrm{d}\Lambda} = \frac{\hat{n}}{\Lambda} \left[\frac{m_0}{\Lambda} H\left(\frac{m_0}{\Lambda}\right) - 1\right]  \;,
\end{equation}
with $H(z) = K_3(z) / K_2(z)$.
Starting from Eq.~\eqref{eq:RS_quant},
the derivatives of $\hat{e}$ and $\hat{P}_l$ with respect to $\xi$ and $\Lambda$ can be obtained as
\begin{align}
 \begin{pmatrix}
  \partial_\xi \hat{e} \\ \partial_\xi \hat{P}_l
 \end{pmatrix} &= -\frac{ge^{\hat{\alpha}} \Lambda^4}{8\pi^2} \int_{-1}^1
\frac{\mathrm{d}v^z\, v_z^2}{(1 + \xi v_z^2)^3} \begin{pmatrix}
  1 \\ v_z^2
 \end{pmatrix} \nonumber\\
& \times \left[\left(4 + \lambda\right) \Gamma(4,\lambda) - 3\lambda \Gamma(3,\lambda)\right]  \;,\nonumber\\
 \begin{pmatrix}
  \partial_\Lambda \hat{e} \\ \partial_\Lambda \hat{P}_l
 \end{pmatrix} &= \frac{ge^{\hat{\alpha}} \Lambda^3}{4\pi^2} \int_{-1}^1 \frac{\mathrm{d}v^z}{(1 + \xi v_z^2)^2} \begin{pmatrix}
  1 \\ v_z^2
 \end{pmatrix} \nonumber\\
 & \times \left[(4 + \lambda) \Gamma(4,\lambda) - 3 \lambda \Gamma(3,\lambda)\right].
\end{align}
In deriving the above results, we employed the relations:
\begin{gather}
 \frac{\partial \Gamma(4, \lambda)}{\partial \lambda} = 3\Gamma(3,\lambda) - \Gamma(4,\lambda), \nonumber\\
 \partial_\xi \lambda = \frac{v_z^2 \lambda}{2(1 + \xi v_z^2)}, \quad
 \partial_\Lambda \lambda = -\frac{\lambda}{\Lambda}  \;.
\end{gather}

In the case where the particle number is not conserved, $\hat{\alpha} = D \hat{\alpha} = 0$, such that only the equations for $\hat{e}$ and $\hat{P}_l$ have to be taken into account from Eq.~\eqref{aH_system_gen}.
Inverting the matrix on the left-hand side of Eq.~\eqref{aH_system_gen} gives
\begin{equation}
 \begin{pmatrix}
  D\xi \\  D\Lambda
 \end{pmatrix} = \frac{\partial(\xi, \Lambda)}{\partial(\hat{e},\hat{P}_l)}
 \begin{pmatrix}
  \partial_\Lambda \hat{P}_l & -\partial_\Lambda \hat{e} \\
  -\partial_\xi \hat{P}_l & \partial_\xi \hat{e}
 \end{pmatrix}
 \begin{pmatrix}
  S_e \\ S_l
 \end{pmatrix} \;.
\end{equation}
The prefactor appearing on the right-hand side of the above equation represents the Jacobian of the transformation from $(\xi,\Lambda)$ to $(\hat{e}, \hat{P}_l)$:
\begin{equation}
 \frac{\partial(\xi, \Lambda)}{\partial(\hat{e}, \hat{P}_l)} = \frac{1}{\partial_\xi \hat{e}\, \partial_\Lambda \hat{P}_l - \partial_\Lambda \hat{e} \, \partial_\xi \hat{P}_l}\;.
\end{equation}

In the case of particle-number conservation, $\hat{n} = n_{\rm eq}$ and $S_n = -\hat{n}/ \tau$, such that the equation for $\hat{n}$ admits the simple solution
\begin{equation}
 \hat{n}(\tau) = \frac{\tau_0 \hat{n}_0}{\tau} \;.
\end{equation}
We can view this equation as fixing the $\hat{\alpha}$ degree of freedom, which we now seek to eliminate.
Using Eq.~\eqref{aH_system_gen}, $D\hat{\alpha}$ can be obtained as
\begin{equation}
 D\hat{\alpha} = -\frac{1}{\tau} - \frac{\hat{n}_\xi }{\hat{n}} D \xi - \frac{\hat{n}_\Lambda}{\hat{n}} D\Lambda \; .
\end{equation}
With the above, Eq.~\eqref{eq:aH_df} becomes
\begin{equation}
 Df = -\frac{f}{\tau} + \Delta_\xi f\, D \xi + \Delta_\Lambda f \, D \Lambda \;,
\end{equation}
where we introduced the following notation:
\begin{equation}
 \Delta_\xi f = \partial_\xi f - \frac{\partial_\xi \hat{n}}{\hat{n}} f, \quad
 \Delta_\Lambda f = \partial_\Lambda f - \frac{\partial_\Lambda \hat{n}}{\hat{n}} f \;.
\end{equation}
We are now left with two degrees of freedom, $\xi$ and $\Lambda$, whose equations of motion read
\begin{equation}
 \begin{pmatrix}
  D\xi \\  D\Lambda
 \end{pmatrix} = \frac{\Delta(\xi, \Lambda)}{\Delta(\hat{e}, \hat{P}_l)}
 \begin{pmatrix}
  \Delta_\Lambda \hat{P}_l & -\Delta_\Lambda \hat{e} \\
  -\Delta_\xi \hat{P}_l & \Delta_\xi \hat{e}
 \end{pmatrix}
 \begin{pmatrix}
  \Delta S_e \\ \Delta S_L
 \end{pmatrix},
\end{equation}
where
\begin{gather}
 \frac{\Delta(\xi, \Lambda)}{\Delta(\hat{e}, \hat{P}_l)} =
 \frac{1}{\Delta_\xi \hat{e} \, \Delta_\Lambda \hat{P}_l - \Delta_\Lambda \hat{e} \, \Delta_\xi \hat{P}_l}\;, \nonumber\\
 \Delta S_e = -\frac{\hat{P}_l}{\tau}\;, \quad
 \Delta S_L = -\frac{2\hat{P}_l - \hat{I}^{\rm RS}_{420}}{\tau} - \frac{\hat{P}_l - P}{\tau_R}\;.
\end{gather}

%%%
\subsection{Runge-Kutta time integrator} \label{app:num:RK}

In this section we give a brief description of the time-stepping algorithm.
In this paper, we employ the explicit third-order total-variation diminishing (TVD) Runge-Kutta algorithm involving two intermediate stages, see Refs.~\cite{Ambrus:2016aub,Shu:1988,Gottlieb:1998} for details.
For the model equation
\begin{equation}
 \frac{\partial f}{\partial \tau} = L[\tau, f] \;,
\end{equation}
this scheme allows $f$ to be advanced from time step $\tau_n$ to $\tau_{n+1} = \tau_n + \delta \tau$ as follows:
\begin{align}
 f_{n+1} &= \frac{1}{3} f_n + \frac{2}{3} f^{(2)}_n + \frac{2}{3} \delta \tau L\left[\tau_n + \frac{\delta \tau}{2}, f^{(2)}_n\right] \;, \nonumber\\
 f^{(2)}_n &= \frac{3}{4} f_n + \frac{1}{4} f^{(1)}_n + \frac{1}{4} \delta \tau L[\tau_n +\delta \tau, f^{(1)}_n] \;,\nonumber\\
f^{(1)}_n &= f_n + \delta \tau L[\tau, f_n] \;.
\label{eq:RK}
\end{align}
The algorithm applies straightforwardly to systems of equations, in particular to the equations of second-order fluid dynamics, those of leading-order anisotropic hydrodynamics, as well as to the Boltzmann equation written in the form \eqref{eq:boltz_Fnj}.

For the Bjorken-flow simulations considered in this paper, we employed an adaptive time step $\delta \tau_n\equiv \delta\tau(\tau_n)$, determined via
\begin{equation}
 \delta\tau_n \equiv {\rm min} (\alpha_\tau \tau_n, \alpha_R \tau_R) \;,
\end{equation}
where we used $\alpha_\tau = 10^{-3}$ and $\alpha_R = 1/2$ in all considered setups (second-order hydrodynamics, leading-order anisotropic hydrodynamics, kinetic theory).

\subsection{Note on code availability}
\label{app:num:code}

The numerical code, raw data, and scripts to generate the plots shown in this manuscript are available as a code capsule on Code Ocean \cite{RLB_bjork_codeocean}.
Note that a key ingredient in this code is the fast and accurate evaluation of the modified Bessel functions $K_n(z)$, as well as of the Bickley function ${\rm Ki}_1(z)$, which is performed using the algorithms derived by D.E.~Amos \cite{AmosBessel,Amos609}.
We are grateful to OpenSpecfun for providing the AMOS package of functions required for the evaluation of the modified Bessel functions $K_n(z)$.\footnote{Source files downloaded from \texttt{https://github.com/JuliaMath/openspecfun}, commit number \texttt{70239b8d1fe351042ad3321e33ae97923967f7b9}.}

%%%%%%%%%%%%%%% +++ tables with L
\begin{table*}
	\begin{tabular}{|c|c|c|}
		\hline
		& Basis-free coefficients
		& DNMR coefficients  \\ \hline \hline
		& \multicolumn{2}{|c|}{}\\[-7pt]
		$\zeta/\tau_\Pi$
		& \multicolumn{2}{|c|}{${\displaystyle P \frac{z^4}{54} \left[1 - \tfrac{3\pi z}{2} + \left(\tfrac{29}{12} - \tfrac{9}{2} L\right)z^2 - \tfrac{3\pi z^3}{8} + \left(\tfrac{43}{36} - \tfrac{27}{8} L - \tfrac{3}{2} L^2 \right) z^4 \right]}$} \\[7pt]
		$\bar{\zeta}/\tau_\Pi$
		& \multicolumn{2}{|c|}{${\displaystyle P\frac{5z^4}{108} \left[1 - \tfrac{3\pi z}{5} + \left(\tfrac{149}{120} - \tfrac{3}{5} L\right)z^2 - \tfrac{3\pi z^3}{20} + \left(\tfrac{791}{2880} - \tfrac{7}{80} L\right)z^4 \right]}$}  \\[7pt]\hline
		& & \\[-7pt]
		$\delta_{\Pi\Pi}/\tau_\Pi$
		& ${\displaystyle \frac{2}{3} + \frac{\pi z}{4} + \left(\frac{10}{9} + \frac{3\pi^2}{8} + 2L\right)z^2 + \frac{3\pi}{16} (2 + 3\pi^2 + 22L) z^3 }$
		& ${\displaystyle
			\frac{2}{3} - \left(\frac{29}{9} + 2 L\right)z^2 + \frac{5\pi z^3}{3}}$  \\
		& ${\displaystyle - \left(\frac{263}{216} + \frac{\pi^2}{4} - \frac{27\pi^4}{32} - \frac{50 + 189\pi^2}{24} L - 10L^2\right) z^4 }$
		& ${\displaystyle + \left(\frac{445}{216} + \frac{157}{12} L + 6 L^2\right) z^4 }$  \\[7pt]
		$\bar{\delta}_{\Pi\Pi}/\tau_\Pi$
		& ${\displaystyle \frac{2}{3} + \frac{\pi z}{10} + \left(\frac{54\pi^2 - 5}{180} + 2L \right) \frac{z^2}{5} + \left(\frac{24\pi^2 - 135}{200} + L\right) \frac{3\pi z^3}{10}} $
		& ${\displaystyle \frac{2}{3} - \left(\frac{113}{36} + 2L\right)z^2 + \frac{5\pi z^3}{3} }$ \\
		& ${\displaystyle + \left(\frac{1879}{864} - \frac{181\pi^2}{40} + \frac{27\pi^4}{50} + \left(\frac{27\pi^2}{5} - \frac{65}{6} \right) L + 6L^2\right)\frac{z^4}{25} }$
		& ${\displaystyle + \left(\frac{1807}{864} + \frac{79}{6} L + 6L^2 \right) z^4 }$ \\[7pt]
		$\ell_{\Pi V}/\tau_\Pi$
		& ${\displaystyle  -\frac{2z^4 T}{3} \left[\frac{23}{12} + L - \frac{9\pi z}{8} + (193 + 51L) \frac{z^2}{36} - (135 + 48L) \frac{\pi z^3}{32}\right. }$
		& ${\displaystyle \frac{z^2 T}{9} \left[1 + 3(1 + L) z^2 - \frac{5\pi z^3}{4}\right.}$ \\[7pt]
		& ${\displaystyle  \left. + \left(\frac{46543}{432} + 9\pi^2 + \frac{959}{36}L - 15L^2\right) \frac{z^4}{8} \right] }$
		& ${\displaystyle \left. + \left(\frac{263}{24} + 7L\right) \frac{z^4}{4} \right]}$ \\[7pt]
		$\tau_{\Pi V}/\tau_\Pi$
		& ${\displaystyle  \frac{8z^4 T}{3} \left[\frac{23}{12} + L - \frac{9\pi z}{8} + \left(\frac{515}{3} + \frac{113}{2} L\right) \frac{z^2}{24} - (143 + 48L) \frac{3\pi z^3}{64}\right. }$
		& ${\displaystyle -\frac{z^2 T}{9} \left[1 + (13 + 6L) \frac{z^2}{2} - 5\pi z^3\right.}$ \\[7pt]
		& ${\displaystyle  \left. + \left(\frac{42089}{1728} + \frac{63\pi^2}{32} + \frac{805}{144} L - 3L^2 \right) z^4 \right] }$
		& ${\displaystyle \left. + \left(\frac{1259}{24} - 21L\right) \frac{z^4}{4} \right]}$ \\[7pt]
		$\lambda_{\Pi V}/\tau_\Pi$
		& ${\displaystyle  - \frac{T z^2}{12} \left[1 + \left(\frac{125}{8} + 9 L\right)z^2 - 9\pi z^3 + \left(\frac{3531}{4} + 301 L\right) \frac{z^4}{16}\right]}$
		& ${\displaystyle  - \frac{T z^2}{18}\left[1 - \frac{11z^2}{4} + \frac{5\pi z^3}{2} - \left(\frac{407}{4} - 69 L\right) \frac{z^4}{16} \right]}$ \\[7pt]
		$\lambda_{\Pi \pi}/\tau_\Pi$
		& ${\displaystyle -\frac{z^2}{36} \left[1 + \left(\frac{19}{8} + 3L\right) z^2 - \frac{5\pi z^3}{8} + \left(\frac{367}{576} + \frac{41}{16} L\right)z^4 \right]}$
		& ${\displaystyle -\frac{7 z^2}{180} \left[1 + \left(\frac{181}{168} + \frac{15}{7} L\right) z^2 - \left(\frac{163}{2124} - L\right) \frac{295 z^4}{112}\right]}$ \\[7pt]
		$ \bar{\lambda}_{\Pi \pi}/\tau_\Pi$
		& ${\displaystyle \frac{z^2}{18} \left[1 - \frac{5z^2}{8} + \frac{5\pi z^3}{16} - \left(\frac{275}{576} - \frac{7L}{16} \right) z^4 \right]}$
		& ${\displaystyle \frac{2 z^2}{45} \left[1 - \frac{23z^2}{96} - \left(\frac{5}{36} + L\right)\frac{5z^4}{32} \right]}$ \\[7pt] \hline
	\end{tabular}
	\caption{The small-$z$ expansion of the transport coefficients from Eqs.~(\ref{bf_zeta})--(\ref{bf_lambda_Pi_pi}) and Eqs.~(\ref{bf_zeta_bar})--(\ref{bf_lambda_bar_Pi_pi}), with $L = \ln(e^\gamma z / 2)$ and $\gamma \simeq 0.577$ being the Euler-Mascheroni constant.
		The left column corresponds to the expansion of the coefficients obtained using the basis-free method, the right column corresponds to the DNMR coefficients.
		\label{tbl:tcoeffs_Pi_L}}
\end{table*}

\begin{table*}
	\begin{tabular}{|c|c|c|}
		\hline
		& Basis-free coefficients
		& DNMR coefficients \\ \hline \hline
		& \multicolumn{2}{|c|}{}\\[-7pt]
		$\kappa/\tau_V$
		& \multicolumn{2}{|c|}{${\displaystyle \frac{n}{12} \left[1 - \tfrac{13 z^2}{8} + \pi z^3 - \left(\tfrac{131}{64} - \tfrac{19}{16}L \right) z^4\right]}$} \\[7pt]\hline
		& & \\[-7pt]
		$\delta_{VV}/\tau_V$
		& ${\displaystyle 1 + \tfrac{z^2}{2} - \tfrac{\pi z^3}{2} + \left(\tfrac{4}{3} - \tfrac{5L}{4} \right) z^4 }$
		& ${\displaystyle 1 + \tfrac{z^2}{6} + \left(\tfrac{14}{15} + L\right)\tfrac{5 z^4}{12} }$ \\[7pt]
		$\ell_{V \Pi}/\tau_V$
		& ${\displaystyle 3\beta \left[ \tfrac{23}{12} + L + (\tfrac{7}{6} + L) \tfrac{3\pi}{2}z +
			\left(\tfrac{252 \pi^2 - 229}{96} + \tfrac{9\pi^2 + 24}{4} L + \tfrac{9}{2} L^2\right)z^2 \right. }$
		& ${\displaystyle -\frac{11 \beta}{4} \left[1 - \tfrac{12 \pi}{11} z + \left(\tfrac{227}{24} - 56L\right)\tfrac{z^2}{11} \right.}$ \\
		& ${\displaystyle + \left(\tfrac{63\pi^2 - 121}{8} + \tfrac{27\pi^2 + 105}{4} L + 27 L^2\right) \tfrac{\pi}{2} z^3}$
		& ${\displaystyle + (29 + 72L) \tfrac{\pi}{22} z^3}$ \\
		& ${\displaystyle \left. + \left(\tfrac{189 \pi^4 - 545\pi^2}{8} + \tfrac{9925}{576} + \left(\tfrac{27\pi^4 + 142\pi^2}{4} - \tfrac{399}{16}\right) 3L + \tfrac{486\pi^2 + 347}{4} L^2 + 87 L^3\right)\tfrac{z^4}{4}\right]  }$
		& ${\displaystyle \left. + \left(\tfrac{1117}{192} + \tfrac{337}{16} L + 168L^2\right) \tfrac{z^4}{11} \right]}$ \\[7pt]
		$\ell_{V \pi}/\tau_V$
		& ${\displaystyle \frac{\beta z^2}{48} \left[1 + \left(\tfrac{31}{12} + 3L\right)z^2 - \tfrac{5\pi z^3}{4} + \left(\tfrac{877}{72} + L\right) \tfrac{z^4}{8} \right]}$
		& ${\displaystyle \frac{\beta}{20} \left[1 - \tfrac{7z^2}{24} - \left(\tfrac{5}{12} + L\right)\tfrac{5z^4}{16} \right]}$  \\[7pt]
		$\tau_{V \Pi}/\tau_V$
		& ${\displaystyle 3 \beta \left[ \tfrac{35}{12} + L + \left(\tfrac{5}{3} + L\right)3\pi z + \left(\tfrac{972 \pi^2 - 109}{96} + \tfrac{4 + \pi^2}{4} 27 L + \tfrac{27L^2}{2} \right)z ^2 \right.}$
		& ${\displaystyle -\frac{11\beta}{4} \left[1 - \tfrac{24 \pi z}{11} - \left(\tfrac{223}{8} + 168 L\right) \tfrac{z^2}{11} \right.}$ \\
		& ${\displaystyle + \left(\tfrac{153\pi^2 - 137}{8} + \tfrac{53 + 9\pi^2}{2} 3L + 54L^2\right)\pi z^3 +
			\left(\tfrac{1107 \pi^4}{32}  - \tfrac{1873\pi^2}{32} + \tfrac{6533}{2304} \right.
		}$
		& ${\displaystyle + (94 + 144L) \tfrac{\pi z^3}{11} }$ \\
		& ${\displaystyle \left. \left. + \left(\tfrac{405\pi^4}{16} + \tfrac{1551\pi^2}{8} - \tfrac{9623}{192}\right)L + \tfrac{2430 \pi^2 + 2779}{16} L^2 + \tfrac{435 L^3}{4} \right)z^4\right]}$
		& ${\displaystyle \left. + \left(\tfrac{9629}{192} + \tfrac{7057}{16} L + 840L^2\right)\tfrac{z^4}{11}\right] }$ \\[7pt]
		$\tau_{V \pi}/\tau_V$
		& ${\displaystyle \frac{\beta z^2}{8} \left[1 + \tfrac{191 + 216 L}{72} z^2 - \tfrac{5\pi z^3}{4} + \left(\tfrac{719}{432} + \tfrac{L}{3}\right)z^4\right]}$
		& ${\displaystyle \frac{\beta}{20} \left[1 + \tfrac{3z^2}{8} - (37 + 12L) \tfrac{5z^4}{192} \right]}$ \\[7pt]
		$\lambda_{VV}/\tau_V$
		& ${\displaystyle \frac{3}{5} \left[1 + z^2 - \pi z^3 + \left(\tfrac{16}{15} - L\right) \tfrac{5 z^4}{2} \right]}$
		& ${\displaystyle \frac{3}{5} \left[1 + \frac{z^2}{3} + \left(\tfrac{14}{15} +L \right) \tfrac{5z^4}{6} \right]}$ \\[7pt]
		$\lambda_{V \Pi}/\tau_V$
		& ${\displaystyle - \frac{3\beta}{4} \left[
			\tfrac{17}{6} + L + \left(\tfrac{5}{3} + L\right)3\pi z + \left(\tfrac{81 \pi^2}{4} - \tfrac{71}{24} + \tfrac{54\pi^2 + 215}{4} L + 27 L^2 \right) \tfrac{z^2}{2}\right. }$
		& ${\displaystyle \frac{3\beta}{4} \left[1 - 2\pi z - \left(\tfrac{59}{24} + 14 L\right) z^2 \right.}$ \\
		& ${\displaystyle + \left(\tfrac{153\pi^2 - 142}{8} + \tfrac{36\pi^2 + 211}{8} 3L + 54 L^2\right)\pi z^3 +
			\left(\tfrac{1107\pi^4}{32} - \tfrac{3827\pi^2}{64} + \tfrac{1739}{576} \right. }$
		& ${\displaystyle + \left(\tfrac{97}{12} + 12L\right)\pi z^3 }$ \\
		& ${\displaystyle \left. \left. + \left(\tfrac{810 \pi^4 + 6177 \pi^2}{32} - \tfrac{10301}{192} \right)L + \tfrac{810 \pi^2 + 917}{16} 3L^2 + \tfrac{435L^3}{4} \right) z^4 \right]}$
		& ${\displaystyle \left. + \left(\tfrac{2585}{576} + \tfrac{615L}{16} + 70 L^2 \right)z^4\right]}$ \\[7pt]
		$\lambda_{V \pi}/\tau_V$
		& ${\displaystyle \frac{\beta}{16} \left[1 - \tfrac{3z^2}{4} - \left(\tfrac{701}{576} + \tfrac{13L}{8} \right)z^4 \right]}$
		& ${\displaystyle \frac{\beta}{20} \left[1 - \tfrac{3z^2}{8} + \left(\tfrac{59}{192} - \tfrac{L}{16} \right)z^4 \right]}$ \\[7pt]\hline
		%%%
	\end{tabular}
	\caption{The small-$z$ expansion of the transport coefficients from Eqs.~(\ref{bf_kappa})--(\ref{bf_lambda_V_pi}) and Eq.~(\ref{bf_lambda_bar_pi_Pi}).
		The left-column corresponds to the expansion of the coefficients obtained using the basis-free method, the right-column corresponds to the DNMR coefficients.
		\label{tbl:tcoeffs_V}}
\end{table*}

\begin{table*}
	\begin{tabular}{|c|c|c|}
		\hline
		& Basis-free coefficients
		& DNMR coefficients \\ \hline \hline
		& \multicolumn{2}{|c|}{}\\[-7pt]
		$\eta/\tau_\pi$
		& \multicolumn{2}{|c|}{${\displaystyle \frac{4P}{5} \left[1 + \tfrac{z^2}{24} + \left(\tfrac{2}{3} + L\right) \tfrac{z^4}{16} \right]}$} \\[7pt]\hline
		& & \\[-7pt]
		$\delta_{\pi\pi}/\tau_\pi$
		& ${\displaystyle \frac{4}{3}  + \frac{z^2}{36} - \frac{25 z^4}{864}}$
		& ${\displaystyle \frac{4}{3}  + \frac{z^2}{60} - \frac{7z^4}{1440}}$ \\[7pt]
		$\tau_{\pi\pi}/\tau_\pi$
		& ${\displaystyle \frac{10}{7} + \frac{z^2}{21} - \frac{25z^4}{504} }$
		& ${\displaystyle \frac{10}{7} + \frac{z^2}{35} - \frac{z^4}{120} }$ \\[7pt]
		$\lambda_{\pi \Pi}/\tau_\pi$
		& ${\displaystyle \frac{6}{5} \left[1 + \tfrac{\pi z}{4} + \tfrac{28 + 9\pi^2 + 48L}{24} z^2 + (2 + 3\pi^2 + 22L) \tfrac{3\pi z^3}{16} \right.}$
		& ${\displaystyle \frac{6}{5} \left[1 - \left(\tfrac{19}{6} + 2L\right)z^2 + \tfrac{5\pi z^3}{3} \right.}$ \\
		& ${\displaystyle \left. \left(\tfrac{27\pi^4 - 8\pi^2}{32} - \tfrac{85}{72} + \left(\tfrac{63\pi^2}{8} + \tfrac{13}{6} \right)L + 10L^2 \right) z^4 \right]}$
		& ${\displaystyle \left. + \left(\tfrac{151}{72} + \tfrac{79}{6} L + 6 L^2\right)z^4 \right]}$ \\[7pt]
		$\bar{\lambda}_{\pi \Pi}/\tau_\pi$
		& ${\displaystyle \frac{6}{5} \left[1 + \tfrac{\pi z}{10} + \tfrac{9\pi^2 - 5 + 60L}{150} z^2 + \tfrac{24\pi^2 - 135 + 200 L}{2000} 3\pi z^3 \right. }$
		& ${\displaystyle \frac{6}{5} \left[1 - \left(\tfrac{19}{6} + 2L\right)z^2 + \tfrac{5\pi z^3}{3} \right.}$ \\
		& ${\displaystyle + \left. \left(\tfrac{167}{72} - \tfrac{181 \pi^2}{40} + \tfrac{27\pi^4}{50} + \tfrac{162\pi^2 - 325}{30} L + 6L^2\right)\tfrac{z^4}{25} \right]}$
		& ${\displaystyle \left. + \left(\tfrac{151}{72} + \tfrac{79}{6} L + 6 L^2\right)z^4 \right]}$ \\[7pt]
		$\tau_{\pi V}/\tau_\pi$
		& ${\displaystyle -\frac{2z^2}{5\beta} \left[ 1 + \left(\tfrac{7}{4} + L\right) 9z^2 - 9 \pi z^3 + \left(\tfrac{457}{8} + 20L\right) z^4 \right]}$
		& ${\displaystyle -\frac{4z^2}{15 \beta} \left[1 - \tfrac{21z^2}{8} + \tfrac{5 \pi z^3}{2} - \left(\tfrac{429}{8} - 35 L \right) \tfrac{z^4}{8}\right]}$ \\[7pt]
		$\ell_{\pi V}/\tau_\pi$
		& ${\displaystyle  -\frac{2z^2}{5\beta} \left[1 + \left(\tfrac{17}{12} + L\right)3 z^2 - \tfrac{9\pi}{4} z^3 + \left(\tfrac{85}{32} + L \right) 4z^4 \right]}$
		& ${\displaystyle -\frac{4z^2}{15 \beta}\left[1 - \tfrac{7z^2}{8} + \tfrac{5\pi z^3}{8} - \left(\tfrac{97}{64} - \tfrac{7L}{8}\right)z^4 \right]}$ \\[7pt]
		$\lambda_{\pi V}/\tau_\pi$
		& ${\displaystyle  -\frac{z^2}{10\beta} \left[1 + \left(\tfrac{125}{72} + L\right)9 z^2 - 9\pi z^3 + \left(\tfrac{3531}{64} + \tfrac{301 L}{16} \right) z^4 \right]}$
		& ${\displaystyle -\frac{z^2}{15 \beta}\left[1 - \tfrac{11z^2}{4} + \tfrac{5\pi z^3}{2} - \left(\tfrac{407}{64} - \tfrac{69L}{16}\right)z^4 \right]}$ \\[7pt]\hline
	\end{tabular}
	\caption{The small-$z$ expansion of the transport coefficients from Eqs.~(\ref{bf_eta})--(\ref{bf_lambda_pi_V}) and Eq.~(\ref{bf_lambda_bar_pi_Pi}).
		The left-column corresponds to the expansion of the coefficients obtained using the basis-free method, the right-column corresponds to the DNMR coefficients.
		\label{tbl:tcoeffs_pi_L}}
\end{table*}

\begin{table*}
	\begin{tabular}{|c|c|c|}
		\hline
		& Basis-free coefficients
		& DNMR coefficients \\ \hline \hline
		& \multicolumn{2}{|c|}{}\\[-7pt]
		$\delta_{VE} / \tau_V$
		& \multicolumn{2}{|c|}{${\displaystyle \frac{P}{12T^2} \left[1 - \tfrac{13z^2}{8} + \pi z^3 - \left(131- 76 L\right) \tfrac{z^4}{64} \right]}$} \\[7pt]\hline
		& & \\[-7pt]
		$\delta_{\Pi V E}/\tau_\Pi$
		& ${\displaystyle \frac{13z^2}{36} \left[1 - \tfrac{18\pi}{13}z -\tfrac{5(1+360L)}{312} z^2 - \tfrac{63\pi}{52} z^3 \right.}$
		& ${\displaystyle \frac{z^2}{18} \left[1 + \left(\tfrac{109}{12} + 6L\right)z^2 - \tfrac{11\pi}{2} z^3 \right.}$
		\\[7pt]
		& ${\displaystyle \left. - \left(\tfrac{30185}{7488} - \tfrac{18\pi^2}{13} + \tfrac{2567L}{208}\right) z^4 \right]}$
		& ${\displaystyle \left. + \left(\tfrac{7331}{576} - \tfrac{37L}{16} \right) z^4 \right]}$
		\\[7pt]
		$\delta_{V \Pi E}/\tau_V$
		& ${\displaystyle -\frac{2}{m_0^2} \left[1 +
			\tfrac{29 + 6L}{16} z^2 + (1 + L) \tfrac{9\pi}{8} z^3  \right.}$
		& ${\displaystyle -\frac{2}{m_0^2} \left[1 - \tfrac{z^2}{8} + \tfrac{3\pi }{4} z^3 + \left(\tfrac{289}{192} + \tfrac{47}{8}L \right) z^4\right]}$
		\\[7pt]
		& ${\displaystyle \left. + \left(\tfrac{185 + 486 \pi^2}{128} + (175+54\pi^2) \tfrac{3L}{64} + \tfrac{81L^2}{16} \right) z^4 \right]}$
		&
		\\[7pt]
		$\delta_{V\pi E} / \tau_V$
		& ${\displaystyle \frac{\beta^2}{48} \left[1 - \tfrac{23}{12}z^2 + \tfrac{5\pi}{2} z^3 - \left(\tfrac{197}{72} - 69 L \right) \tfrac{z^4}{8}\right] }$
		& ${\displaystyle \frac{\beta^2}{240} \left[1 - \left(\tfrac{7}{4} + L\right) 3z^2 + \frac{5\pi z^3}{2} - \left(\tfrac{215}{8} - 9L\right) \frac{z^4}{8} \right]}$ \\[7pt]
		$\delta_{\pi V E} / \tau_\pi$
		& ${\displaystyle \frac{8}{5} \left[1 + \tfrac{5z^2}{16} - \tfrac{3\pi}{8} z^3 + \left(\tfrac{1}{64} - L\right)\tfrac{3z^4}{2}\right]}$
		& ${\displaystyle \frac{8}{5} \left[1 + \tfrac{z^2}{12} + (13 + 10 L) \tfrac{z^4}{32} \right]}$ \\[7pt]
		$\delta_{V B} / \tau_V$
		& ${\displaystyle \frac{3 \beta}{4} \left[1 + \left(\tfrac{137}{24} + 4L\right) z^2 - 3\pi z^3 + \left(\tfrac{2719}{192} + \tfrac{257 L}{48} \right) z^4 \right]}$
		& ${\displaystyle \frac{5 \beta}{12} \left[1 - \tfrac{53z^2}{40} + \pi z^3 - \left(\tfrac{779}{320} - \tfrac{23L}{16} \right) z^4 \right]}$\\[7pt]
		$\delta_{\pi B} / \tau_\pi$
		& ${\displaystyle \frac{\beta}{2} \left[1 - \tfrac{5z^2}{24} - (23 + 36 L) \tfrac{5z^4}{576} \right]}$
		& ${\displaystyle \frac{2\beta}{5} \left(1 - \tfrac{z^2}{12} + \tfrac{5z^4}{96} \right)}$\\[7pt]
		\hline
	\end{tabular}
	\caption{The small-$z$ expansion of the transport coefficients from Eqs.~(\ref{bf_delta_V_E})--(\ref{bf_delta_pi_B}) arising from the coupling of a charged fluid with external electric and magnetic fields.
		The left column corresponds to the expansion of the coefficients obtained using the basis-free method, while the right-column corresponds to the DNMR coefficients.
		\label{tbl:tcoeffs_mhd}}
\end{table*}
%%%%%%%%%%%%%%% --- tables with L

%%%%

\bibliography{RTA_bulk}

\end{document}